\pdfoutput=1
\documentclass{article}
\usepackage[main, final]{neurips_2026}

\usepackage[utf8]{inputenc}
\usepackage[T1]{fontenc}
\usepackage[pdfpagelabels=false,hypertexnames=false]{hyperref}
\usepackage{url}
\usepackage{booktabs}
\usepackage{amsfonts}
\usepackage{amsmath}
\usepackage{amsthm}
\usepackage{amssymb}
\usepackage{mathtools}
\usepackage{nicefrac}
\usepackage{microtype}
\usepackage{xcolor}
\usepackage{algorithm}
\usepackage{algorithmic}
\usepackage{graphicx}
\usepackage{pdflscape}
\usepackage{multirow}
\usepackage{makecell}

\newtheorem{theorem}{Theorem}[section]
\newtheorem{corollary}[theorem]{Corollary}

\newtheorem{proposition}[theorem]{Proposition}
\newtheorem{definition}[theorem]{Definition}
\newtheorem{assumption}[theorem]{Assumption}
\newtheorem{remark}[theorem]{Remark}

\newcommand{\Mman}{\mathcal{M}}
\newcommand{\Tan}[1]{T_{#1}\Mman}
\newcommand{\Exp}[1]{\mathrm{Exp}_{#1}}
\newcommand{\Log}[1]{\mathrm{Log}_{#1}}
\newcommand{\PT}[2]{\Gamma_{#1 \to #2}}
\newcommand{\FM}[1]{\bar{#1}}%
\DeclareMathOperator{\inj}{inj}
\newcommand{\dM}{d_{\Mman}}
\newcommand{\norm}[1]{\left\|#1\right\|}
\newcommand{\inner}[2]{\langle #1,\, #2\rangle}
\newcommand{\R}{\mathbb{R}}
\newcommand{\E}{\mathbb{E}}
\newcommand{\bb}[1]{\boldsymbol{#1}}
\newcommand{\Var}{\mathrm{Var}}

\DeclareMathOperator*{\argmin}{arg\,min}

\title{Riemannian Ordinary Least Squares}

\author{%
  Xiaoyu Chen$^{1}$\thanks{Corresponding author: \texttt{xchen325@buffalo.edu}.}$^{\phantom{*}}$, Yujing (Zipan) Huang$^{1}$, Yingyan Zeng$^{2}$ \\
  $^{1}$University at Buffalo \quad $^{2}$University of Cincinnati
}

\begin{document}

\maketitle

\begin{abstract}
Many scientific responses take values in Riemannian manifolds, with predictors that are scalar- and/or manifold-valued.
A regression analysis in such settings must provide accurate prediction with hypothesis testing for individual coefficients.
Yet no current manifold-regression framework accommodates both predictor types and enables per-coefficient hypothesis tests.
We propose Riemannian Ordinary Least Squares (ROLS) as a tangent space conditional-mean model, which is the ordinary least squares (OLS) analogue for Riemannian manifolds.
ROLS is well-defined under an explicit injectivity condition that keeps the logarithm maps single-valued, with cut-locus conditions for branch-sensitive cases outside it.
Within the local geometric setting it admits a closed-form estimator, an $O(n^{-1})$ excess prediction-error bound with explicit curvature bias, and asymptotic $t$- and Wald tests for individual coefficients.
With synthetic simulations and three real-data case studies, ROLS is shown to be competitive with the state-of-the-art baselines. In addition, the per-coefficient tests agree with effects reported in literature.
We envision ROLS to be a standard tool for manifold-valued regression with coefficient-level inference to support scientific discovery, which can be interpreted in the same way as Euclidean OLS. The source code is available at \url{https://github.com/XiaoyuChenPub/ROLS}.
\end{abstract}

\section{Introduction}
\label{sec:intro}

In scientific applications, regression analysis is used not only to model predictor-response relationships, but also to determine which covariates have statistically significant effects.
For Euclidean data, ordinary least squares (OLS) is the standard tool, because the estimator is closed form and each coefficient can be tested.
Analogous questions arise when responses lie on Riemannian manifolds, including diffusion tensors in $\mathrm{Sym}^+(3)$ for brain microstructure~\citep{Pennec2006,Brooks2019}, protein backbone dihedral angles on the flat torus $\mathbb T^2$, and wind directions on $S^2$ in geophysics.
In such applications, the covariates are typically heterogeneous, combining manifold-valued and scalar predictors, yet scientific interpretation still requires the same tractable estimation and coefficient-level inference that OLS provides in Euclidean regression.

Existing manifold-regression frameworks each address only a strict subset of these estimation and inference requirements.
Geodesic regression~\citep{Fletcher2013,Hinkle2014}, multivariate general linear models (MGLM) on Riemannian manifolds~\citep{Kim2014MGLM}, and symmetric-space regression~\citep{Shi2009,Cornea2017} provide inference for scalar covariates~\citep{Fletcher2013,Kim2014MGLM,Cornea2017}, but none of them accommodates manifold-valued predictors. Symmetric-space regression further restricts the response to a Riemannian symmetric space, excluding spaces such as the Wasserstein space.
Fr\'echet regression~\citep{PetersenMuller2019frechet,LinMuller2021} and its extensions~\citep{BhattacharjeeMueller2023,LinMuellerPark2023,HanSong2023,ZhangLiXue2024} accommodate general metric-space responses. However, they yield neither closed-form estimators nor individual coefficient tests, aside from specialized $F$-tests for Bures-Wasserstein responses~\citep{XuLi2025}.
Intrinsic Gaussian-process regression~\citep{Mallasto2018} similarly focuses on prediction without providing coefficient-level inference.
Conversely, methods that accept manifold-valued or non-Euclidean predictors~\citep{PetersenMuller2019wasserstein,ZhouMuller2022,Capitaine2024} are either domain-specific or nonparametric. In particular, distribution-in-distribution-out (DIDO) regression~\mbox{\citep{ChenDIDO2025}} is restricted to $\mathcal{P}_2(\R)$ without scalar covariates, while nonlinear Fr\'echet (NL-Fr\'echet) regression~\mbox{\citep{BhattacharjeeLiXue2025}} is purely predictive.
Consequently, no existing framework simultaneously provides (i)~mixed manifold-valued and scalar covariates, (ii)~individual coefficient tests, including tests for manifold-predictor coefficients, and (iii)~explicit injectivity conditions ensuring that the tangent-space linearization is well defined (Table~\ref{tab:method_comparison} summarizes (i) and (ii)).

\begingroup
\setlength{\textfloatsep}{6pt}
\begin{table}[t]
\centering
\caption{Comparison of regression methods for manifold-valued responses.}
\label{tab:method_comparison}
\small
\setlength{\tabcolsep}{3.5pt}
\begin{tabular}{lcccc}
\toprule
Method & \makecell{Manifold \\ predictors} & \makecell{Scalar \\ predictors} & \makecell{Individual \\ coeff. tests} & \makecell{Response} \\
\midrule
Geodesic reg.\ \citep{Fletcher2013}$^{\dagger}$ & $\times$ & \checkmark & \checkmark & General manifolds \\
Polynomial reg.\ \citep{Hinkle2014} & $\times$ & \checkmark & $\times$ & General manifolds \\
MGLM \citep{Kim2014MGLM} & $\times$ & \checkmark & \checkmark & General manifolds \\
\citet{Cornea2017} & $\times$ & \checkmark & \checkmark & Sym.\ sp. \\
Fr\'echet reg.\ \citep{PetersenMuller2019frechet} & $\times$ & \checkmark & $\times$ & Metric sp. \\
Single-idx Fr.\ \citep{BhattacharjeeMueller2023} & $\times$ & \checkmark & $\times$ & Metric sp. \\
Additive mfld.\ \citep{LinMuellerPark2023} & $\times$ & \checkmark & $\times$ & Lie grp. \\
NL Fr\'echet \citep{BhattacharjeeLiXue2025} & \checkmark & \checkmark & $\times$ & Metric sp. \\
DIDO reg.\ \citep{ChenDIDO2025} & \checkmark & $\times$ & $\times$ & $\mathcal{P}_2(\R)$ \\
\midrule
\textbf{Riemannian OLS (this paper)} & \checkmark & \checkmark & \checkmark & General manifolds \\
\bottomrule
\end{tabular}
\\[2pt]
{\scriptsize $^{\dagger}$The tangent-space formulation of geodesic
regression (base point fixed at the Fr\'echet mean) is a special case of
this paper's method with $p{=}0,\,q{=}1$;
this paper additionally supplies an asymptotic Wald test for the
geodesic-regression coefficient.}
\end{table}
\endgroup

To address these gaps, we propose Riemannian Ordinary Least Squares (ROLS), a regression framework based on tangent-space linearization at the Fr\'echet mean.
ROLS lifts each manifold-valued variable to the tangent space at its Fr\'echet mean via the Riemannian logarithm map, parallel transports the lifted manifold predictors to the response tangent space, and fits a linear conditional mean model for the lifted response.
Because the logarithm map and geodesic parallel transport are only locally unique, we formalize explicit support and injectivity conditions in Section~\ref{sec:topo} to separate the well-posed local setting from the branch-sensitive case near the cut locus.
Under these local conditions, the resulting block normal equations admit a closed-form solution, and the linear tangent-space structure yields asymptotic $t$-tests for manifold predictors and Wald tests for scalar predictors (Sections~\ref{sec:model} and~\ref{sec:stats}).
ROLS thus plays the same role for manifold-valued responses that classical OLS plays in Euclidean regression, serving as both an interpretable inferential tool and a natural baseline.

Specifically, this paper makes four contributions:
\begin{itemize}
\item We develop a unified, closed-form regression estimator for manifold-valued responses with mixed manifold and scalar predictors, which recovers tangent-space geodesic regression, DIDO regression, and Euclidean Fr\'echet regression as special cases (Appendix~\ref{sec:S5}).
\item We clarify when tangent-space linearization is geometrically valid by characterizing the local support and injectivity conditions required for estimation and inference, and by quantifying how cut-locus branching compromises identifiability outside this neighborhood (Theorem~\ref{thm:cutlocus}).
\item We enable coefficient-level hypothesis testing via asymptotic $t$- and Wald tests backed by a consistent pairs bootstrap that accounts for estimated Fr\'echet means and transports (Corollary~\ref{cor:bootstrap-validity}). We also establish an $O(n^{-1})$ excess prediction-error bound that explicitly isolates the curvature-induced bias of tangent-space linearization (Theorem~\ref{thm:consistency}).
\item Across five simulated manifolds and three real-data applications on $\mathrm{Sym}^+(3)$, $\mathbb T^2$, and $S^2$, we show that ROLS achieves superior predictive accuracy over state-of-the-art baselines (Tables~\ref{tab:comp-sigma-q} and~\ref{tab:casestudy}) while uncovering domain-consistent covariate effects that purely predictive methods cannot test.
\end{itemize}

\section{Notation and Preliminaries}
\label{sec:prelim}

Throughout the paper, $i=1,\ldots,n$ indexes observations, $j=1,\ldots,p$ indexes manifold-valued covariates, and $k=1,\ldots,q$ indexes scalar covariates.
The response is $Y\in\Mman$, the manifold covariates are $X_1,\ldots,X_p\in\Mman$, and the scalar covariates are $z_1,\ldots,z_q\in\R$.
The response and all manifold-valued covariates are assumed to lie in the same Riemannian manifold $\Mman$, since the model uses parallel transport between their Fr\'echet means.
We write $\FM{Y}$ and $\FM{X}_j$ for population Fr\'echet means, $\hat{\bar Y}$ and $\hat{\bar X}_j$ for their empirical analogues, $\bar z_k=\E z_k$, and $\tilde z_k=z_k-\bar z_k$.
All logarithm, exponential, and transport maps are written with their base points as explicit subscripts.
Norms and inner products are taken in the tangent space indicated by context, usually $T_{\FM{Y}}\Mman$ after linearization.
The ROLS parameter is $\bb\beta=(\alpha_1,\ldots,\alpha_p,v_1,\ldots,v_q)$, with scalar coefficients $\alpha_j\in\R$ for manifold covariates and tangent-vector coefficients $v_k\in T_{\FM{Y}}\Mman$ for scalar covariates.

Let $(\Mman, g)$ be a complete, connected Riemannian manifold of dimension $d$ with geodesic distance $\dM(p,q)$.
The tangent space $\Tan{p}$ at each point $p$ is a $d$-dimensional inner product space under the Riemannian metric $g_p$.
The exponential map $\Exp{p}\colon \Tan{p} \to \Mman$ sends a tangent vector $v$ to the endpoint of the unit-time geodesic starting at $p$ with initial velocity $v$, and its local inverse is the logarithm map $\Log{p}$.
Let $\inj_p$ denote the injectivity radius at $p$ and $\inj(\Mman)=\inf_{p\in\Mman}\inj_p$ the global injectivity radius.
On the geodesic ball $B(p,\inj_p)$, $\Log{p}$ is single-valued and smooth.
The cut locus $\mathrm{Cut}(p)$ is the set where geodesics from $p$ cease to minimize length or cease to be unique; equivalently, it marks the boundary beyond which $\Log{p}$ may become multi-valued.

Parallel transport relates tangent vectors at different base points.
For a curve $\gamma$ from $p$ to $q$, we write $\PT{p}{q}^{\gamma}\colon \Tan{p} \to \Tan{q}$ for parallel transport along $\gamma$ via the Levi-Civita connection; this map is a linear isometry.
When $\dM(p,q)<\min(\inj_p,\inj_q)$ and the minimizing geodesic is unique, we use that geodesic and abbreviate $\PT{p}{q}^{\gamma}$ to $\PT{p}{q}$.
If a unique minimizing geodesic is not available, ROLS requires a fixed transport convention specified before fitting.

For a random variable $X$ on $\Mman$ with finite second moment, the Fr\'echet mean is any minimizer of $p \mapsto \E[\dM^2(X, p)]$~\citep{BhattacharyaPatrangenaru2003,BhattacharyaPatrangenaru2005}.
Under the Afsari condition (the support of $X$ lies in a geodesic ball of radius $\rho < \tfrac12\min(\inj_p,\,\pi/\sqrt{\kappa_+})$ around some center $p$, where $\kappa_+$ is an upper bound on the sectional curvature), the Fr\'echet mean exists uniquely~\citep{Afsari2011} and satisfies a central limit theorem (CLT)~\citep{BhattacharyaLin2017}; we denote it $\FM{X}$.

Curvature controls the distortion between tangent-space and manifold geometry.
Let $\kappa_- \leq \mathrm{Sec}(\Mman) \leq \kappa_+$ denote sectional curvature bounds.
The Hessian and Rauch comparison theorems~\citep{CorderoMcCannSchmuckenschlager2001} introduce two distortion factors (explicit formulas in Appendix~\ref{sec:distortion}): $\eta(\kappa_+, r) \in (0,1]$, which contracts distances on positively curved manifolds, and $\tilde\eta(\kappa_-, r) \geq 1$, which expands distances on negatively curved ones.
On non-negatively curved manifolds ($S^2$, $\mathrm{SO}(3)$), neither factor inflates ($\eta \leq 1$, $\tilde\eta = 1$); on $\mathrm{Sym}^+(k)$ ($\kappa_- < 0$), $\tilde\eta > 1$.

\section{Riemannian Ordinary Least Squares}
\label{sec:model}

We formulate the ROLS model and derive its closed-form estimator, then characterize the cut-locus condition under which the estimator is identifiable, together with the branch-sensitive failure mode.

\subsection{ROLS Model and Estimation}
\label{sec:model-estimation}

\begin{figure}[!htbp]
  \centering
  \includegraphics[width=0.92\textwidth]{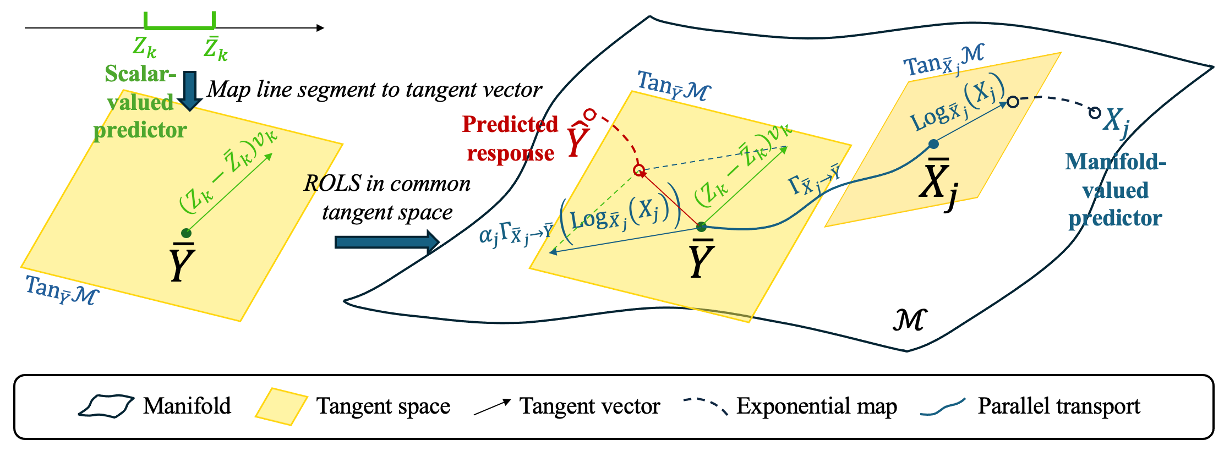}
  \caption{Geometric overview of ROLS.
  Each manifold predictor $X_j$ is log-mapped to $\Tan{\FM{X}_j}$ and parallel-transported to the common tangent space $\Tan{\FM{Y}}$, where it is weighted by the scalar coefficient $\alpha_j$.
  Each scalar predictor $z_k$ contributes a centered tangent direction $(z_k-\bar z_k)\,v_k$.
  The predicted response is $\hat Y=\Exp{\FM{Y}}\bigl(\sum_j\alpha_j u_j+\sum_k(z_k-\bar z_k)v_k\bigr)$, where $u_j$ is the transported predictor.}
  \label{fig:rols-illustration}
\end{figure}

ROLS operates in the tangent space at the response Fr\'echet mean $\Tan{\FM{Y}}$, see the construction in Figure~\ref{fig:rols-illustration}.
For each observation $i$, every manifold-valued covariate is first mapped to the tangent space at its own mean via the logarithm map, then parallel-transported to $\Tan{\FM{Y}}$, giving $u_{ij} \coloneqq \PT{\FM{X}_j}{\FM{Y}}\bigl(\Log{\FM{X}_j}(X_{ij})\bigr) \in \Tan{\FM{Y}}$ for $j=1,\ldots,p$.
The response is lifted to the tangent space as $w_i \coloneqq \Log{\FM{Y}}(Y_i) \in \Tan{\FM{Y}}$.

\begin{definition}[Riemannian Ordinary Least Squares]
ROLS specifies a conditional mean model after lifting all variables to the common tangent space $T_{\FM{Y}}\Mman$:
\begin{equation}\label{eq:model}
  \Log{\FM{Y}}(Y)
  = \sum_{j=1}^{p} \alpha_j \PT{\FM{X}_j}{\FM{Y}}\bigl(\Log{\FM{X}_j}(X_j)\bigr)
  + \sum_{k=1}^{q} (z_k - \bar z_k)\, v_k
  + \varepsilon,
\end{equation}
where $\varepsilon \in \Tan{\FM{Y}}$ is the residual at the population Fr\'echet least-squares (FLS) minimizer.
\end{definition}

\begin{remark}[Special cases]\label{rem:special-cases}
Several regression models are recovered by restricting the spaces or the covariate classes.
In Euclidean space, ROLS becomes a centered vector-valued linear model with scalar block coefficients for vector covariates, and it reduces to ordinary centered multiple linear regression when $d=1$.
With no manifold predictors and one scalar predictor, it gives the tangent-space form of geodesic regression.
Appendix~\ref{sec:S5} gives special cases.
\end{remark}

To estimate the parameters, define the FLS objective as the tangent-space quadratic
\begin{equation}\label{eq:FLS}
  L(\bb{\alpha}, \bb{v})
  \coloneqq \frac{1}{2}\E\!\left[
    \norm{w - m_{\bb\beta}(X,z)}^2_{g_{\FM{Y}}}
  \right],
\end{equation}
where $m_{\bb\beta}(X,z) = \sum_j \alpha_j u_j + \sum_k \tilde{z}_k v_k$ is the conditional mean prediction in $\Tan{\FM{Y}}$.
On flat manifolds $L$ coincides with the manifold Fr\'echet risk $\tfrac{1}{2}\E[\dM^2(Y, \Exp{\FM{Y}}(m_{\bb\beta}(X,z)))]$; on curved manifolds the two are related by a multiplicative factor at most $\tilde\eta(\kappa_-, r_0)^2$, where $r_0$ denotes the data spread $\sup_i \dM(Y_i,\FM{Y})$ (Remark~\ref{rem:bias}).
Since $L$ is quadratic in $(\bb\alpha, \bb v)$, the optimal coefficients satisfy a block linear system analogous to the Euclidean normal equations.
Define the design operator for observation $i$ as the $(p+q)$-tuple $\bb{Z}_i = (u_{i1},\ldots,u_{ip},\, \tilde{z}_{i1}\,\mathrm{id},\ldots,\tilde{z}_{iq}\,\mathrm{id})$, and stack the parameter vector $\bb{\beta} = (\alpha_1,\ldots,\alpha_p, v_1,\ldots,v_q)^\top$.

\begin{theorem}[Generalized Normal Equations on the mixed design space]\label{thm:normal}
Fix an orthonormal frame of $\Tan{\FM{Y}}$ and identify it with $\R^d$.
On the mixed Hilbert design space $\mathcal{H} \coloneqq \R^p \oplus (\Tan{\FM{Y}})^q \cong \R^{p+qd}$, the population FLS minimizer $\bb{\beta}^* = (\alpha_1^*,\ldots,\alpha_p^*, v_1^*,\ldots,v_q^*) \in \mathcal{H}$ satisfies the block normal equations
\begin{equation}\label{eq:normal}
  \Sigma_{\bb{Z}\bb{Z}}\, \bb{\beta}^* = C_{\bb{Z}Y},
\end{equation}
where $\Sigma_{\bb{Z}\bb{Z}} \in \R^{(p+qd)\times(p+qd)}$ is the self-adjoint block Gram matrix
\begin{equation}
  \Sigma_{\bb{Z}\bb{Z}} =
  \begin{pmatrix}
    \bigl[\E\, g_{\FM{Y}}(u_j, u_\ell)\bigr]_{j,\ell=1}^{p} &
    \bigl[\E\,\tilde z_k\, u_j^{\top}\bigr]_{j,k} \\[4pt]
    \bigl[\E\,\tilde z_k\, u_j\bigr]_{k,j} &
    \bigl[\E\,\tilde z_k\tilde z_\ell\, I_d\bigr]_{k,\ell=1}^{q}
  \end{pmatrix},
\end{equation}
and $C_{\bb{Z}Y} \in \R^{p+qd}$ has components $\E\, g_{\FM{Y}}(u_j, w) \in \R$ and $\E\,\tilde z_k\, w \in \R^d$.
When $\Sigma_{\bb{Z}\bb{Z}} \succ 0$, the unique minimizer is $\bb{\beta}^* = \Sigma_{\bb{Z}\bb{Z}}^{-1} C_{\bb{Z}Y}$.
\end{theorem}

The block structure in Theorem~\ref{thm:normal} is the direct analogue of Euclidean multiple regression.
The upper-left block records inner products between transported manifold predictors, the lower-right block is the usual scalar-covariate covariance matrix repeated across tangent coordinates, and the off-diagonal blocks are the cross-covariances between scalar and manifold covariates.
Solving~\eqref{eq:normal} therefore estimates both sets of parameters jointly within a single linear system, adjusting each effect for the collinearity between manifold and scalar predictors.
In this parameterization, each scalar covariate is associated with a directional coefficient vector $v_k \in T_{\FM{Y}}\Mman$, whereas each manifold predictor is weighted by a scalar coefficient $\alpha_j \in \R$ multiplying its transported tangent displacement.
The proof follows by reducing the FLS objective to a quadratic in the Hilbert space $(\Tan{\FM{Y}}, g_{\FM{Y}})$ and differentiating, see Appendix~\ref{sec:S1}.

In practice, $\FM{Y}$ and $\FM{X}_j$ are approximated by empirical Fr\'echet means, and expectations by sample averages.
The point estimator is therefore closed-form.
We compute $\hat{\bar Y}$ and $\hat{\bar X}_j$, transport predictors $\hat u_{ij} = \PT{\hat{\bar X}_j}{\hat{\bar Y}}^{\gamma_j}\bigl(\Log{\hat{\bar X}_j}(X_{ij})\bigr)$, center $\tilde z_{ik}$, log-map $w_i = \Log{\hat{\bar Y}}(Y_i)$, and solve the block normal equations for $\hat{\bb\beta} = (\hat\alpha, \hat v)$.
The maps are evaluated only at the $p+1$ estimated means, so a fit needs $O(np)$ logarithm and transport evaluations (timings are reported in Appendix~\ref{sec:S7-cost}).

\subsection{Geometric Identifiability}
\label{sec:topo}

The basic operations of ROLS are meaningful only when the logarithm maps at the predictor and response Fr\'echet means are single-valued on the observed supports.
In Euclidean space this condition is automatic.
However, on a curved manifold, the logarithm map can become multi-valued at the cut locus. 
So the tangent design may depend on the chosen log branch, and Fr\'echet means near the cut locus can obey non-standard (smeary) limit theorems~\citep{EltznerHuckemann2019}.
The next theorem isolates the geometric condition under which the estimator and statistical theory operate, then records the branch-sensitive failure mode outside it.
See proof in Appendix~\ref{sec:S2}.

\begin{theorem}[Cut-locus identifiability]\label{thm:cutlocus}
Assume $\Sigma_{\bb{Z}\bb{Z}} \succ 0$.

\textup{(a)} If every manifold predictor and the response are supported in strongly convex normal balls around their Fr\'echet means,
\begin{equation}\label{eq:inj-cond}
  \mathrm{supp}(X_j)\subset B(\FM{X}_j,r_j),\quad r_j<\mathrm{inj}_{\FM{X}_j},
  \qquad
  \mathrm{supp}(Y)\subset B(\FM{Y},r_Y),\quad r_Y<\mathrm{inj}_{\FM{Y}},
\end{equation}
then all log maps used by ROLS are single-valued and smooth, and the population FLS objective has a unique minimizer.

\textup{(b)} If a predictor $X_j$ admits two measurable log branches $\ell,\ell'$ on a positive-probability set $S$, write $\bb Z^{(\ell)}$ for the design matrix built using branch $\ell$ in slot $j$ and define $\Sigma^{(\ell)}=\E[\bb Z^{(\ell)}(\bb Z^{(\ell)})^\top]$, $C^{(\ell)}=\E[\bb Z^{(\ell)} W]$ with $W=\Log{\FM Y}(Y)$ (analogously for $\ell'$), $\Delta\Sigma=\Sigma^{(\ell)}-\Sigma^{(\ell')}$, $\Delta C=C^{(\ell)}-C^{(\ell')}$, and $\Delta Z_{ij}=\PT{\FM{X}_j}{\FM{Y}}\{\ell(X_{ij})-\ell'(X_{ij})\}\mathbf 1\{X_{ij}\in S\}$.
Let $P_j^\perp$ denote the projection in $L^2(\Tan{\FM Y})$ orthogonal to the remaining design (Appendix~\ref{sec:S2}).
The branch-induced coefficient gap obeys the exact identity $\bb\beta^*_{(\ell)} - \bb\beta^*_{(\ell')} = (\Sigma^{(\ell')})^{-1}\bigl(\Delta C - \Delta\Sigma\,\bb\beta^*_{(\ell)}\bigr)$.
\end{theorem}

\begin{remark}[Predictively non-trivial gaps]\label{rem:cutlocus-gap}
When $P_j^\perp\Delta Z_j = 0$ in $L^2$, the branch perturbation is absorbed by the remaining covariates, so $\bb\beta^*_{(\ell)}$ and $\bb\beta^*_{(\ell')}$ parameterize the same population predictor even when they differ.
A gap that changes the fitted predictor requires both $P_j^\perp\Delta Z_j \neq 0$ in $L^2$ and $\Delta C - \Delta\Sigma\,\bb\beta^*_{(\ell)} \neq 0$ in $\R^{p+qd}$ (Appendix~\ref{sec:S2}).
\end{remark}

\section{Statistical Properties}
\label{sec:stats}

Under the conditions of Subsection~\ref{sec:topo}, the closed-form ROLS estimator reduces to ordinary least squares in the product tangent space $\R^p\times(T_{\FM{Y}}\Mman)^q$, with three manifold-specific modifications.
Curvature distorts the relation between tangent residuals and geodesic errors, tangent linearization can introduce approximation bias, and the tangent spaces themselves are anchored at empirical Fr\'echet means.
The theory below follows the estimator's order of operations.
First we show that the fitted conditional mean attains the expected parametric prediction behavior.
We then pass from prediction to coefficient error, and finally to the asymptotic distribution needed for individual tests.

\begin{assumption}\label{ass:main}
(i) The Fr\'echet means $\FM{Y}$ and each $\FM{X}_j$ exist uniquely, the variables are supported within the corresponding Afsari ball, and they have finite second moments there. (ii) The injectivity condition \eqref{eq:inj-cond} holds almost surely, and the predictions $\Exp{\FM{Y}}(m_{\hat{\bb\beta}}(X,z))$ lie within radius $r_0$ of $\FM{Y}$. (iii) The block Gram matrix $\Sigma_{\bb Z\bb Z}$ has smallest eigenvalue $\lambda_*>0$, excluding degenerate mixed-design configurations. (iv) The population residual $\varepsilon = w - \sum_j \alpha_j^* u_j - \sum_k \tilde z_k v_k^*$ at the FLS minimizer $\bb\beta^*$ has finite second moment, decomposed as $\E[\norm{\varepsilon}^2] = \sigma^2 + b^2$, where $\sigma^2$ is the irreducible noise variance and $b^2\geq 0$ is the squared tangent-linear misspecification bias.
Under correct specification $\E[\varepsilon\mid X_1,\ldots,X_p,z_1,\ldots,z_q]=0$ (invoked in Theorem~\ref{thm:inference}), $b=0$ and $\sigma^2 = \E[\norm{\varepsilon}^2]$.
Population orthogonality $\E[\bb Z\,\varepsilon]=0$ holds automatically at $\bb\beta^*$.
The observations are independent and identically distributed.
\end{assumption}

\subsection{Consistency and Convergence Rate}
\label{sec:stats-consistency}

The first question is whether the closed-form normal equations in Theorem~\ref{thm:normal} lead to a statistically stable predictor.
In Euclidean OLS, given a full-rank design, the leading prediction-error term is the familiar variance contribution $\sigma^2(p+qd)/n$.
On a manifold the tangent residual and the geodesic residual differ by a Rauch factor, and a nonlinear relation in normal coordinates adds a deterministic bias.
Theorem~\ref{thm:consistency} states the Euclidean rate, the curvature distortion, and the bias term explicitly.

\begin{theorem}[Consistency and convergence rate]\label{thm:consistency}
Under Assumption~\ref{ass:main} and finite fourth moments, let $\bb\beta^*=(\alpha_1^*,\ldots,\alpha_p^*,v_1^*,\ldots,v_q^*)$ be the population FLS minimizer and write $b^2 \coloneqq \E[\|w - \sum_j \alpha_j^* u_j - \sum_k \tilde z_k v_k^*\|^2] - \sigma^2$ for the squared tangent-linear misspecification bias.
For a new observation $(X,z,Y)$, independent of the sample, the empirical FLS estimator satisfies
\[
  \E\bigl[\dM^2(\hat Y,Y)\bigr]
  \le \tilde\eta(\kappa_-,r_0)^2
  \left\{ \frac{C\sigma^2(p+qd)}{n\lambda_*}+\sigma^2+b^2\right\},
\]
where $C$ depends on fourth moments and $\tilde\eta$ is the Rauch distortion factor.
If $b=0$, then $\hat{\bb\beta}_n\xrightarrow{P}\bb\beta^*$ and the estimation-driven excess prediction term is $O((p+qd)/n)$ in expectation.
\end{theorem}

Theorem~\ref{thm:consistency} decomposes the prediction error into estimation variance, curvature distortion, and linearization bias.
The leading term $C\sigma^2(p+qd)/(n\lambda_*)$ matches the Euclidean OLS variance in the mixed product tangent space and depends on the ratio $n/(p+qd)$. The term $b^2$ is the approximation error from higher-order curvature effects.
Ambient curvature scales the squared-error bound by $\tilde\eta^2$. The root mean squared error (RMSE) bound has the factor $\tilde\eta$ from~\eqref{eq:eta-tilde}, and the parametric rate is unchanged. In particular, $\tilde\eta=1$ on flat and non-negatively curved manifolds ($\mathbb T^2$, $S^2$, $\mathrm{SO}(3)$) and remains finite on non-positively curved manifolds ($\mathrm{Sym}^+(3)$, $\mathbb H^3$) under bounded support.

The prediction bound controls the fitted conditional mean.
For inference we also need the coefficients themselves to concentrate in the product tangent space.

\begin{corollary}[Parameter estimation rate]\label{cor:beta-rate}
Under the same conditions, $\|\hat{\bb\beta}_n-\bb\beta^*\|^2=O_P((p+qd)/n)$ in the product tangent space, with constants depending on $\lambda_*$ and the fourth moments $M_4$.
\end{corollary}

\begin{remark}[Curvature distortion and bias]\label{rem:bias}
When the true relationship is genuinely geodesic on the manifold rather than linear in $T_{\FM{Y}}\Mman$, normal-coordinate expansions give a leading curvature correction of order $K r_0^3$ in tangent radius \citep{doCarmo1992,Pennec2006}. The constant $K=\max(|\kappa_-|,\kappa_+)$ bounds the absolute sectional curvature.
Squaring gives $b^2=O(K^2r_0^6)$.
The Afsari support condition therefore keeps the bias controlled by restricting the data to a normal neighborhood where higher-order curvature terms remain small.
On flat manifolds, including $\mathcal P_2(\R)$, $\tilde\eta=1$ and the curvature contribution to $b$ is zero. ROLS then reduces to the Euclidean FLS estimator.
\end{remark}

Corollary~\ref{cor:beta-rate} transfers the prediction-error bound to the parameter level in the product tangent space.
The coefficient tests in the next subsection use this scale.

\subsection{Asymptotic Normality and Wald Tests}
\label{sec:stats-inference}

The final step is coefficient-level inference.
In Euclidean OLS this follows from a CLT for the score.
In ROLS, the score is still the ordinary tangent-space score, but it is evaluated after estimating the Fr\'echet means that define the log maps and transports.
Those means fluctuate at order $n^{-1/2}$, so their influence functions appear in the same limit as the score.
The next theorem records the resulting asymptotic normality and the tests it supports.

\begin{theorem}[Asymptotic normality and statistical inference]\label{thm:inference}
Assume Assumption~\ref{ass:main}, finite fourth moments, correct specification ($b=0$), and $C^2$ dependence of the log and transport maps on the Fr\'echet-mean base points in a geodesic neighborhood.
Assume also that the Fr\'echet functions of $Y$ and $X_j$ are twice continuously differentiable near their means with nonsingular Hessians, so that the empirical Fr\'echet means admit asymptotically linear representations.
Writing $O_i=(Y_i,X_{i1},\ldots,X_{ip},z_{i1},\ldots,z_{iq})$ for the $i$-th observation tuple,
\begin{equation}
  \sqrt n(\hat{\bb\beta}_n-\bb\beta^*) = n^{-1/2}\sum_{i=1}^n \mathrm{IF}(O_i)+o_P(1) \xrightarrow{d} \mathcal N(0,\Sigma_{\bb\beta}), \label{eq:asymp-normal}
\end{equation}
where $\mathrm{IF}$ is the influence function~\eqref{eq:full-if}, $\Sigma_{\bb\beta}=\E[\mathrm{IF}(O)\mathrm{IF}(O)^\top]$, and the difference $\hat v_k - v_k^*$ is taken after parallel-transporting the frame from $\bar Y$ to $\hat{\bar Y}$.
Consequently, $t$-tests for manifold coefficients $H_0:\alpha_j=0$ and Wald $\chi^2_d$ tests for scalar-predictor tangent coefficients $H_0:v_k=0$ are valid for any consistent estimator of $\Sigma_{\bb\beta}$, provided the corresponding scalar variance $[\Sigma_{\bb\beta}]_{jj}$ or tangent-vector block $\Sigma_{\bb\beta}^{kk}$ is nonsingular.
The proof is in Appendix~\ref{sec:S4}.
\end{theorem}

\begin{corollary}[Pairs bootstrap validity]\label{cor:bootstrap-validity}
Under the assumptions of Theorem~\ref{thm:inference}, let $\hat{\bb\beta}_n^{*}$ denote the ROLS estimator computed on a pairs bootstrap resample, including recomputation of the empirical Fr\'echet means, log maps, transports, and block normal equations.
Conditional on the data, $\sqrt n(\hat{\bb\beta}_n^{*}-\hat{\bb\beta}_n) \xrightarrow{d_P} \mathcal N(0,\Sigma_{\bb\beta}),$ and the second-moment matrix of this difference, truncated at a level that grows sufficiently slowly, consistently estimates $\Sigma_{\bb\beta}$.
Thus the resulting standard errors and studentized coefficient tests are asymptotically valid.
\end{corollary}

Appendix~\ref{sec:S4-boot} proves Corollary~\ref{cor:bootstrap-validity} by treating the Fr\'echet means and $\bb\beta$ as one $Z$-estimator. The same truncation produces a covariance estimator for the full $\Sigma_{\bb\beta}$.
Plugging any such covariance into the tests of Theorem~\ref{thm:inference} gives $T_j=\hat\alpha_j/\hat{\mathrm{se}}(\hat\alpha_j)$ for $H_0:\alpha_j=0$ and $W_k=n\hat v_k^\top(\hat\Sigma_{\bb\beta}^{kk})^{-1}\hat v_k$ for $H_0:v_k=0$.
For the standard errors reported in the simulations and Section~\ref{sec:casestudies}, we use the model-based estimator of the tangent-space component of $\Sigma_{\bb\beta}$.
With tangent design $\bb Z_i=[u_{i1},\ldots,u_{ip},\tilde z_{i1}I_d,\ldots,\tilde z_{iq}I_d]^\top\in\R^{(p+qd)\times d}$, Gram matrix $\hat A=\sum_i\bb Z_i\bb Z_i^\top$, and residual $r_i=w_i-\bb Z_i^\top\hat{\bb\beta}$, this estimator is $n\hat\sigma^2\hat A^{-1}$, where $\hat\sigma^2=\sum_i\|r_i\|^2/\nu$ and $\nu=nd-p-qd-d$ removes $d$ degrees of freedom for the estimated response Fr\'echet mean.
Thus $T_j$ is compared with $t_\nu$, while $W_k$ is compared with $\chi^2_d$.
The $\chi^2_d$ approximation over-rejects when $n/d$ is small, because $\hat A$ then estimates a Gram matrix of dimension $p+qd$ (Appendix~\ref{sec:S7-highdim}).
A pairs-bootstrap calibration of these same statistics recomputes the model-based covariance on each resample, and parallel-transports $\hat v_k^{*}$ to $T_{\hat{\bar Y}}\Mman$ before forming $\hat v_k^{*}-\hat v_k$ (pairs variant of Algorithm~\ref{alg:bootstrap}).

\textbf{Misspecification-robust standard errors.}
The covariance $n\hat\sigma^2\hat A^{-1}$ pools the tangent residuals into one variance.
When $b>0$, $\bb\beta^*$ is the best tangent-linear approximation at the Fr\'echet mean and the residuals vary with the design, so this matrix does not estimate their contribution to $\Sigma_{\bb\beta}$.
Under the same tangent equation $w_i=\bb Z_i^\top\bb\beta+\varepsilon_i$, the sensitivity analysis replaces it by the sandwich $n\hat A^{-1}\hat\Omega\hat A^{-1}$, where $\hat\Omega=\sum_i\bb Z_i r_ir_i^\top\bb Z_i^\top$.
The heteroskedasticity-consistent estimator HC3~\citep{MacKinnonWhite1985} forms $\hat\Omega$ from the leverage-adjusted residuals $(I_d-H_{ii})^{-1}r_i$, with $H_{ii}=\bb Z_i^\top\hat A^{-1}\bb Z_i$.
Either matrix is substituted into the same $t_\nu$ and $\chi^2_d$ statistics.
Appendix~\ref{sec:S7-misspec} checks their level under quadratic alternatives, and Appendix Table~\ref{tab:hc3-casestudy} reports the HC3 $p$-values for the case studies.

\section{Simulation Studies}
\label{sec:sims}

We evaluate ROLS on five manifolds, $\mathbb S^2$, $\mathrm{SO}(3)$, $\mathrm{Sym}^+(3)$, the flat torus $\mathbb T^2$, and the Poincar\'e ball $\mathbb H^3$, which include positive, zero, and negative curvature and trivial vs.\ non-trivial topology.
Throughout, we use the canonical metrics, namely the round metric on $\mathbb S^2$, the flat metric on $\mathbb T^2$, the bi-invariant metric on $\mathrm{SO}(3)$, the affine-invariant metric on $\mathrm{Sym}^+(3)$, and the Poincar\'e metric on $\mathbb H^3$. Explicit formulas are summarized in Appendix~\ref{sec:S5}.
Six experiments test, in turn, the cut-locus condition of Theorem~\ref{thm:cutlocus}, the parameter-error rate of Corollary~\ref{cor:beta-rate}, the calibration of the asymptotic tests of Theorem~\ref{thm:inference}, their power, predictive accuracy under varying noise and sample size, and robustness to tangent-space misspecification.
Data follow Model~\eqref{eq:model} with $\alpha$ taken from a fixed alternating-sign pool with leading entries $(0.5,-0.3,0.2,-0.15,\ldots)$, $\|v_k\|=0.2$ in random tangent directions at $\FM Y$, $z_{ik}\stackrel{\mathrm{iid}}{\sim}\mathcal N(0,1)$, and isotropic tangent noise of scale $\tau$.
The default configuration is $n{=}200$, $p{=}2$ manifold-valued and $q{=}3$ scalar predictors, $\tau{=}0.15$.
Prediction error is reported as Fr\'echet RMSE.
See Appendix~\ref{sec:S7} for full data-generating process, including the selected baselines and hyperparameters.

\textbf{Cut-locus condition on $\mathbb S^2$.}
We sweep the data spread $r_0$ over $[0.1, 3.0]$ ($50$ trials per spread, $n=300$).
Test RMSE is stable at $\approx 0.11$ for $r_0 \le 0.8$ and rises sharply once $r_0$ approaches the Karcher convexity radius $\pi/2$, reaching $1.10$ at $r_0$ near the cut locus $\pi$ (Fig.~\ref{fig:headline}(a); detailed sweep in Appendix~Fig.~\ref{fig:cutlocus}).
As $r_0$ approaches $\pi/2$, both the empirical squared-RMSE ratio and $\eta(\kappa_+,r_0)$ from~\eqref{eq:eta} fall toward zero, reflecting the tangent-linearization bias $b^2$ of Theorem~\ref{thm:consistency}.

\textbf{Parameter convergence.}
We regress $\log\|\hat\beta - \beta^*\|$ on $\log n$ over $n \in \{50,100,200,500,1000,2000\}$ ($100$ trials each).
The slopes for $\hat\alpha$ and $\hat v$ on all five manifolds lie in $[-0.54,-0.49]$ with $R^2 \ge 0.98$, consistent with the $-1/2$ rate of Corollary~\ref{cor:beta-rate} (Fig.~\ref{fig:headline}(b); per-manifold slopes in Appendix Table~\ref{tab:conv-slopes}).
On $\mathbb H^3$ the fit uses the Lorentz model with a converged Fr\'echet mean (Appendix~\ref{sec:S7-dgp}).

\textbf{Calibration of the asymptotic tests.}
Across $n \in \{50,100,200,500\}$ and both null hypotheses, all $40$ Type~I rejection rates lie in $[0.026, 0.068]$ at the nominal $0.05$ level ($500$ trials per cell; Fig.~\ref{fig:headline}(c) summarizes power and Type~I error, and Appendix Fig.~\ref{fig:typeI-tau015} gives the full sample-size grid).
The pairs-bootstrap test stays at or below $0.063$ in all $20$ settings of Appendix Table~\ref{tab:bootstrap}.

\textbf{Power.}
The $t$-test for $H_0\colon \alpha=0$ saturates at $\geq 96\%$ rejection by $|\alpha|=0.06$ on $\mathbb T^2$, $\mathrm{SO}(3)$, and $\mathrm{Sym}^+(3)$, by $|\alpha|=0.10$ on $\mathbb S^2$, and requires $|\alpha|\approx 0.25$ on $\mathbb H^3$.
The gap on $\mathbb H^3$ comes from the Poincar\'e conformal factor at the response base point, which inflates the effective tangent-space noise by roughly $2\times$ relative to the predictor variation.
The Wald test for $H_0\colon v=0$ saturates at $\geq 99\%$ by $\|v\|=0.06$ on all five manifolds because the model is linear in $v$ and this coefficient is unaffected by curvature.

\begin{figure}[!htbp]
  \centering
  \includegraphics[width=\textwidth]{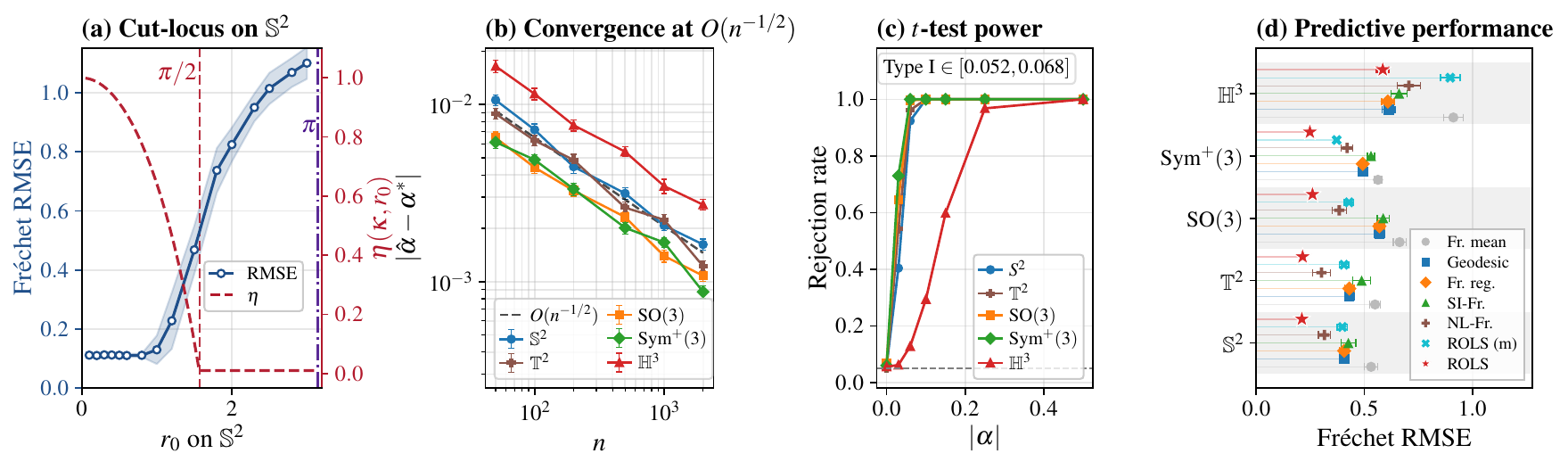}
  \caption{Simulation summary. (a)~Cut-locus on $\mathbb S^2$: test RMSE vs.\ data spread $r_0$; the dashed curve is $\eta(\kappa_+,r_0)$ from~\eqref{eq:eta}. (b)~Parameter convergence: $|\hat\alpha-\alpha^*|$ vs.\ $n$ on five manifolds ($\mathbb H^3$ in the Lorentz model); dashed line is the $O(n^{-1/2})$ rate of Corollary~\ref{cor:beta-rate}. (c)~$t$-test power for $H_0\colon\alpha=0$; dashed line is the nominal $\alpha=0.05$; inset shows the Type~I range at $|\alpha|=0$, $n=200$. (d)~RMSE at the default cell $(n,p,q,\tau){=}(200,2,3,0.15)$ for six baselines on five manifolds; ROLS shown as a red star.
  Full sweeps and per-cell numbers in Appendix~\ref{sec:S7}.}
  \label{fig:headline}
\end{figure}

\textbf{Predictive accuracy.}
We benchmark against six baselines (Fr\'echet mean, geodesic, Fr\'echet regression, single-index Fr\'echet (SI-Fr.), NL-Fr., and manifold-only ROLS\,(m)) across configurations varying $\tau$, $q$, $n$, and $p$ on each manifold, with $100$ trials each (baselines and full grid in Appendix~\ref{sec:S7}).
ROLS has the lowest mean RMSE in $44$ of the $45$ configurations, with a median reduction of $28\%$ over the strongest baseline, which is NL-Fr\'echet in $27$ configurations (Fig.~\ref{fig:headline}(d); sweeps in Appendix~Fig.~\ref{fig:sim-summary}, full table in Appendix Table~\ref{tab:comp-sigma-q}).
The exception is $\mathbb H^3$ at $\tau=0.30$, where Fr\'echet regression is $0.9\%$ lower. At $q=0$, ROLS coincides with ROLS\,(m).
The tangent linear-ridge ablation produces RMSE comparable to ROLS (median $1.2\%$ higher, range $-1.5\%$ to $6.9\%$ across the $45$ cells), so the scalar $\alpha_j$ coefficients show no loss relative to a full $d\times d$ map.

\textbf{Robustness under tangent-space misspecification.}
We add a quadratic distortion $\gamma\|u_i\|^2 \hat n$ to the $\mathbb S^2$ data-generating process.
ROLS reaches RMSE $0.070$ at $\gamma=0$ (a $68\%$ reduction over NL-Fr\'echet's $0.218$) and retains a $31\%$ reduction over NL-Fr\'echet at $\gamma=0.6$ ($0.300$ vs.\ $0.438$).
In this design the omitted term contributes a squared bias of order $\gamma^2$, which is the term $b^2$ in Theorem~\ref{thm:consistency}.
Under this misspecification the model-based test of $\alpha=0$ over-rejects, up to $0.36$ at $\gamma=0.6$. The sandwich test stays at or below $0.084$, and a targeted lack-of-fit diagnostic detects the departure (Appendix~\ref{sec:S7-misspec}).
Base-point sensitivity, tangent dimension up to $d=55$, and computational cost are reported in Appendices~\ref{sec:S7-base}--\ref{sec:S7-cost}.

\section{Real-Data Case Studies}
\label{sec:casestudies}

The real-data studies test whether the simulation findings extend to scientific prediction and inference tasks with mixed covariates and manifold-valued responses.
We evaluate ROLS on three public datasets on three manifolds.
Diffusion tensor imaging of the corpus callosum (DTI-CC, $\mathrm{Sym}^+(3)$) data are extracted from the Autism Brain Imaging Data Exchange (ABIDE-II)~\citep{DiMartino2014}, protein backbone dihedral angles ($\mathbb T^2$) from the Protein Data Bank (PDB)~\citep{Berman2000}, and $850$\,hPa winds ($S^2$) from the National Centers for Environmental Prediction/National Center for Atmospheric Research (NCEP/NCAR) Reanalysis~1~\citep{Kalnay1996}.

We compare the same method classes used in simulation and report Fr\'echet RMSE under cross-validation (CV).
The splitting scheme respects the sampling structure of each dataset, with subject-blocked folds for diffusion tensors, repeated folds for protein residues, and blocked temporal folds for wind.
Dataset construction and preprocessing are reported in Appendix~\ref{sec:S8}.
Table~\ref{tab:casestudy} summarizes the prediction results and Table~\ref{tab:fulldata_inference} the primary coefficient-level inference.

\begin{table}[h]
\centering
\caption{Real-data case studies: Fr\'echet RMSE (mean $\pm$ std across
CV folds). \textbf{Bold}: lowest mean RMSE.}
\label{tab:casestudy}
\scriptsize
\setlength{\tabcolsep}{2pt}
\begin{tabular}{ll@{\hskip 5pt}cccc@{\hskip 8pt}ccc}
\toprule
 & & \multicolumn{4}{c}{\textit{Scalar-only baselines}}
 & \multicolumn{3}{c}{\textit{Manifold-aware methods}} \\
\cmidrule(lr){3-6} \cmidrule(lr){7-9}
\textbf{Dataset} & \textbf{Manifold}
 & \textbf{Fr.\ mean}
 & \textbf{Geodesic}
 & \textbf{Fr.\ reg}
 & \textbf{SI-Fr.}
 & \textbf{NL-Fr.}
 & \textbf{ROLS\,(m)}
 & \textbf{ROLS} \\
\midrule
DTI-CC & $\mathrm{Sym}^+(3)$
 & $1.138_{\pm.011}$
 & $1.129_{\pm.012}$
 & $1.130_{\pm.012}$
 & $1.128_{\pm.012}$
 & $1.134_{\pm.012}$
 & $\mathbf{.400_{\pm.017}}$
 & $\mathbf{.400_{\pm.017}}$ \\
Protein backbone & $\mathbb T^2$
 & $1.779_{\pm.009}$
 & $1.215_{\pm.014}$
 & $1.357_{\pm.014}$
 & $1.220_{\pm.014}$
 & $1.494_{\pm.014}$
 & $1.395_{\pm.016}$
 & $\mathbf{1.177_{\pm.016}}$ \\
Wind & $S^2$
 & $.546_{\pm.114}$
 & $.447_{\pm.040}$
 & $.462_{\pm.071}$
 & $.456_{\pm.041}$
 & $.414_{\pm.078}$
 & $.098_{\pm.009}$
 & $\mathbf{.095_{\pm.005}}$ \\
\bottomrule
\end{tabular}
\end{table}

ROLS improves on the strongest scalar-only baseline and the manifold-aware NL-Fr\'echet baseline across all three datasets (Table~\ref{tab:casestudy}) and provides individual coefficient tests that the nonparametric baselines do not.
Per-dataset numbers are in Appendix~\ref{sec:S8}.

\begin{figure}[h]
  \centering
  \includegraphics[width=\textwidth]{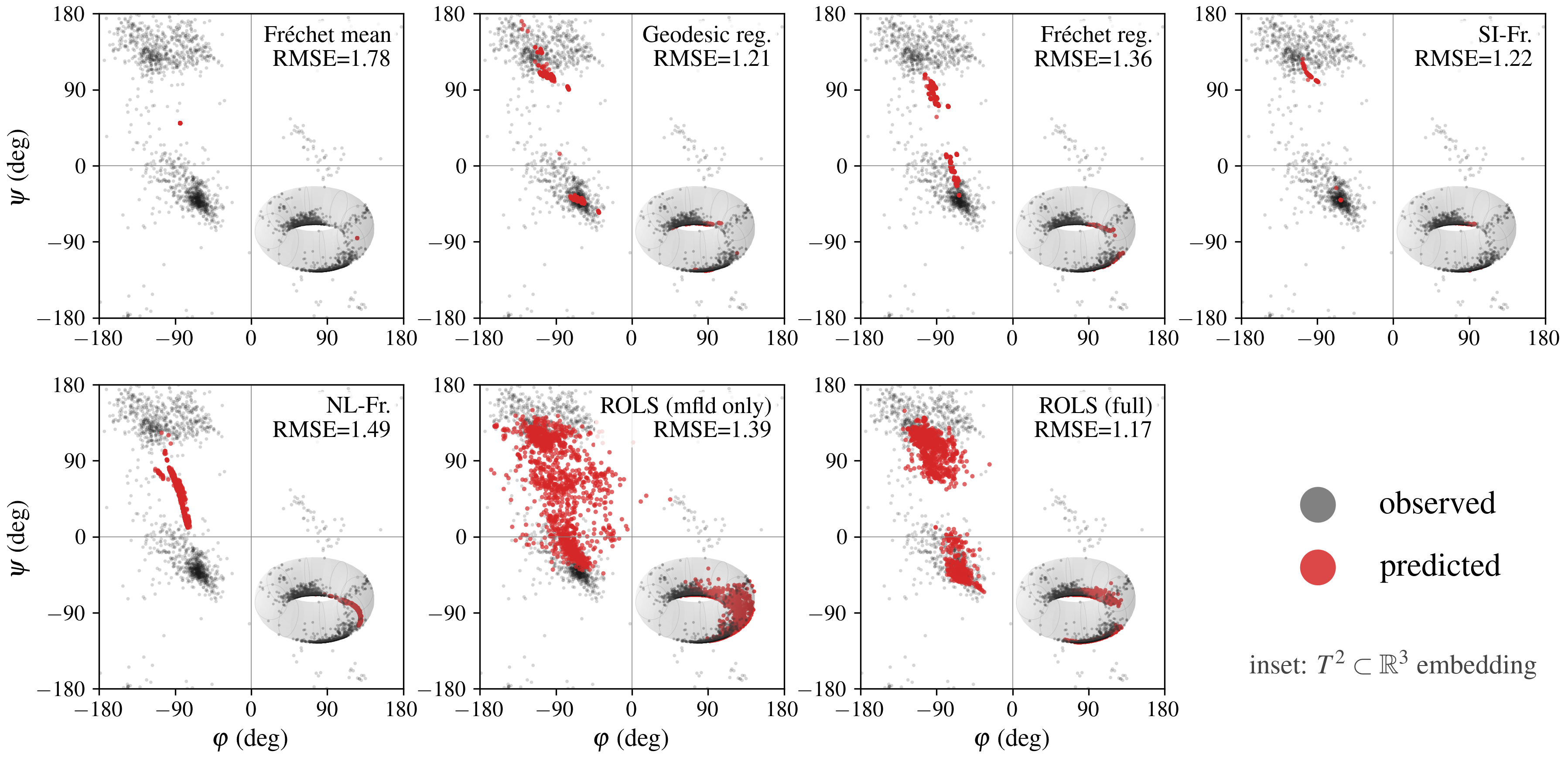}
  \caption{Ramachandran predictions for seven estimators on PDB.
  Each panel overlays predicted (red) and observed (grey) $(\varphi,\psi)$ on the flat torus $[-180^\circ,180^\circ]^2$, with $T^2\subset\R^3$ in the inset. }
  \label{fig:protein_ramachandran_main}
  \end{figure}

Figure~\ref{fig:protein_ramachandran_main} shows how the estimators differ on the protein-backbone torus, where the Ramachandran plot is the canonical check.
Observed $(\varphi,\psi)$ pairs form two geometrically distant basins, the $\alpha$-helix near $(-60^\circ,-45^\circ)$ and the $\beta$-sheet near $(-120^\circ,+120^\circ)$, so a useful predictor has to place each residue in the correct basin.
Scalar-only baselines collapse to a point or a narrow cluster.
NL-Fr\'echet places predictions across the $\varphi<0$ band because the kernel averages residues from both basins, and manifold-only ROLS spreads predictions over the $\varphi<0$ half-plane without separating the two basins.
Full ROLS is the only estimator that recovers separate $\alpha$- and $\beta$-clusters in both basins. This is consistent with complementary information in the neighboring dihedrals and the scalar descriptors.

Beyond prediction, the coefficient tests in Table~\ref{tab:fulldata_inference} agree with effects reported in the literature.
In DTI-CC, the along-tract tensor neighbors and the genu-versus-body contrast are significant, in agreement with the anterior-posterior microstructural gradient of the corpus callosum~\citep{HoferFrahm2006}.
Autism spectrum disorder (ASD) diagnosis, age, sex, and full-scale intelligence quotient (FIQ) are not significant once subject-level clustering is accounted for, consistent with later quality-controlled reanalyses that revised early autism-DTI claims downward~\citep{Koldewyn2014}.
On the protein backbone, $(\varphi,\psi)$ at offsets $-2,-1,+1$ dominate, recovering the local-context sequence-structure rule used in fragment-based folding~\citep{Simons1997}.
The significance of the first Kidera principal component (PC1) agrees with the hydrophobicity-driven secondary-structure preference of~\citet{ChouFasman1974}, as quantified by the leading factor of~\citet{Kidera1985}.
The designed null, Kidera PC3, remains non-significant.
In the wind study, the three neighboring grid points dominate, consistent with synoptic transport, and sea-level pressure is not significant.
Where the tangent-linear model does not fit exactly, $\alpha_j$ and $v_k$ are the coefficients of the best-fitting tangent-linear approximation at the Fr\'echet mean.
The lack-of-fit diagnostic in Appendix~\ref{sec:S7-misspec} attributes about $1.6\%$ of the residual variation on the protein data and $10\%$ on DTI-CC to omitted quadratic terms.
Every manifold-predictor coefficient has $p<10^{-5}$ under both the model-based and the heteroskedasticity-robust Wald tests, and Benjamini--Hochberg adjustment changes no conclusion (Appendix~\ref{sec:S8}).

\begin{table}[h]
  \centering
  \caption{Coefficient-level tests for three case studies.}
  \label{tab:fulldata_inference}
  \newsavebox{\inferencetab}%
  \sbox{\inferencetab}{%
    \footnotesize
    \setlength{\tabcolsep}{2.5pt}%
    \begin{tabular}{@{}lcc@{\hskip 12pt}lcc@{\hskip 12pt}lcc@{}}
    \toprule
    \multicolumn{3}{c}{\textbf{DTI-CC}}
     & \multicolumn{3}{c}{\textbf{Protein backbone}}
     & \multicolumn{3}{c}{\textbf{Wind}} \\
    \cmidrule(r){1-3} \cmidrule(lr){4-6} \cmidrule(l){7-9}
    Predictor & Type & $p$
     & Predictor & Type & $p$
     & Predictor & Type & $p$ \\
    \midrule
    Tensor neighbor 1 & Manifold & $0.002$
     & Offset $-2$ & Manifold & $<\!10^{-5}$
     & Grid point 1 & Manifold & $<\!10^{-5}$ \\
    Tensor neighbor 2 & Manifold & $0.002$
     & Offset $-1$ & Manifold & $<\!10^{-5}$
     & Grid point 2 & Manifold & $<\!10^{-5}$ \\
    CC subregion & Scalar & $0.002$
     & Offset $+1$ & Manifold & $<\!10^{-5}$
     & Grid point 3 & Manifold & $<\!10^{-5}$ \\
    Age & Scalar & $0.301$
     & Kidera PC1 & Scalar & $<\!10^{-5}$
     & Pressure & Scalar & $0.734$ \\
    Sex & Scalar & $0.291$
     & Kidera PC2 & Scalar & $<\!10^{-5}$
     & Temperature & Scalar & $<\!10^{-5}$ \\
    ASD diagnosis & Scalar & $0.549$
     & Kidera PC3 & Scalar & $0.345$
     & Season & Scalar & $0.008$ \\
    FIQ & Scalar & $0.447$
     & $\alpha$-helix & Scalar & $<\!10^{-5}$
     & & & \\
     & & & Solvent-accessibility & Scalar & $2.8{\times}10^{-4}$ & & & \\
    \bottomrule
    \end{tabular}%
  }%
  \usebox{\inferencetab}\\[2pt]
  \parbox{\wd\inferencetab}{\scriptsize\raggedright
  $p$: subject-blocked permutation (DTI-CC), Wald test on the full-data fit (protein), Benjamini--Hochberg-adjusted Wald test (wind).\par}
\end{table}
\section{Conclusions}
\label{sec:conclusion}

ROLS is a closed-form tangent-space analogue of Euclidean OLS for manifold-valued responses with mixed manifold and scalar predictors.
Under explicit geometric regularity conditions, ROLS admits parametric-rate estimation and individual coefficient tests, and the cut-locus analysis identifies when the tangent-space design is well defined.
Across simulations and three case studies, ROLS shows superior predictive performance and enables coefficient-level inference.

This study opens many future directions.
First, the scalar coefficients can be extended to a matrix- or operator-valued linear map, so that a transported tangent vector can be rotated for directional estimation.
Second, ROLS restricts manifold predictors to lie on the response manifold.
Allowing predictors from different manifolds would make ROLS applicable in more complex case studies.
For example, \citet{ChoiPark2025} use a Hilbert--Schmidt operator to map predictors from other manifolds to the response manifold.
Third, the exponential, logarithm, and parallel-transport maps can be replaced by a second-order retraction and a vector transport, which applies to manifolds where those maps are unavailable, including tensor responses of fixed multilinear rank~\citep{Kressner2014}.
Fourth, strongly separated modes, or modes that cross the cut locus, call for cluster-specific centers or a mixture of local ROLS models.
Fifth, the present theory fixes the dimension, and sparse designs in which $d$ grows with $n$ remain open.
Finally, each new batch now requires a full refit.
Recursive updates are available for Fr\'echet means~\citep{Ho2013,Cheng2016}, and an online version of ROLS would need a separate analysis.

\begin{ack}
This work was in part supported by the National Science Foundation [CAREER Award CMMI-2543434, EAGER Award CMMI-2430998] and the American Heart Association Collaborative Science [Award 23CSA1052735].
\end{ack}

\bibliographystyle{plainnat}
\bibliography{neurips2026}

\newpage
\appendix

This appendix collects the supporting material in the following order.
Appendix~\ref{sec:distortion} records the Hessian and Rauch comparison factors used throughout the proofs.
Appendices~\ref{sec:S1}--\ref{sec:S4} contain the complete proofs of Theorems~\ref{thm:normal}--\ref{thm:inference}. Appendix~\ref{sec:S4-boot} proves the pairs bootstrap by embedding the Fr\'echet means and $\bb\beta$ in one $Z$-estimator.
Appendix~\ref{sec:S5} derives the special cases summarized in Remark~\ref{rem:special-cases} and records the manifold-specific operations used in the experiments.
Appendix~\ref{sec:S6} gives the fixed-base residual bootstrap as a within-tangent-space diagnostic.
Appendix~\ref{sec:S7} provides additional simulation results, and Appendix~\ref{sec:S8} provides the full real-data case-study details.
We use the same notation as the main text throughout.
Standard references for Riemannian geometry are \citet{doCarmo1992}, \citet{CheegerEbin1975}, and \citet{Berger2003}.

\section{Curvature Distortion Factors}
\label{sec:distortion}

The Hessian comparison theorem for $\tfrac{1}{2}\dM^2(p,\cdot)$ \citep{CorderoMcCannSchmuckenschlager2001} yields the contraction factor
\begin{equation}\label{eq:eta}
  \eta(\kappa_+, r) =
  \begin{cases}
    \sqrt{\kappa_+}\,r\,/\,\tan(\sqrt{\kappa_+}\,r) & \kappa_+ > 0,\\
    1 & \kappa_+ \leq 0,
  \end{cases}
\end{equation}
defined for $r < \pi/(2\sqrt{\kappa_+})$ when $\kappa_+ > 0$ and for all $r \geq 0$ otherwise.
This equals the minimum eigenvalue of $\mathrm{Hess}(\tfrac{1}{2}\dM^2(p,\cdot))$ at distance $r$, with $\eta \in (0,1]$.
The Rauch comparison theorem gives the expansion factor
\begin{equation}\label{eq:eta-tilde}
  \tilde\eta(\kappa_-, r) =
  \begin{cases}
    \sinh(\sqrt{|\kappa_-|}\,r)\,/\,(\sqrt{|\kappa_-|}\,r) & \kappa_- < 0,\\
    1 & \kappa_- \geq 0,
  \end{cases}
\end{equation}
with $\tilde\eta \geq 1$.

\section{Proof of the Generalized Normal Equations (Theorem~\ref{thm:normal})}
\label{sec:S1}

We restate the theorem for convenience.

\medskip
\noindent\textbf{Theorem~\ref{thm:normal}.}
\textit{Fix an orthonormal frame of $\Tan{\FM{Y}}$ and identify it with $\R^d$.
On the mixed Hilbert design space $\mathcal{H} \coloneqq \R^p \oplus (\Tan{\FM{Y}})^q \cong \R^{p+qd}$, the population FLS minimizer $\bb{\beta}^* = (\alpha_1^*,\ldots,\alpha_p^*,v_1^*,\ldots,v_q^*) \in \mathcal{H}$ satisfies the block normal equations $\Sigma_{\bb{Z}\bb{Z}}\,\bb{\beta}^* = C_{\bb{Z}Y}$, where $\Sigma_{\bb{Z}\bb{Z}} \in \R^{(p+qd)\times(p+qd)}$ is the self-adjoint block Gram matrix with blocks $\bigl[\E\,g_{\FM{Y}}(u_j,u_\ell)\bigr]_{j,\ell}$, $\bigl[\E\,\tilde z_k\,u_j^{\top}\bigr]_{j,k}$, $\bigl[\E\,\tilde z_k\,u_j\bigr]_{k,j}$, $\bigl[\E\,\tilde z_k\tilde z_\ell\,I_d\bigr]_{k,\ell}$, and $C_{\bb{Z}Y}$ has entries $\E\,g_{\FM{Y}}(u_j,w)$ and $\E\,\tilde z_k\,w$.
When $\Sigma_{\bb{Z}\bb{Z}} \succ 0$, the unique minimizer is $\bb{\beta}^* = \Sigma_{\bb{Z}\bb{Z}}^{-1} C_{\bb{Z}Y}$.}

\begin{proof}
Recall that $w = \Log{\FM{Y}}(Y) \in \Tan{\FM{Y}}$ is the lifted response and $\hat{w} = \sum_{j=1}^p \alpha_j u_j + \sum_{k=1}^q \tilde{z}_k v_k$ is the model prediction in $\Tan{\FM{Y}}$.
The manifold Fr\'echet risk is $\frac{1}{2}\E[\dM^2(Y,\,\Exp{\FM{Y}}(\hat{w}))]$.
Under the injectivity condition~\eqref{eq:inj-cond}, the logarithm map $\Log{\FM{Y}}$ is a diffeomorphism on the geodesic ball $B(\FM{Y},\inj_{\FM{Y}})$, and every observation $Y_i$ and every prediction $\Exp{\FM{Y}}(\hat{w}_i)$ lie in this ball.
Because the ROLS model \eqref{eq:model} postulates a tangent-space linear relationship, the natural working objective is the tangent-space quadratic
\begin{equation}
  L(\bb{\alpha},\bb{v})
  \coloneqq \frac{1}{2}\E\bigl[\norm{w - \hat{w}}^2_{g_{\FM{Y}}}\bigr].
\end{equation}
On a flat manifold this equals $\frac{1}{2}\E[\dM^2(Y,\Exp{\FM{Y}}(\hat{w}))]$ exactly.
On a curved manifold the two objectives differ by a factor controlled by the Rauch comparison theorem (see Theorem~\ref{thm:consistency} and its proof in Appendix~\ref{sec:S3}).
The manifold prediction error is bounded above by $\tilde\eta(\kappa_-, r_0)^2$ times $2L$.
Since $L$ is a quadratic functional in $(\bb{\alpha},\bb{v})$, the normal equations derived below are exact for $L$.

Since $\Tan{\FM{Y}}$ equipped with $g_{\FM{Y}}$ is a finite-dimensional real Hilbert space, the objective $L$ is a quadratic functional in the parameters $(\bb{\alpha}, \bb{v})$.
We compute its derivatives by differentiating under the expectation sign, which is justified by the square-integrability assumption $\E[\norm{w}^2] < \infty$.

\medskip
We begin by differentiating $L$ with respect to the scalar coefficients.
For any fixed $j^* \in \{1,\ldots,p\}$,
\begin{align}
  \frac{\partial L}{\partial \alpha_{j^*}}
  &= \frac{\partial}{\partial \alpha_{j^*}}
     \frac{1}{2}\E\bigl[\norm{w - \hat{w}}^2_{g_{\FM{Y}}}\bigr] \notag\\
  &= \E\bigl[g_{\FM{Y}}\!\bigl(w - \hat{w},\;
     -\tfrac{\partial \hat{w}}{\partial \alpha_{j^*}}\bigr)\bigr] \notag\\
  &= -\E\bigl[g_{\FM{Y}}\!\bigl(w - \hat{w},\; u_{j^*}\bigr)\bigr],
\end{align}
since $\frac{\partial \hat{w}}{\partial \alpha_{j^*}} = u_{j^*}$.
Setting this to zero and expanding $\hat{w}$:
\begin{align}
  0 &= \E\bigl[g_{\FM{Y}}\!\bigl(w - \hat{w},\; u_{j^*}\bigr)\bigr] \notag\\
    &= \E\bigl[g_{\FM{Y}}(w, u_{j^*})\bigr]
       - \E\!\left[g_{\FM{Y}}\!\left(\sum_{j=1}^p \alpha_j u_j
         + \sum_{k=1}^q \tilde{z}_k v_k,\; u_{j^*}\right)\right].
\end{align}
Using bi-linearity of $g_{\FM{Y}}(\cdot,\cdot)$ and linearity of expectation:
\begin{align}
  \E\bigl[g_{\FM{Y}}(w, u_{j^*})\bigr]
  &= \sum_{j=1}^p \alpha_j \E\bigl[g_{\FM{Y}}(u_j, u_{j^*})\bigr]
   + \sum_{k=1}^q \E\bigl[\tilde{z}_k\, g_{\FM{Y}}(v_k, u_{j^*})\bigr].
   \label{eq:S1-deriv-alpha}
\end{align}
The coefficient $v_k \in \Tan{\FM{Y}}$ does not depend on the observation, so $\E[\tilde{z}_k\, g_{\FM{Y}}(v_k, u_{j^*})] = g_{\FM{Y}}(v_k,\, \E[\tilde{z}_k\, u_{j^*}]) = g_{\FM{Y}}(v_k,\, \mathrm{cov}(z_k, X_{j^*}))$.

Identifying $\mathrm{cov}(X_j, X_{j^*}) = \E[g_{\FM{Y}}(u_j, u_{j^*})]$, $\mathrm{cov}(X_{j^*}, Y) = \E[g_{\FM{Y}}(u_{j^*}, w)]$, and $\mathrm{cov}(z_k, X_{j^*}) = \E[\tilde{z}_k\, u_{j^*}] \in \Tan{\FM{Y}}$, equation \eqref{eq:S1-deriv-alpha} reads as the $j^*$-th row of the block normal equation
\begin{equation}
  \sum_{j=1}^p \mathrm{cov}(X_j,X_{j^*})\, \alpha_j
  + \sum_{k=1}^q g_{\FM{Y}}\!\bigl(v_k,\, \mathrm{cov}(z_k,X_{j^*})\bigr)
  = \mathrm{cov}(X_{j^*}, Y).
  \label{eq:row-alpha}
\end{equation}

\medskip
We next compute the functional derivative with respect to the tangent-vector parameters.
For any fixed $k^* \in \{1,\ldots,q\}$ and any direction $h \in \Tan{\FM{Y}}$, the directional (Gateaux) derivative of $L$ at $\bb{v}$ in direction $e_{k^*} \otimes h$ (i.e., perturbing $v_{k^*}$ by $h$) is
\begin{align}
  \frac{d}{d\varepsilon}\bigg|_{\varepsilon=0}
  L(\bb{\alpha},\, v_{k^*}+\varepsilon h,\, \{v_k\}_{k\neq k^*})
  &= -\E\bigl[g_{\FM{Y}}\!\bigl(w - \hat{w},\; \tilde{z}_{k^*}\, h\bigr)\bigr]
   = -\E\bigl[\tilde{z}_{k^*}\, g_{\FM{Y}}(w - \hat{w},\, h)\bigr].
\end{align}
Setting this to zero for all $h \in \Tan{\FM{Y}}$, by the non-degeneracy of $g_{\FM{Y}}$,
\begin{align}
  0 &= \E\bigl[\tilde{z}_{k^*}(w - \hat{w})\bigr] \notag\\
    &= \E[\tilde{z}_{k^*} w]
       - \sum_{j=1}^p \alpha_j \E[\tilde{z}_{k^*}\, u_j]
       - \sum_{k=1}^q \E[\tilde{z}_{k^*}\,\tilde{z}_k]\, v_k,
       \label{eq:S1-deriv-v}
\end{align}
where in the last line we used bi-linearity of $g_{\FM{Y}}$ and the fact that $v_k$ is a deterministic parameter, so $\E[\tilde{z}_{k^*}\tilde{z}_k\, v_k] = \E[\tilde{z}_{k^*}\tilde{z}_k]\,v_k$.
Identifying $\mathrm{cov}(z_{k^*}, Y) = \E[\tilde{z}_{k^*}\, w] \in \Tan{\FM{Y}}$, $\mathrm{cov}(z_{k^*}, X_j) = \E[\tilde{z}_{k^*}\, u_j] \in \Tan{\FM{Y}}$, and $\mathrm{cov}(z_{k^*}, z_k) = \E[\tilde{z}_{k^*}\tilde{z}_k] \in \R$, equation \eqref{eq:S1-deriv-v} is the $k^*$-th row of the block normal equation
\begin{equation}
  \sum_{j=1}^p \alpha_j\, \mathrm{cov}(z_{k^*}, X_j)
  + \sum_{k=1}^q \mathrm{cov}(z_{k^*}, z_k)\, v_k
  = \mathrm{cov}(z_{k^*}, Y).
  \label{eq:row-v}
\end{equation}

\medskip
Collecting equations \eqref{eq:row-alpha} for $j^* = 1,\ldots,p$ and \eqref{eq:row-v} for $k^* = 1,\ldots,q$ into matrix form gives $\Sigma_{\bb{Z}\bb{Z}}\,\bb{\beta} = C_{\bb{Z}Y}$, with the block Gram matrix
\begin{equation}
  \Sigma_{\bb{Z}\bb{Z}} =
  \begin{pmatrix}
    \bigl[\mathrm{cov}(X_j,X_\ell)\bigr]_{j,\ell=1}^p &
    \bigl[\mathrm{cov}(z_k, X_j)^\top\bigr]_{j,k} \\[4pt]
    \bigl[\mathrm{cov}(z_k, X_j)\bigr]_{k,j} &
    \bigl[\mathrm{cov}(z_k, z_\ell)\,I_d\bigr]_{k,\ell=1}^q
  \end{pmatrix}
\end{equation}
and right-hand side $C_{\bb{Z}Y}$ with components $\mathrm{cov}(X_j, Y)$ ($j=1,\ldots,p$) and $\mathrm{cov}(z_k, Y)$ ($k=1,\ldots,q$).
The scalar block is repeated on each of the $d$ tangent coordinates. The matrix has order $p+qd$ as in Theorem~\ref{thm:normal}.

\medskip
It remains to verify positive definiteness.
The matrix $\Sigma_{\bb{Z}\bb{Z}}$ is a block Gram matrix of second moments of random elements in $\Tan{\FM{Y}} \cong \R^d$ and $\R$.
Hence $\Sigma_{\bb{Z}\bb{Z}}$ is symmetric positive semi-definite by construction.
Under the positive-definiteness hypothesis ($\Sigma_{\bb{Z}\bb{Z}} \succ 0$), the block normal equation has the unique solution $\bb{\beta}^* = \Sigma_{\bb{Z}\bb{Z}}^{-1} C_{\bb{Z}Y}$.

\medskip
Finally, since $L(\bb{\alpha},\bb{v})$ is a quadratic (convex) functional in the finite-dimensional parameter $\bb{\beta} = (\bb{\alpha},\bb{v}) \in \R^p \times (\Tan{\FM{Y}})^q \cong \R^{p+qd}$, the critical point found above is a global minimum.
When $\Sigma_{\bb{Z}\bb{Z}} \succ 0$, the Hessian of $L$ (viewed as a $(p+qd)\times(p+qd)$ matrix acting on the parameter space $\R^{p+qd}$) equals $\Sigma_{\bb{Z}\bb{Z}}$, which is positive definite, confirming the unique minimizer.
\end{proof}

\section{Proof of Cut-Locus Identifiability (Theorem~\ref{thm:cutlocus})}
\label{sec:S2}

We restate the theorem for convenience.

\medskip
\noindent\textbf{Theorem~\ref{thm:cutlocus}.}
\textit{Assume $\Sigma_{\bb{Z}\bb{Z}} \succ 0$.
\textup{(a) Safe case.}
If for each $j$ the support of $X_j$ sits inside a strongly convex normal ball $B(\FM{X}_j, r_j)$ with $r_j < \mathrm{inj}_{\FM{X}_j}$ (preferably $r_j < \mathrm{conv}_{\FM{X}_j}$) and $\mathrm{supp}(Y)\subset B(\FM{Y}, r_Y)$ with $r_Y<\mathrm{inj}_{\FM{Y}}$, where $\mathrm{inj}_p$ and $\mathrm{conv}_p$ denote the pointwise injectivity and convexity radii at $p$, then each $\Log{\FM{X}_j}$ is single-valued and smooth on $\mathrm{supp}(X_j)$, $L$ is a strictly convex quadratic on $\R^{p+qd}$, and the population minimizer $\bb{\beta}^*$ is unique.
\textup{(b) Conditional branch sensitivity.}
Suppose on a measurable set $S$ with $\mathbb{P}(S)>0$ there exist two admissible measurable log-branches $\ell,\ell'\colon S\to\Tan{\FM{X}_j}$ with $\Exp{\FM{X}_j}(\ell(x)) = \Exp{\FM{X}_j}(\ell'(x)) = x$ and $\|\ell(x)\| = \|\ell'(x)\| = \dM(\FM{X}_j, x)$, and set $\Delta Z_{ij}=\PT{\FM{X}_j}{\FM{Y}}\{\ell(X_{ij})-\ell'(X_{ij})\}\mathbf 1\{X_{ij}\in S\}$.
Let $P_j^\perp$ denote the orthogonal projection in $L^2(\Tan{\FM Y})$ onto the complement of the closed linear span of the remaining design $\{u_l\}_{l\neq j}\cup\{\tilde z_k e_a\}_{k,a}$.
Then $\bb\beta^*_{(\ell)} - \bb\beta^*_{(\ell')} = (\Sigma^{(\ell')})^{-1}\bigl(\Delta C - \Delta\Sigma\,\bb\beta^*_{(\ell)}\bigr)$.}

\begin{proof}

\noindent\textbf{Part (a): Safe case via the inverse function theorem.}

Let $p \in \Mman$ and let $\mathrm{inj}_p$ denote the pointwise injectivity radius at $p$.
By a standard result in Riemannian geometry \citep[Chapter~3, Proposition~3.6]{doCarmo1992}, $\Exp{p}\colon B(0,\mathrm{inj}_p)\subset\Tan{p} \to B(p,\mathrm{inj}_p)\subset\Mman$ is a diffeomorphism onto the open geodesic ball of radius $\mathrm{inj}_p$.
Equivalently, the differential $d\Exp{p}(v)$ is nonsingular for all $\|v\|<\mathrm{inj}_p$ and no two points in $B(0,\mathrm{inj}_p)$ are sent to the same image, so by the inverse function theorem the logarithm $\Log{p}\colon B(p,\mathrm{inj}_p)\to\Tan{p}$ is a smooth diffeomorphism.
When additionally $r<\mathrm{conv}_p$ (the convexity radius), the ball $B(p,r)$ is strongly convex, and any two of its points are joined by a unique minimizing geodesic lying entirely in $B(p,r)$.
The radii here are the pointwise injectivity and convexity radii at the base point.

Under the safe-case hypothesis, $\mathrm{supp}(X_j)\subset B(\FM{X}_j, r_j)$ with $r_j<\mathrm{inj}_{\FM{X}_j}$, and $\mathrm{supp}(Y)\subset B(\FM{Y}, r_Y)$ with $r_Y<\mathrm{inj}_{\FM{Y}}$.
Hence $\Log{\FM{X}_j}(X_{ij})$ and $\Log{\FM{Y}}(Y_i)$ are each uniquely defined and vary smoothly with the observations.

Moreover, parallel transport $\PT{\FM{X}_j}{\FM{Y}}\colon\Tan{\FM{X}_j}\to\Tan{\FM{Y}}$ along the (unique) minimal geodesic is a linear isometry \citep[Chapter~2]{doCarmo1992}, hence injective.
Therefore $X_{ij}\mapsto u_{ij} = \PT{\FM{X}_j}{\FM{Y}}\bigl(\Log{\FM{X}_j}(X_{ij})\bigr)$ is the composition of a smooth diffeomorphism with a linear isometry, a smooth injection on $\mathrm{supp}(X_j)$.

We now confirm that $\Sigma_{\bb{Z}\bb{Z}}\succ 0$ yields an ordinary strictly convex quadratic objective.
If $\Sigma_{\bb{Z}\bb{Z}}\bb{c}=0$ for some nonzero $\bb{c}=(a_1,\ldots,a_p, b_1,\ldots,b_q)\in\R^{p+qd}$, then
\begin{align}
  0 &= \bb{c}^\top\Sigma_{\bb{Z}\bb{Z}}\bb{c}
     = \E\!\left[\norm{\sum_{j=1}^p a_j u_j
                + \sum_{k=1}^q \tilde z_k\, b_k}^2_{g_{\FM{Y}}}\right],
\end{align}
which forces an a.s.\ linear dependence, contradicting $\Sigma_{\bb{Z}\bb{Z}}\succ 0$.
Hence $L(\bb\beta)=\tfrac12\bb\beta^\top\Sigma_{\bb{Z}\bb{Z}}\bb\beta - C_{\bb{Z}Y}^\top\bb\beta + \mathrm{const}$ is a strictly convex quadratic and the population minimizer $\bb\beta^* = \Sigma_{\bb{Z}\bb{Z}}^{-1} C_{\bb{Z}Y}$ is unique.

\medskip
\noindent\textbf{Part (b): Conditional branch sensitivity.}

Assume the hypotheses of Part~(b): $\Log{\FM{X}_j}$ admits two distinct measurable branches $\ell, \ell'\colon S\to\Tan{\FM{X}_j}$ on a set $S$ with $\mathbb{P}(S)>0$, each returning minimal geodesics, with $\Delta Z_j$ as defined in Theorem~\ref{thm:cutlocus}.
Off $S$ the two branches agree, so the induced tangent-space design differs in the $j$-th block only.
Write $\bb{Z}^{(\ell)}$ and $\bb{Z}^{(\ell')}$ for the two design matrices in $\R^{(p+qd)\times d}$ obtained by using branch $\ell$ or $\ell'$ in the $j$-th slot, and $W = \Log{\FM{Y}}(Y)$ for the response tangent vector.
The induced Gram and cross-covariance matrices are $\Sigma^{(\ell)} = \E[\bb{Z}^{(\ell)}(\bb{Z}^{(\ell)})^\top]$ and $C^{(\ell)} = \E[\bb{Z}^{(\ell)}\,W]$, with analogous definitions for $\ell'$, and population minimizers $\bb\beta^*_{(\ell)} = (\Sigma^{(\ell)})^{-1}C^{(\ell)}$ and $\bb\beta^*_{(\ell')} = (\Sigma^{(\ell')})^{-1}C^{(\ell')}$.
Let $\Delta\Sigma = \Sigma^{(\ell)} - \Sigma^{(\ell')}$ and $\Delta C = C^{(\ell)} - C^{(\ell')}$, both supported in the $j$-th row/column through $\Delta Z_j$.

Subtracting the two normal equations $\Sigma^{(\ell)}\bb\beta^*_{(\ell)} = C^{(\ell)}$ and $\Sigma^{(\ell')}\bb\beta^*_{(\ell')} = C^{(\ell')}$ yields the exact identity
\begin{equation}
  \bb\beta^*_{(\ell)} - \bb\beta^*_{(\ell')}
  = (\Sigma^{(\ell')})^{-1}\bigl(\Delta C - \Delta\Sigma\,\bb\beta^*_{(\ell)}\bigr).
  \label{eq:branch-diff-exact}
\end{equation}
Abbreviating $\Sigma = \Sigma^{(\ell')}$ and $\bb\beta^* = \bb\beta^*_{(\ell)}$, the $j$-th coordinate of \eqref{eq:branch-diff-exact} is
\begin{equation}
  \alpha^*_{j,(\ell)} - \alpha^*_{j,(\ell')}
  = \bb{e}_j^\top\,\Sigma^{-1}\bigl(\Delta C - \Delta\Sigma\,\bb\beta^*\bigr),
  \label{eq:branch-diff-j}
\end{equation}
where $\bb{e}_j$ selects the $j$-th manifold-predictor slot of the concatenated design.

The non-degeneracy condition (ND), $P_j^\perp\Delta Z_j \neq 0$ in $L^2$, ensures that the branch perturbation $\Delta Z_j$ is not absorbed by the remaining design at the $L^2$ level, so $\Delta\Sigma$ contributes a non-trivial perturbation to the Gram matrix.
Conversely, when (ND) fails so that $\Delta Z_j$ lies in the $L^2$-span of the remaining design, the perturbation block of $\Delta\Sigma$ is rank-deficient and the two branch conventions become observationally equivalent.

Applying Cauchy--Schwarz in $\R^{p+qd}$ directly to \eqref{eq:branch-diff-j} yields the parameter-space upper bound
\begin{equation}
  \bigl|\alpha^*_{j,(\ell)} - \alpha^*_{j,(\ell')}\bigr|
  \;\leq\;
  \bigl\|(\Sigma^{(\ell')})^{-1}\bb{e}_j\bigr\|_2
  \;\bigl\|\Delta C - \Delta\Sigma\,\bb\beta^*_{(\ell)}\bigr\|_2,
  \label{eq:cutlocus-gap-ub-derived}
\end{equation}
where $\|(\Sigma^{(\ell')})^{-1}\bb{e}_j\|_2$ is the $j$-th column norm of $(\Sigma^{(\ell')})^{-1}$ and quantifies the sensitivity of the $j$-th coefficient to perturbations of the right-hand side.
The vector $\Delta C - \Delta\Sigma\,\bb\beta^*_{(\ell)}$ is zero when the response is uncorrelated with the branch perturbation in the $\Sigma$-weighted sense.
The two branches then share the same population minimizer even though (ND) holds.
A gap that changes the predictor requires both (ND) and $\Delta C - \Delta\Sigma\,\bb\beta^*_{(\ell)} \neq 0$.
\end{proof}

\begin{remark}[On the role of non-degeneracy]
Without condition \textup{(ND)} the branch perturbation $\Delta Z_j$ lies in the $L^2$-span of the remaining design, $P_j^\perp\Delta Z_j$ vanishes, and the two branch conventions are observationally equivalent at the population level, with the branch swap on $S$ absorbable into a redefinition of the remaining coefficients.
The two branches then induce the same population predictor, and the $j$-th coefficient is no longer separately identified.
\end{remark}

\begin{remark}[Relation to global injectivity]
Two log branches at $\FM{X}_j$ appear at distance $\inj_{\FM{X}_j}$.
This radius can exceed the global infimum $\inj(\Mman)=\inf_p\inj_p$, so a point at distance $\inj(\Mman)$ from $\FM{X}_j$ can still lie inside the injectivity ball of $\FM{X}_j$.
Part~(a) therefore uses the pointwise radius.
\end{remark}

\section{Proof of Consistency and Convergence Rate (Theorem~\ref{thm:consistency})}
\label{sec:S3}

We restate the theorem for convenience.

\medskip
\noindent\textbf{Theorem~\ref{thm:consistency}.}
\textit{Under Assumption~\ref{ass:main} (Afsari condition, injectivity condition, $\Sigma_{\bb{Z}\bb{Z}} \succ 0$ with smallest eigenvalue $\lambda_*>0$, and finite residual second moment) and finite fourth moments, let $\bb\beta^*$ denote the population FLS minimizer and $b^2 \coloneqq \E[\norm{w - \sum_j \alpha_j^* u_j - \sum_k \tilde z_k v_k^*}^2] - \sigma^2$ the squared model misspecification bias.
Then the empirical FLS estimator $\hat{\bb{\beta}}_n$ satisfies
\begin{equation}
  \E\bigl[\dM^2(\hat Y, Y)\bigr]
  \leq \tilde\eta(\kappa_-, r_0)^2
  \left(\frac{C\,\sigma^2(p+qd)}{n\lambda_*} + \sigma^2 + b^2\right),
\end{equation}
where $r_0 = \sup_i \dM(Y_i, \FM{Y})$, $\kappa_- \leq \mathrm{Sec}(\Mman)$ is the sectional curvature lower bound, $C > 0$ depends on the fourth moments of the covariates, and
\begin{equation}
  \tilde\eta(\kappa_-, r) =
  \begin{cases}
    \dfrac{\sinh(\sqrt{|\kappa_-|}\,r)}{\sqrt{|\kappa_-|}\,r} & \kappa_- < 0,\\[4pt]
    1 & \kappa_- \geq 0.
  \end{cases}
  \label{eq:eta-tilde-appendix}
\end{equation}
When the model is correctly specified ($b = 0$), $\hat{\bb{\beta}}_n \xrightarrow{P} \bb{\beta}^*$ and $\E[\dM^2(\hat Y, Y)] \leq \tilde\eta^2\,\sigma^2 + O(n^{-1})$.}

\begin{proof}

Under the injectivity condition, the response $Y_i$ lies in $B(\FM{Y}, \inj_{\FM{Y}})$ and $\hat Y = \Exp{\FM{Y}}(\hat{w})$ lies in the same ball by Assumption~\ref{ass:main}(ii).

We require two-sided comparison geometry.
Let $\kappa_- \leq \mathrm{Sec}(\Mman) \leq \kappa_+$ denote two-sided sectional curvature bounds.
For points $p, q, q' \in \Mman$ with $r = \max\{\dM(p,q), \dM(p,q')\}$ satisfying $r < \min(\inj_p, \pi/(2\sqrt{\kappa_+}))$ (with the convention $\pi/(2\sqrt{\kappa_+}) = \infty$ when $\kappa_+ \leq 0$), we use two comparison bounds.

For the \textbf{lower bound}, the Hessian comparison theorem for $\tfrac{1}{2}\dM^2(p,\cdot)$ (see \citet{CorderoMcCannSchmuckenschlager2001}; \citet[Chapter~8]{doCarmo1992}) gives
\begin{equation}
  \dM^2(q, q')
  \geq \eta(\kappa_+, r)\,\norm{\Log{p}(q) - \Log{p}(q')}^2_{g_p},
  \label{eq:rauch-lower}
\end{equation}
where $\eta(\kappa_+, r) \in (0,1]$ is defined in \eqref{eq:eta}.
The factor $\eta = \sqrt{\kappa_+}\,r/\tan(\sqrt{\kappa_+}\,r)$ is the minimum eigenvalue of $\mathrm{Hess}(\tfrac{1}{2}\dM^2(p,\cdot))$ at distance $r$ from $p$; positive curvature contracts the exponential map, so the tangent-space norm overestimates the geodesic distance by this factor.
For the \textbf{upper bound}, the Rauch comparison theorem (Jacobi field divergence under negative curvature; see \citet[Theorem~1.28]{CheegerEbin1975}) gives
\begin{equation}
  \dM^2(q, q')
  \leq \tilde\eta(\kappa_-, r)^2\,\norm{\Log{p}(q) - \Log{p}(q')}^2_{g_p},
  \label{eq:rauch-upper}
\end{equation}
where $\tilde\eta(\kappa_-, r) \geq 1$ is defined in \eqref{eq:eta-tilde}.
If $\kappa_+ \leq 0$, then $\eta = 1$ and $\dM \geq \norm{\Log{p}(q) - \Log{p}(q')}$.
If $\kappa_- \geq 0$, as on $S^2$ and $\mathrm{SO}(3)$, then $\tilde\eta = 1$ and \eqref{eq:rauch-upper} reads $\dM^2 \leq \norm{\Log{p}(q) - \Log{p}(q')}^2$.
If $\kappa_- < 0$, as on $\mathrm{Sym}^+(k)$ and $\mathbb{H}^d$, then $\tilde\eta > 1$ and \eqref{eq:rauch-upper} exceeds the tangent norm.

Let $(X,z,Y)$ be a new observation, independent of the sample.
Assumption~\ref{ass:main}(ii) bounds both $\dM(Y,\FM Y)$ and $\dM(\hat Y,\FM Y)$ by $r_0$.
Write $\Log{\FM Y}(\hat Y)=\hat w+\delta$, where $\hat w=\bb Z(\theta^*)^\top\hat{\bb\beta}_n$ is the tangent prediction at the population base points.
Applying \eqref{eq:rauch-upper} with $p = \FM{Y}$, $q = Y$, and $q' = \hat Y$:
\begin{equation}
  \E[\dM^2(\hat Y, Y)]
  \leq \tilde\eta(\kappa_-, r_0)^2\,\E\bigl[\norm{w - \hat{w}-\delta}^2_{g_{\FM{Y}}}\bigr].
  \label{eq:dist-to-tang}
\end{equation}

\begin{remark}
The lower bound \eqref{eq:rauch-lower} uses $\kappa_+$ and bounds $\dM^2$ from below; the upper bound \eqref{eq:rauch-upper} uses $\kappa_-$ and bounds $\dM^2$ from above.
These are different directions of the comparison and require different curvature bounds.
On non-negatively curved manifolds ($\kappa_-\geq 0$), $\tilde\eta=1$, so the tangent-space residual bounds the geodesic residual from above.
On non-positively curved manifolds with $\kappa_-<0$, $\tilde\eta>1$, and the geodesic residual can exceed the tangent-space residual.
\end{remark}

\medskip
We now work entirely in the Hilbert space $(\Tan{\FM{Y}}, g_{\FM{Y}})$.
Define $\hat{w}^*_i = \sum_j \alpha_j^* u_{ij} + \sum_k \tilde{z}_{ik} v_k^*$ as the population best linear predictor (the projection of $w$ onto the linear span of the covariates in $\Tan{\FM{Y}}$), and define $\varepsilon_i = w_i - \hat{w}^*_i$ as the residual.
By the definition of $\bb\beta^*$ as the population FLS minimizer, $\hat{w}^*$ is the orthogonal projection of $w$ onto the linear model space, so $\E[g_{\FM{Y}}(\varepsilon, f)] = 0$ for every function $f$ in the span of the covariates.

Under misspecification, $\E[\varepsilon \mid X, Z]$ may be nonzero.
The excess second moment is the bias $b^2 = \E[\norm{\varepsilon}^2] - \sigma^2$, where $\sigma^2$ is the irreducible noise variance in Assumption~\ref{ass:main}(iv).
Under correct specification, $b = 0$ and $\varepsilon$ is that noise term.

The empirical FLS estimator satisfies $\hat{\bb{\beta}}_n = \hat\Sigma_{\bb{Z}\bb{Z}}^{-1}\hat C_{\bb{Z}Y}$.
The tangent-space prediction error decomposes as
\begin{align}
  \E\bigl[\norm{w - \hat{w}}^2_{g_{\FM{Y}}}\bigr]
  &= \E\bigl[\norm{(\hat{w}^* - \hat{w}) + \varepsilon}^2\bigr] \notag\\
  &= \underbrace{\E\bigl[\norm{\hat{w}^* - \hat{w}}^2\bigr]}_{\text{estimation variance}}
  + \underbrace{\E[\norm{\varepsilon}^2]}_{\sigma^2 + b^2},
  \label{eq:bv-decomp}
\end{align}
Conditional on the sample, the cross term equals $(\bb\beta^*-\hat{\bb\beta}_n)^\top\E[\bb Z\varepsilon]$.
It is zero because the new observation is independent of $\hat{\bb\beta}_n$, and the population normal equations give $\E[\bb Z\varepsilon]=0$.
This holds under correct specification and under misspecification.

The fitted prediction uses the empirical base points $\hat\theta$.
The difference $\delta=\Log{\FM Y}(\hat Y)-\hat w$ in~\eqref{eq:dist-to-tang} is a smooth function of $\hat\theta-\theta^*$.
The mean-square bound for the Fr\'echet means, proved below, gives $\E\|\delta\|^2=O(1/n)$.
In the expansion of $\|w-\hat w-\delta\|^2$, the Cauchy--Schwarz inequality bounds the cross term in $\hat w^*-\hat w$ by $O(1/n)$.
The cross term in $\varepsilon$ has expectation $\E\langle\E(\varepsilon\mid X,z),\E(\delta\mid X,z)\rangle$.
This expectation is zero under correct specification.
Under misspecification it is $O(1/n)$.
The expansion $\hat\theta-\theta^*=n^{-1}\sum_i\psi_i+r_n$ has $\E\psi_i=0$ and $\|r_n\|\le C\|\hat\theta-\theta^*\|^2$, so $\|\E(\hat\theta-\theta^*)\|\le C\,\E\|\hat\theta-\theta^*\|^2$.
These bounds give $\E\|w-\hat w-\delta\|^2=\E\|w-\hat w\|^2+O(1/n)$.

\medskip
The variance term measures how well $\hat\Sigma_{\bb{Z}\bb{Z}}$ and $\hat C_{\bb{Z}Y}$ approximate their population counterparts.
By the law of large numbers, $\hat\Sigma_{\bb{Z}\bb{Z}} \xrightarrow{\mathrm{a.s.}} \Sigma_{\bb{Z}\bb{Z}}$ and $\hat C_{\bb{Z}Y} \xrightarrow{\mathrm{a.s.}} C_{\bb{Z}Y}$.

For the convergence rate, consider the error $\hat{\bb{\beta}}_n - \bb{\beta}^* = \hat\Sigma_{\bb{Z}\bb{Z}}^{-1}\hat C_{\bb{Z}Y} - \Sigma_{\bb{Z}\bb{Z}}^{-1}C_{\bb{Z}Y}$.
Writing $\hat\Sigma = \Sigma + \Delta$ where $\Delta = \hat\Sigma - \Sigma$, on the event $E_n=$$\{\norm{\Delta}_{\mathrm{op}} < \lambda_*/2\}$, which has probability tending to one by Chebyshev's inequality and~\eqref{eq:gram-rate}, the Neumann series $\hat\Sigma^{-1} = (\Sigma + \Delta)^{-1} = (\mathrm{id} + \Sigma^{-1}\Delta)^{-1}\Sigma^{-1}$ converges and $\norm{\hat\Sigma^{-1}}_{\mathrm{op}} \leq 2/\lambda_*$, giving
\begin{align}
  \norm{\hat{\bb{\beta}}_n - \bb{\beta}^*}
  &\leq \norm{\hat\Sigma^{-1}}_{\mathrm{op}}\,\norm{\hat C - C}
       + \norm{\hat\Sigma^{-1} - \Sigma^{-1}}_{\mathrm{op}}\,\norm{C} \notag\\
  &\leq \frac{2}{\lambda_*}\norm{\hat C - C}
       + \frac{2}{\lambda_*^2}\norm{\Delta}_{\mathrm{op}}\,\norm{C},
\end{align}
on the same event.

The empirical Gram matrix $\hat\Sigma$ and cross-covariance $\hat C$ are constructed using design vectors $\hat Z_i$ that depend on the empirical Fr\'echet means $\hat{\bar Y}, \hat{\bar X}_j$, which are computed from all $n$ observations.
We decompose $\hat\Sigma - \Sigma = (\hat\Sigma_{\rm orcl} - \Sigma) + (\hat\Sigma - \hat\Sigma_{\rm orcl})$, where $\hat\Sigma_{\rm orcl} = n^{-1}\sum_i Z_i(\theta^*) Z_i(\theta^*)^\top$ uses the population Fr\'echet means $\theta^* = (\bar Y, \bar X_j)$.
The oracle term has independent and identically distributed summands.
The Frobenius norm and the second moments of the entries give $\E\norm{\hat\Sigma_{\rm orcl} - \Sigma}^2_{\rm op} \leq (p+qd)^2 M_4/n$.
For fixed $p$, $q$, and $d$, $(p+qd)M_4$ is constant and is included in $C_1$ below.
The base-point correction $\hat\Sigma - \hat\Sigma_{\rm orcl}$ is bounded by a Taylor expansion in $\theta$ around $\theta^*$ and by the mean-square Fr\'echet-mean rate $\E\|\hat\theta - \theta^*\|^2 = O(1/n)$.
Under the Afsari condition the Fr\'echet functions are strongly convex on the support balls.
On these compact supports the empirical Fr\'echet functions are averages of bounded functions, and the bounded-differences inequality gives this mean-square rate.
The design is Lipschitz in the base points, because the supports lie strictly inside the injectivity balls.
Both contributions have mean square $O(1/n)$, so
\begin{align}
  \label{eq:gram-rate}
  \E\bigl[\norm{\hat\Sigma - \Sigma}^2_{\mathrm{op}}\bigr]
  &\leq \frac{C_1\,(p+qd)}{n}, &
  \E\bigl[\norm{\hat C - C}^2\bigr]
  &\leq \frac{C_1'\,(p+qd)\,M_4}{n},
\end{align}
where $M_4 = \max(\E[\norm{u_j}^4], \E[\tilde{z}_k^4], \E[\norm{w}^4])$ is the largest fourth moment of the covariates and the response, and $C_1, C_1'$ are constants depending on the curvature of the manifold and the smoothness of the log and transport maps in the base points.
Therefore, on $E_n$,
\begin{equation}
  \E\bigl[\norm{\hat{\bb{\beta}}_n - \bb{\beta}^*}^2\,\mathbf 1_{E_n}\bigr]
  \leq \frac{C_2\,M_4\,(p+qd)}{n\lambda_*^2}
  \label{eq:beta-error}
\end{equation}
for a constant $C_2 > 0$.
By Chebyshev's inequality, $P(E_n^c) \le 4\,\E\norm{\Delta}^2_{\mathrm{op}}/\lambda_*^2 \le 4C_1(p+qd)/(n\lambda_*^2)$.
This is the bound in Corollary~\ref{cor:beta-rate}. Equation~\eqref{eq:beta-error} and $P(E_n^c)\to 0$ give $\|\hat{\bb\beta}_n - \bb\beta^*\|^2 = O_P((p+qd)/n)$.

On $E_n$, the variance term in \eqref{eq:bv-decomp} satisfies $\E[\norm{\hat{w}^* - \hat{w}}^2\mathbf 1_{E_n}] \leq \norm{\Sigma_{\bb{Z}\bb{Z}}}_{\mathrm{op}}\, \E[\norm{\hat{\bb{\beta}}_n - \bb{\beta}^*}^2\mathbf 1_{E_n}]$, since $(X,z,Y)$ is independent of the sample. Combining this bound with \eqref{eq:beta-error} gives
\begin{equation}
  \E\bigl[\norm{\hat{w}^* - \hat{w}}^2\mathbf 1_{E_n}\bigr]
  \leq \frac{C_3\,M_4\,(p+qd)}{n\lambda_*},
\end{equation}
where $C_3$ absorbs $\norm{\Sigma_{\bb{Z}\bb{Z}}}_{\mathrm{op}}/\lambda_*$.
Under the Afsari condition, $M_4$ is bounded by a constant depending on the relevant injectivity radius and the covariate distribution.

\medskip
By Assumption~\ref{ass:main}(ii), $\dM(\hat Y,Y)\le 2r_0$, so $\E[\dM^2(\hat Y,Y)] \le \E[\dM^2(\hat Y,Y)\mathbf 1_{E_n}] + 4r_0^2\,P(E_n^c)$.
The second term, the $O(1/n)$ correction for $\delta$, and the variance term are each $O((p+qd)/n)$ and are included in $C_3$.
Substituting into \eqref{eq:dist-to-tang} and using \eqref{eq:bv-decomp} on $E_n$ gives
\begin{align}
  \E\bigl[\dM^2(\hat Y, Y)\bigr]
  &\leq \tilde\eta(\kappa_-, r_0)^2\,\E\bigl[\norm{w - \hat{w}-\delta}^2_{g_{\FM{Y}}}\mathbf 1_{E_n}\bigr]
     + 4r_0^2\,P(E_n^c)
     \notag\\
  &\leq \tilde\eta(\kappa_-, r_0)^2\left(
     \frac{C_3\,M_4\,(p+qd)}{n\lambda_*} + \sigma^2 + b^2
  \right).
\end{align}
Setting $C \coloneqq C_3 M_4 / \sigma^2$ (finite under the Afsari condition and the finite fourth-moment hypothesis) collects the fourth-to-second moment ratio and recovers the $\tilde\eta^2(C\sigma^2(p+qd)/(n\lambda_*) + \sigma^2 + b^2)$ form stated in Theorem~\ref{thm:consistency}.

As $n \to \infty$ the right-hand side tends to $\tilde\eta(\kappa_-, r_0)^2(\sigma^2 + b^2)$, so under correct specification ($b = 0$) the excess risk is bounded by $\tilde\eta^2\sigma^2 + O(n^{-1})$, and \eqref{eq:beta-error} with $P(E_n^c)\to0$ gives $\hat{\bb{\beta}}_n \xrightarrow{P} \bb{\beta}^*$.
The variance contribution therefore decays at the parametric $n^{-1}$ rate, with a purely geometric factor $\tilde\eta^2 \geq 1$ in front.
When $\kappa_- < 0$, as on $\mathbb{H}^d$, the factor is strictly larger than one.
When $\kappa_- \geq 0$, as on $S^2$ and $\mathrm{SO}(3)$, $\tilde\eta = 1$ and the excess-risk rate is the Euclidean $O(n^{-1})$.
\end{proof}

\section{Proof of Asymptotic Normality (Theorem~\ref{thm:inference})}
\label{sec:S4}

The main claim of Theorem~\ref{thm:inference} is the asymptotic-linearity expansion $\sqrt{n}(\hat{\bb\beta}_n - \bb\beta^*) = n^{-1/2}\sum_i \mathrm{IF}(O_i) + o_P(1)$ for an influence function $\mathrm{IF}$ built from the oracle tangent-space score and the Fr\'echet-mean influence functions.
The limiting covariance is the sandwich $\Sigma_{\bb\beta}=\E[\mathrm{IF}(O)\mathrm{IF}(O)^\top]$.
The tangent-space summands may be heteroskedastic.
Under condition (D2) in the proof, the sandwich reduces to $\Sigma_{\mathrm{tan}}+\Sigma_{\mathrm{curv}}$.

\begin{proof}
Recall $d=\dim\Mman$ from Section~\ref{sec:prelim}.
Fix an orthonormal basis of $T_{\FM{Y}}\Mman$ and identify the product tangent space $\mathcal{H} = \R^p \times (T_{\FM{Y}}\Mman)^q$ with $\R^{p+qd}$.

We first establish a central limit theorem for the oracle FLS estimator $\tilde{\bb{\beta}}_n$ computed with the true Fr\'echet means $\FM{Y}$ and $\FM{X}_j$.
By the decomposition $w_i = \bb{Z}_i^\top\bb{\beta}^* + \varepsilon_i$, the oracle estimation error is
\begin{equation}\label{eq:inf-oracle-err}
  \tilde{\bb{\beta}}_n - \bb{\beta}^*
  = \hat\Sigma_{\bb{Z}\bb{Z}}^{-1}\,\frac{1}{n}\sum_{i=1}^n \bb{Z}_i\varepsilon_i.
\end{equation}
Define $\xi_i \coloneqq \bb{Z}_i\varepsilon_i \in \R^{p+qd}$.
The summands $\xi_1,\ldots,\xi_n$ are independent and identically distributed with $\E[\xi_i] = 0$ (by the population normal equations $\E[\bb{Z}\varepsilon] = 0$) and covariance
\begin{equation}\label{eq:inf-xi-cov}
  \Omega
  \coloneqq
  \E\bigl[\xi_i\xi_i^\top\bigr]
  = \E\bigl[\bb{Z}_i\,\varepsilon_i\varepsilon_i^\top\,\bb{Z}_i^\top\bigr],
\end{equation}
which allows heteroskedastic and design-dependent tangent-space noise.
The finite fourth-moment hypothesis guarantees $\E[\|\xi_i\|^4] < \infty$ and hence the Lindeberg condition.
By the multivariate central limit theorem,
\begin{equation}\label{eq:inf-clt-score}
  \frac{1}{\sqrt{n}}\sum_{i=1}^n \xi_i
  \;\xrightarrow{d}\;
  \mathcal{N}(0,\Omega).
\end{equation}
By the strong law of large numbers, $\hat\Sigma_{\bb{Z}\bb{Z}} \xrightarrow{\mathrm{a.s.}} \Sigma_{\bb{Z}\bb{Z}}$, and $\Sigma_{\bb{Z}\bb{Z}}$ is invertible by Assumption~\ref{ass:main}(iii).
Slutsky's lemma applied to \eqref{eq:inf-oracle-err} and \eqref{eq:inf-clt-score} yields
\begin{equation}\label{eq:inf-oracle-clt}
  \sqrt{n}\bigl(\tilde{\bb{\beta}}_n - \bb{\beta}^*\bigr)
  \;\xrightarrow{d}\;
  \mathcal{N}\!\bigl(0,\;
  \Sigma_{\bb{Z}\bb{Z}}^{-1}\Omega\Sigma_{\bb{Z}\bb{Z}}^{-1}\bigr).
\end{equation}

To bridge from this oracle result to the plug-in estimator, we record a Neyman-orthogonality property of the FLS score with respect to the base-point nuisance parameters.
In practice, the true Fr\'echet means are replaced by their empirical counterparts $\hat\theta_n = (\hat{\bar Y}, \hat{\bar X}_1,\ldots,\hat{\bar X}_p)$.
Write $\theta^* = (\FM{Y}, \FM{X}_1,\ldots,\FM{X}_p)$.
For a base-point configuration $\theta$ in a geodesic neighborhood of $\theta^*$, let $\bb{Z}_i(\theta)$, $w_i(\theta)$, $\Sigma(\theta)$, $C(\theta)$ denote the $\theta$-dependent design vector, response lift, Gram matrix, and cross-covariance.
The population minimizer at $\theta$ is $\bb{\beta}^*(\theta) = \Sigma(\theta)^{-1}C(\theta)$, and the population residual is $\varepsilon_i(\theta) = w_i(\theta) - \bb{Z}_i(\theta)^\top\bb{\beta}^*(\theta)$.

Since $\bb{\beta}^*(\theta)$ minimizes the FLS objective at base points $\theta$, the population normal equations give
\begin{equation}\label{eq:inf-pop-ortho}
  \E\bigl[\bb{Z}_i(\theta)\,\varepsilon_i(\theta)\bigr] = 0
  \quad\text{for all } \theta \text{ near } \theta^*.
\end{equation}
Under the Afsari support condition, the logarithmic map and parallel transport are smooth in the base point (see \citet[Chapter~1]{CheegerEbin1975}), so the left-hand side of \eqref{eq:inf-pop-ortho} is a differentiable function of $\theta$.
Since \eqref{eq:inf-pop-ortho} holds identically in $\theta$, differentiation yields
\begin{equation}\label{eq:inf-neyman}
  \partial_\theta\,\E\bigl[\bb{Z}_i(\theta)\,\varepsilon_i(\theta)\bigr]
  \Big|_{\theta = \theta^*} = 0.
\end{equation}
Equation~\eqref{eq:inf-neyman} is the Neyman-orthogonality condition of the FLS score with respect to the nuisance parameter $\theta$~\citep{Chernozhukov2018}.

With \eqref{eq:inf-neyman} in hand, we can now show that the plug-in and oracle scores are asymptotically equivalent.
Define the empirical score $S_n(\theta) = n^{-1}\sum_{i=1}^n \bb{Z}_i(\theta)\,\varepsilon_i(\theta)$.
The plug-in estimator satisfies
\begin{equation}\label{eq:inf-plugin}
  \hat{\bb{\beta}}_n - \bb{\beta}^*(\hat\theta_n)
  = \hat\Sigma(\hat\theta_n)^{-1}\,S_n(\hat\theta_n).
\end{equation}
Working in normal coordinates around $\theta^*$, a first-order Taylor expansion of $\sqrt{n}\,S_n(\hat\theta_n)$ gives
\begin{equation}\label{eq:inf-taylor}
  \sqrt{n}\,S_n(\hat\theta_n)
  = \sqrt{n}\,S_n(\theta^*)
  + \bigl[\partial_\theta S_n(\theta^*)\bigr]\,
    \sqrt{n}(\hat\theta_n - \theta^*)
  + R_n.
\end{equation}
The smoothness of the geometric maps under the Afsari condition together with the finite fourth-moment hypothesis yields a uniform bound on $\partial_\theta^2 S_n$ in a neighborhood of $\theta^*$.
Combined with the Fr\'echet mean estimation rate $\|\hat\theta_n - \theta^*\| = O_P(n^{-1/2})$ \citep{BhattacharyaPatrangenaru2005}, this gives $R_n = O_P(\sqrt{n}\,\|\hat\theta_n - \theta^*\|^2) = O_P(n^{-1/2}) = o_P(1)$.
For the first-order term, the law of large numbers gives
\[
  \partial_\theta S_n(\theta^*) \;\xrightarrow{P}\;
  \partial_\theta\,\E\bigl[\bb{Z}(\theta)\varepsilon(\theta)\bigr]\big|_{\theta^*} = 0
\]
by the Neyman condition \eqref{eq:inf-neyman}, where the interchange of differentiation and expectation is justified by the dominated convergence theorem under the bounded support guaranteed by the Afsari condition.
Since $\sqrt{n}(\hat\theta_n - \theta^*) = O_P(1)$, the first-order term is $o_P(1) \cdot O_P(1) = o_P(1)$.
Substituting into \eqref{eq:inf-taylor} yields the score equivalence
\begin{equation}\label{eq:inf-score-equiv}
  \sqrt{n}\,S_n(\hat\theta_n) = \sqrt{n}\,S_n(\theta^*) + o_P(1).
\end{equation}
Since $\hat\Sigma(\hat\theta_n) \xrightarrow{P} \Sigma_{\bb{Z}\bb{Z}}$ by the law of large numbers and the continuity of the geometric operations in $\theta$, combining \eqref{eq:inf-plugin} and \eqref{eq:inf-score-equiv} gives
\begin{equation}\label{eq:inf-plugin-oracle}
  \sqrt{n}\bigl(\hat{\bb{\beta}}_n - \bb{\beta}^*(\hat\theta_n)\bigr)
  = \sqrt{n}\bigl(\tilde{\bb{\beta}}_n - \bb{\beta}^*\bigr) + o_P(1).
\end{equation}

It remains to incorporate the base-point sensitivity $\bb{\beta}^*(\hat\theta_n) - \bb{\beta}^*$ into the limit distribution.
Write $J_\theta = \partial_\theta\bb{\beta}^*(\theta^*)$.
Since $\bb{\beta}^*(\theta) = \Sigma(\theta)^{-1}C(\theta)$, the chain rule yields
\[
  J_\theta
  = \Sigma_{\bb{Z}\bb{Z}}^{-1}\bigl[\partial_\theta C(\theta^*)
  - \bigl(\partial_\theta\Sigma(\theta^*)\bigr)\,\bb{\beta}^*\bigr].
\]
The Fr\'echet mean centering property $\E[\bb{Z}(\theta^*)] = 0$ and $\E[w(\theta^*)] = 0$ cancels the leading terms in $\partial_\theta\Sigma(\theta^*)$ and $\partial_\theta C(\theta^*)$.
The remainders are curvature-driven.
Let $K=\max(|\kappa_-|,\kappa_+)$. This constant bounds the absolute sectional curvature.
Expanding $\Log{\theta+\delta}$ and the associated parallel transport in normal coordinates around $\theta^*$ to second order \citep{Pennec2006} and averaging against the distribution of $(\bb{u}, \tilde{\bb{z}})$, each surviving contribution is the product of a component of the Riemann tensor with a second-order or third-order moment of the covariates, so
\begin{equation}\label{eq:inf-J-form}
  J_\theta
  \;=\; \mathcal{R}\!\bigl(\mu_2,\,\mu_3^{(\tilde z)},\,\mu_3^{(\bb u)}\bigr)
  \;+\; O\!\bigl(K^2 r_0^4 \lambda_*^{-1}\bigr),
\end{equation}
where $\mu_2$ denotes the cross moments of each $u_j$ with the remaining design columns. The terms $\mu_3^{(\tilde z)}$ and $\mu_3^{(\bb u)}$ denote the third cumulants of the scalar covariates and of the transported manifold-predictor log maps, and $\mathcal{R}$ is the linear operator contracting them against the Riemann curvature at $\theta^*$, with operator norm $\|\mathcal R\| = O(K)$.
The second-order terms are the holonomy of $\PT{\FM X_j}{\FM Y}$ as the base point varies, and they vanish when $\FM X_j=\FM Y$ for every $j$.
Consequently $J_\theta = 0$ whenever the Riemann tensor vanishes, equivalently when $\kappa_+ = \kappa_- = 0$.
On $\mathrm{Sym}^+(k)$ one has $\kappa_+=0$ and $\kappa_-<0$.
The Riemann tensor can then be nonzero, and so can $J_\theta$.
As a concrete illustration, on $\mathbb S^2$ with $p=0$, $q=1$, and scalar $\tilde z$ with third cumulant $\mu_3 \coloneqq \E[\tilde z^3]$, \eqref{eq:inf-J-form} reduces at leading order in $r_0$ to $\partial_{\FM{Y}}\bb\beta^*(\theta^*) = \bigl(\mu_3/(3\Var(\tilde z))\bigr)\bigl(\|v\|^2 I - vv^\top\bigr)$, which is non-zero whenever $\mu_3 \ne 0$ and $v \ne 0$.

By the nonsingular-Hessian condition in Theorem~\ref{thm:inference}, the joint empirical Fr\'echet mean admits the asymptotically linear representation
\[
  \sqrt n(\hat\theta_n-\theta^*)
  = n^{-1/2}\sum_{i=1}^n \psi_\theta(O_i)+o_P(1),
  \qquad
  \E[\psi_\theta(O)]=0,
\]
with covariance $\Sigma_\theta=\E[\psi_\theta(O)\psi_\theta(O)^\top]$ \citep{BhattacharyaPatrangenaru2005,BhattacharyaLin2017}.
Taylor expansion therefore gives
\begin{equation}\label{eq:inf-bp-shift}
  \sqrt{n}\bigl(\bb\beta^*(\hat\theta_n) - \bb\beta^*\bigr)
  \;=\; n^{-1/2}\sum_{i=1}^n J_\theta\psi_\theta(O_i) + o_P(1).
\end{equation}

Combining \eqref{eq:inf-plugin-oracle} and \eqref{eq:inf-bp-shift}, the full plug-in estimator has influence function
\begin{equation}\label{eq:full-if}
  \mathrm{IF}(O)
  =
  \Sigma_{\bb{Z}\bb{Z}}^{-1}\bb{Z}\varepsilon
  + J_\theta\psi_\theta(O).
\end{equation}
The multivariate CLT applied to this joint summand gives the sandwich limit
\begin{equation}\label{eq:inf-augmented}
  \sqrt{n}\bigl(\hat{\bb\beta}_n - \bb\beta^*\bigr)
  \;=\; n^{-1/2}\sum_{i=1}^n \mathrm{IF}(O_i)+o_P(1)
  \;\xrightarrow{d}\;
  \mathcal{N}(0, \Sigma_{\bb\beta}),
\end{equation}
where
\begin{equation}\label{eq:full-sandwich}
  \Sigma_{\bb\beta}
  =
  \E\!\left[
  \left\{\Sigma_{\bb{Z}\bb{Z}}^{-1}\bb{Z}\varepsilon
  + J_\theta\psi_\theta(O)\right\}
  \left\{\Sigma_{\bb{Z}\bb{Z}}^{-1}\bb{Z}\varepsilon
  + J_\theta\psi_\theta(O)\right\}^{\top}
  \right].
\end{equation}
Equivalently,
\[
  \Sigma_{\bb\beta}
  =
  \Sigma_{\bb{Z}\bb{Z}}^{-1}\Omega\Sigma_{\bb{Z}\bb{Z}}^{-1}
  + J_\theta\Sigma_\theta J_\theta^\top
  + \Sigma_{\bb{Z}\bb{Z}}^{-1}\E[\bb{Z}\varepsilon\,\psi_\theta(O)^\top]J_\theta^\top
  + J_\theta\E[\psi_\theta(O)\varepsilon^\top\bb{Z}^\top]\Sigma_{\bb{Z}\bb{Z}}^{-1}.
\]
The final two terms are the cross-covariances between the oracle tangent score and the Fr\'echet-mean influence function.
They vanish under (D2), which requires the oracle tangent score to be uncorrelated with the Fr\'echet-mean influence functions.
Under the correct specification in Theorem~\ref{thm:inference}, (D2) holds when the conditional covariance of $\varepsilon$ given $(X,z)$ does not depend on $(X,z)$.
The identity $\E[\bb Z\varepsilon\,w^\top]=\E[\bb Z\,\E(\varepsilon\varepsilon^\top\mid X,z)]$ and the centering $\E[\bb Z]=0$ then give $\E[\bb Z\varepsilon\,w^\top]=0$.
The covariance is then $\Sigma_{\mathrm{tan}}+\Sigma_{\mathrm{curv}}$, with $\Sigma_{\mathrm{tan}}=\Sigma_{\bb{Z}\bb{Z}}^{-1}\Omega\Sigma_{\bb{Z}\bb{Z}}^{-1}$ and $\Sigma_{\mathrm{curv}}=J_\theta\Sigma_\theta J_\theta^\top$.
Appendix~\ref{sec:S6} uses two further conditions.
Under (D1) the tangent noise is isotropic, and $\Sigma_{\mathrm{tan}}$ is proportional to $\Sigma_{\bb Z\bb Z}^{-1}$.
Under (D3) the Fr\'echet-mean covariance is block diagonal, and $\Sigma_{\mathrm{curv}}$ is a sum over the $p+1$ means.

The stated $t$- and Wald limits now follow by projection.
For a tested manifold coefficient, the marginal distribution of $\sqrt n(\hat\alpha_j-\alpha_j^*)$ has variance $[\Sigma_{\bb\beta}]_{jj}$.
If $[\Sigma_{\bb\beta}]_{jj}>0$ and $[\hat\Sigma_{\bb\beta}]_{jj}\to_P[\Sigma_{\bb\beta}]_{jj}$, Slutsky's lemma gives
\[
  T_j
  =
  \frac{\hat\alpha_j-\alpha_j^*}
  {\{[\hat\Sigma_{\bb\beta}]_{jj}/n\}^{1/2}}
  \xrightarrow{d} N(0,1).
\]
For a scalar-predictor tangent coefficient, the sub-vector $\hat v_k$ has asymptotic covariance $\Sigma_{\bb\beta}^{kk}/n$.
If $\Sigma_{\bb\beta}^{kk}$ is nonsingular and $\hat\Sigma_{\bb\beta}^{kk}\to_P\Sigma_{\bb\beta}^{kk}$, the continuous mapping theorem gives
\[
  W_k
  =
  n(\hat v_k-v_k^*)^\top(\hat\Sigma_{\bb\beta}^{kk})^{-1}(\hat v_k-v_k^*)
  \xrightarrow{d}\chi^2_d.
\]

Corollary~\ref{cor:bootstrap-validity} is the pairs-bootstrap counterpart of~\eqref{eq:asymp-normal}. It follows from the $\bb\beta$-block of Proposition~\ref{prop:pairs-boot} below.
\end{proof}

\subsection{Validity of the Pairs Bootstrap}
\label{sec:S4-boot}

Proposition~\ref{prop:pairs-boot} is the $Z$-estimator statement behind Corollary~\ref{cor:bootstrap-validity}.
Conditions~(ii), (iii), and~(v) are the Hessian, differentiability, and fourth-moment hypotheses of Theorem~\ref{thm:inference}, and condition~(iv) is Assumption~\ref{ass:main}(iii).
Condition~(i) is Assumption~\ref{ass:main}(i)--(ii), namely that the Fr\'echet means are unique and the supports lie in the balls~\eqref{eq:inj-cond}, whose closures are compact and strictly inside the injectivity radius.
No further moment or design condition is imposed.

\begin{proposition}[Pairs bootstrap]\label{prop:pairs-boot}
Assume that (i)~$Y$ and each $X_j$ have unique Fr\'echet means, with supports in compact sets strictly inside the injectivity radius of the respective mean; (ii)~the Fr\'echet functions $F_Y(m)=\tfrac12\E\dM^2(Y,m)$ and $F_j(m)=\tfrac12\E\dM^2(X_j,m)$ are twice continuously differentiable near their minimizers with nonsingular Hessians $H_Y$ and $H_j$; (iii)~the logarithm and parallel-transport maps are jointly $C^1$ in the base points near the means and in their arguments; (iv)~$\Sigma_{\bb Z\bb Z}$ is nonsingular; and (v)~the finite fourth moments of Theorem~\ref{thm:inference} hold, in particular $\E\|z\|^4<\infty$.
Let $\hat{\bb\beta}_n^{*}$ be the ROLS estimator on a pairs bootstrap resample, with the Fr\'echet means, logarithm maps, transports, and normal equations recomputed.
Express $\hat v_k^{*}$ and $\hat v_k$ in one fixed orthonormal frame of $\Tan{\FM Y}$ by parallel transport.
On a resample whose normal equations are singular, which has conditional probability tending to zero, define $\hat{\bb\beta}_n^{*}$ arbitrarily.
Then, conditionally on the data and in probability, $\sqrt n(\hat{\bb\beta}_n^{*}-\hat{\bb\beta}_n)\rightsquigarrow \mathcal N(0,\Sigma_{\bb\beta})$ with $\Sigma_{\bb\beta}=\Var(\mathrm{IF}_{\bb\beta}(O))$, where $\mathrm{IF}_{\bb\beta}$ is given in~\eqref{eq:boot-if}.
The second-moment matrix of $\sqrt n(\hat{\bb\beta}_n^{*}-\hat{\bb\beta}_n)$, truncated at a level $\tau_n\to\infty$ that grows sufficiently slowly, converges in probability to $\Sigma_{\bb\beta}$.
\end{proposition}

\begin{proof}
Write $P$ for expectation under the data law and $P_n$ for the empirical measure.
For a pairs bootstrap resample, $P_n^*$ denotes its empirical measure.

We first define the charts to differentiate a map that takes values in a moving tangent space.
Fix orthonormal frames $E_Y\colon\R^d\to\Tan{\FM Y}$ and $E_j\colon\R^d\to\Tan{\FM X_j}$ at the population means, and write $m_Y(a)=\Exp{\FM Y}(E_Ya)$ and $m_j(a)=\Exp{\FM X_j}(E_ja)$ for $a\in\R^d$ near $0$.
A tangent vector at $m_Y(a)$ is expressed in $\R^d$ by $\iota=E_Y^{-1}\circ\PT{m_Y(a)}{\FM Y}$, and a tangent vector at $m_j(a)$ by $\iota_j=E_j^{-1}\circ\PT{m_j(a)}{\FM X_j}$.
By~(i) and~(iii) these identifications are $C^1$ for base points near the means, so $\vartheta=(a_Y,a_1,\ldots,a_p,\mu_z,\bb\beta)$ may be treated as a Euclidean parameter, with true value $\vartheta^*=(0,\ldots,0,\bar z,\bb\beta^*)$.

In these coordinates we stack the Fr\'echet-mean equations and the normal equations into one estimating function.
For base points and centering given by $\vartheta$, let $w_\vartheta$, $u_{j,\vartheta}$, and $\tilde z_\vartheta=z-\mu_z$ be the lifted response, the transported predictors, and the centered covariates in the coordinates $\iota$, and set $r_\vartheta=w_\vartheta-\sum_j\alpha_ju_{j,\vartheta}-\sum_k\tilde z_{\vartheta,k}v_k$.
Define $\psi(O;\vartheta)=(\psi_Y,\psi_1,\ldots,\psi_p,\psi_z,\psi_{\bb\beta})$ by
\begin{gather*}
  \psi_Y=\iota\,\Log{m_Y(a_Y)}(Y),\quad
  \psi_j=\iota_j\,\Log{m_j(a_j)}(X_j),\quad
  \psi_z=z-\mu_z,\\
  \psi_{\bb\beta}=\bigl(\inner{u_{j,\vartheta}}{r_\vartheta}\bigr)_{j\le p}\oplus
  \bigl(\tilde z_{\vartheta,k}\,r_\vartheta\bigr)_{k\le q}.
\end{gather*}
At $\vartheta^*$, $P\psi_Y=-\iota\,\nabla F_Y(\FM Y)=0$, and likewise $P\psi_j=P\psi_z=0$.
The block $P\psi_{\bb\beta}(\cdot;\vartheta^*)=0$ is the population normal equation of Theorem~\ref{thm:normal}, so $P\psi(\cdot;\vartheta^*)=0$.
Let $\Theta_0$ be the neighborhood defined in the next paragraph.
On the event that the sample means lie in $\Theta_0$, the equations $\sum_i\Log{\hat{\bar Y}}(Y_i)=0$ and $\sum_i\Log{\hat{\bar X}_j}(X_{ij})=0$ are interior critical-point equations by the choice of $\Theta_0$, and on the event that the Gram matrix $\hat A$ of Section~\ref{sec:stats-inference} is nonsingular, $P_n\psi_{\bb\beta}=0$ is the ROLS normal equation, so $\hat\vartheta_n$ solves $P_n\psi(\cdot;\hat\vartheta_n)=0$.
On the corresponding event for the resample, $\hat\vartheta_n^*$ solves the same equation with $P_n$ replaced by $P_n^*$.
These events have probability tending to one, as shown below.

The estimating functions form a Donsker class on a neighborhood of $\vartheta^*$.
Let $\Theta_0$ be a compact convex neighborhood of $\vartheta^*$, small enough that the supports in~(i) remain strictly inside the injectivity radius at every base point in $\Theta_0$.
Convexity keeps the segment between any two parameters inside $\Theta_0$.
On this neighborhood, (i) and~(iii) make $w_\vartheta$, $u_{j,\vartheta}$, and their $\vartheta$-derivatives bounded, so $\psi_Y$ and $\psi_j$ are bounded and $\psi_z=O(1+\|z\|)$.
Since $\bb\beta$ ranges over a compact set, $r_\vartheta=O(1+\|z\|)$, and therefore $\inner{u_{j,\vartheta}}{r_\vartheta}=O(1+\|z\|)$ and $\tilde z_{\vartheta,k}r_\vartheta=O(1+\|z\|^2)$.
The same orders hold for the $\vartheta$-derivatives.
The largest is $\partial_{v_{k'}}(\tilde z_{\vartheta,k}r_\vartheta)=-\tilde z_{\vartheta,k}\tilde z_{\vartheta,k'}I_d$.
Hence $\|\psi(O;\vartheta_1)-\psi(O;\vartheta_2)\|\le L(O)\|\vartheta_1-\vartheta_2\|$ and $\sup_{\Theta_0}\|\psi(O;\vartheta)\|\le L(O)$, with $L(O)\le C(1+\|z\|^2)$.
Condition~(v) gives $\E L(O)^2<\infty$.
A class that is Lipschitz in a finite-dimensional parameter, with a square-integrable Lipschitz function, is $P$-Donsker~\citep[Example~19.7]{vanderVaart1998}.

Differentiating $P\psi$ at $\vartheta^*$ gives the Jacobian $-M$.
In the block order $(a_Y,a_1,\ldots,a_p,\mu_z,\bb\beta)$,
\[
  M=\begin{pmatrix}
    H_Y & & & & \\
    & H_1 & & & \\
    & & \ddots & & \\
    & & & I_q & \\
    J_Y & J_1 & \cdots & J_z & \Sigma_{\bb Z\bb Z}
  \end{pmatrix}.
\]
The diagonal blocks of $M$ are $H_Y,H_1,\ldots,H_p$, $I_q$, and $\Sigma_{\bb Z\bb Z}$, because at $\vartheta^*$ the derivatives of $P\psi_Y$, $P\psi_j$, $P\psi_z$, and $P\psi_{\bb\beta}$ in their own parameters are $-H_Y$, $-H_j$, $-I_q$, and $-\Sigma_{\bb Z\bb Z}$, and the derivative of $P\psi$ is $-M$.
Every block strictly above the diagonal is zero, because $\psi_Y$, $\psi_j$, and $\psi_z$ do not depend on $\bb\beta$ or on one another's parameters.
The blocks $J_\bullet=-\partial_\bullet P\psi_{\bb\beta}(\cdot;\vartheta^*)$ exist by~(iii).
In particular $J_z=0$, because $\E w=\E u_j=0$ at the Fr\'echet means and $\E\tilde z=0$.
Thus $M$ is block lower triangular, and it is nonsingular by~(ii) and~(iv).

This nonsingular Jacobian gives the asymptotic linearity of the sample estimator.
After shrinking $\Theta_0$ if necessary, $\vartheta^*$ is the only root of $P\psi(\cdot;\vartheta)=0$ in $\Theta_0$.
At a root, $a_Y$ is a critical point of $F_Y$ near $\FM Y$, which is isolated by~(ii), and likewise each $a_j$, then $\mu_z=\bar z$, and then $\bb\beta=\bb\beta^*$ by~(iv).
The sample Fr\'echet means converge to the population means~\citep{BhattacharyaPatrangenaru2003}, since the means are unique and the supports are compact.
The entries of $n^{-1}\hat A$ and of $n^{-1}\sum_i\bb Z_i w_i$, as functions of the base points and of $\mu_z$, are Lipschitz with envelope $C(1+\|z\|^2)$, so they converge uniformly over $\Theta_0$ by the Glivenko--Cantelli theorem.
At the sample means, $n^{-1}\hat A$ therefore converges to $\Sigma_{\bb Z\bb Z}$, which is nonsingular by~(iv), and $\hat\mu_z$ and $\hat{\bb\beta}_n$ converge to $\bar z$ and $\bb\beta^*$.
Hence the sample normal equations hold and $\hat\vartheta_n$ lies in $\Theta_0$ with probability tending to one, and $\hat\vartheta_n\to\vartheta^*$ in probability.
The $Z$-estimator theorem~\citep[Theorem~5.21]{vanderVaart1998} then yields
\[
  \sqrt n(\hat\vartheta_n-\vartheta^*)
  =n^{-1/2}\sum_{i=1}^n M^{-1}\psi(O_i;\vartheta^*)+o_P(1).
\]
The $\bb\beta$-block of $M^{-1}\psi(\cdot;\vartheta^*)$ is
\begin{equation}\label{eq:boot-if}
  \mathrm{IF}_{\bb\beta}
  =\Sigma_{\bb Z\bb Z}^{-1}\Bigl(\psi_{\bb\beta}
  -\sum_{m\in\{Y,1,\ldots,p\}}J_mH_m^{-1}\psi_m\Bigr).
\end{equation}
At $\vartheta^*$, $\psi_{\bb\beta}=\bb Z\varepsilon$.
The term $H_m^{-1}\psi_m$ is the influence function of the corresponding Fr\'echet mean, and $-\Sigma_{\bb Z\bb Z}^{-1}J_m$ is the corresponding block of $J_\theta$.
Therefore~\eqref{eq:boot-if} is the same influence function as~\eqref{eq:full-if}.
Thus $\Sigma_{\bb\beta}=\Var(\mathrm{IF}_{\bb\beta})$ under the conditions of the proposition.
Condition (D2) of Appendix~\ref{sec:S4} is the further assumption under which this variance splits as $\Sigma_{\bb\beta}=\Sigma_{\mathrm{tan}}+\Sigma_{\mathrm{curv}}$. The leading term of~\eqref{eq:boot-if} then contributes $\Sigma_{\mathrm{tan}}$, and the sum over Fr\'echet means contributes $\Sigma_{\mathrm{curv}}$.

The pairs bootstrap obeys the same expansion conditionally on the data.
Because $\{\psi(\cdot;\vartheta):\vartheta\in\Theta_0\}$ is Donsker, the bootstrap empirical process $\sqrt n(P_n^*-P_n)$ converges conditionally in probability to the same tight Gaussian limit as $\sqrt n(P_n-P)$~\citep[Theorem~3.6.1]{vanderVaartWellner1996}.
A distribution supported in the ball $B(\FM Y,r_Y)$ has every Fr\'echet minimizer in the closed ball $B(\FM Y,2r_Y)$, because outside that ball its Fr\'echet function exceeds $\tfrac12r_Y^2$, which is at least its value at $\FM Y$, and this closed ball is compact because $\Mman$ is complete.
On this ball, $\dM^2(y,\cdot)$ is bounded and Lipschitz uniformly over $y$ in the support, so the resampled Fr\'echet functions converge there uniformly, conditionally in probability, and uniqueness of the mean gives conditional consistency of the resampled mean of $Y$.
The same argument applies to each $X_j$.
The entries of $n^{-1}\hat A$ and of $n^{-1}\sum_i\bb Z_i w_i$ form Donsker classes by the same Lipschitz bound, so Theorem~3.6.1 of \citet{vanderVaartWellner1996} gives their bootstrap convergence, uniformly over $\Theta_0$ and conditionally in probability.
Hence $\hat\mu_z^{*}$ and $\hat{\bb\beta}_n^{*}$ are conditionally consistent, the resampled normal equations are nonsingular and $\hat\vartheta_n^*$ lies in $\Theta_0$ with conditional probability tending to one in probability, and $\hat\vartheta_n^*\to\vartheta^*$ conditionally in probability.
The derivative $-M$ is nonsingular, and the Lipschitz bound implies the $L^2$-continuity $P\|\psi(\cdot;\vartheta)-\psi(\cdot;\vartheta^*)\|^2\to0$ as $\vartheta\to\vartheta^*$.
The bootstrap $Z$-estimator theorem~\citep{WellnerZhan1996,ArconesGine1992} therefore gives, conditionally in probability,
\[
  \sqrt n(\hat\vartheta_n^*-\hat\vartheta_n)
  \rightsquigarrow
  \mathcal N\bigl(0,\,M^{-1}\Var\{\psi(O;\vartheta^*)\}M^{-\top}\bigr).
\]
The $\bb\beta$-block of this covariance is $\Sigma_{\bb\beta}$.
For each fixed $\tau$, conditional weak convergence implies that the second-moment matrix of $\sqrt n(\hat{\bb\beta}_n^{*}-\hat{\bb\beta}_n)$, truncated at squared norm $\tau$, converges in probability to the same truncation of the Gaussian covariance, because the limit puts no mass on the sphere of radius $\sqrt\tau$.
A diagonal argument extends this to some sequence $\tau_n\to\infty$.
The second moment without truncation can fail to converge: a resample of identical observations has positive conditional probability, the normal equations are then singular, and $\hat{\bb\beta}_n^{*}$ was defined arbitrarily on that event.
Section~\ref{sec:stats-inference} instead compares $\hat v_k^{*}$ with $\hat v_k$ after transporting $\hat v_k^{*}$ to $\Tan{\hat{\bar Y}}$.
The two differences differ by the holonomy around a geodesic triangle with sides $O_P(n^{-1/2})$.
That holonomy is $O_P(n^{-1})$, so the $\sqrt n$-scaled statistics differ by $O_P(n^{-1/2})$ and have the same limit law in any orthonormal frame at $\hat{\bar Y}$ that converges to a frame at $\FM Y$, and the Wald statistic does not depend on the frame.
\end{proof}

\section{Special Cases and Manifold-Specific Derivations}
\label{sec:S5}

This appendix gives the formulas behind the special cases summarized in Remark~\ref{rem:special-cases}, namely Euclidean linear regression, geodesic regression, parametric Fr\'echet regression, and DIDO regression.
It then records the manifold-specific operations used in the experiments.

\subsection{Euclidean Linear Regression ($\Mman = \R^d$)}

When $\Mman = \R^d$ with the standard inner product, all geometric operations collapse to linear algebra: $\Log{p}(q) = q - p$, $\Exp{p}(v) = p + v$, $\PT{p}{q} = \mathrm{id}$, and the Fr\'echet mean coincides with the ordinary expectation.
The model~\eqref{eq:model} becomes
\[
  Y_i - \bar Y
  = \sum_{j=1}^{p} \alpha_j (X_{ij} - \bar X_j)
  + \sum_{k=1}^{q} \tilde z_{ik}\, v_k + \varepsilon_i,
\]
which is a centered Euclidean vector-valued linear model with scalar block coefficients for the vector covariates.
When $d=1$, this is ordinary centered multiple linear regression.
The block normal equation~\eqref{eq:normal} reduces to the ordinary normal equations $\hat\Sigma_{\bb{Z}\bb{Z}} \hat\beta = \hat C_{\bb{Z}Y}$ with the usual sample covariance matrix.
In the correctly specified homoskedastic Gaussian case at fixed $d$, the Wald tests of Theorem~\ref{thm:inference} are the exact $F$-test for vector coefficients and the exact $t$-test for scalar coefficients.
Outside that case the reference distributions are asymptotic.
When $n/d$ is small, the $\chi^2_d$ reference over-rejects relative to the $F$ reference (Appendix~\ref{sec:S7-highdim}).

\subsection{Geodesic Regression ($p=0$, Single Scalar Predictor)}

When $p = 0$ and $q = 1$, the model has no manifold predictors and one scalar covariate $z$, yielding
\[
  \Log{\FM{Y}}(Y_i)
  = (z_i - \bar z)\, v + \varepsilon_i.
\]
The normal equation gives the closed-form solution $\hat v = \mathrm{cov}(z, Y) / \mathrm{var}(z)$, where $\mathrm{cov}(z, Y) = \frac{1}{n}\sum_i \tilde z_i\, w_i \in \Tan{\FM{Y}}$.
The predicted geodesic is $\hat Y(z) = \Exp{\FM{Y}}\bigl((z - \bar z)\,\hat v\bigr)$, which is a single geodesic ray from $\FM{Y}$ in the direction $\hat v$, precisely the tangent-space formulation of geodesic regression~\citep{Fletcher2013}.

More generally, in Euclidean space for $p = 0$ and $q \geq 2$, ROLS reduces to the parametric version of Fr\'echet regression~\citep{PetersenMuller2019frechet} in which the weight function $w_i(z)$ is linear in $z$.
The local version of Fr\'echet regression uses kernel weights $w_i(z) = K_h(z - z_i)/\sum_\ell K_h(z - z_\ell)$, making it nonparametric.
On a curved or periodic manifold, Fr\'echet regression with linear weights minimizes a weighted sum of squared manifold distances and so differs from ROLS with $p=0$.

\subsection{DIDO Regression ($\Mman = \mathcal{P}_2(\R)$)}

Distribution-in-distribution-out (DIDO) regression~\citep{ChenDIDO2025} places both predictors and responses in the $2$-Wasserstein space $\mathcal{P}_2(\R)$.
Via the quantile embedding $Q\colon \mathcal{P}_2(\R) \to L^2([0,1])$, the Wasserstein distance becomes the $L^2$ norm on quantile functions, and the tangent space at $\FM{Y}$ is identified with a closed subspace of $L^2([0,1])$.

On $\mathcal{P}_2(\R)$ the sectional curvature vanishes (the space is flat under the quantile embedding) and the optimal-transport map between any two quantile functions is the identity rearrangement.
Therefore the parallel transport along any geodesic connecting $\FM{X}_j$ to $\FM{Y}$ reduces to the identity map in the quantile representation.
Substituting into~\eqref{eq:model}, the ROLS model becomes
\[
  Q_{Y_i} - Q_{\FM{Y}}
  = \sum_{j=1}^p \alpha_j \bigl(Q_{X_{ij}} - Q_{\FM{X}_j}\bigr)
  + \sum_{k=1}^q \tilde z_{ik}\, v_k + \varepsilon_i,
\]
which is the linear model in the quantile Hilbert space studied by~\citep{ChenDIDO2025}, broadened here to admit scalar covariates alongside distributional predictors.

\begin{remark}[Discretization]
The quantile embedding is flat: curvature vanishes, the cut locus is empty, and parallel transport is the identity, so the model is linear regression in $L^2([0,1])$.
In the computations, quantile functions are discretized on a grid of $m$ points.
The fit is then ordinary linear regression in $\R^m$, and Theorems~\ref{thm:consistency}--\ref{thm:inference} apply to that discretization.
\end{remark}

\subsection{Manifold-Specific Derivations}

We record the explicit forms of the geometric operations for each manifold used in the experiments and for $\mathcal P_2(\R)$.

\textbf{Sphere $S^{d} \subset \R^{d+1}$.} $\Exp{p}(v) = \cos(\|v\|)\,p + \sin(\|v\|)\,v/\|v\|$; $\Log{p}(q) = \theta\,(q - \langle p,q\rangle p)/\|q - \langle p,q\rangle p\|$ where $\theta = \arccos(\langle p,q\rangle)$.
Parallel transport along the geodesic from $p$ to $q$: $\PT{p}{q}(v) = v - \frac{\langle \Log{p}(q), v\rangle}{\|\Log{p}(q)\|^2} \bigl(\Log{p}(q) + \Log{q}(p)\bigr)$.
Sectional curvature $\kappa = 1$ and $\inj(S^{d}) = \pi$.

\textbf{$\mathrm{SO}(3)$.}
Identified with unit quaternions modulo sign. $\Log{R}(S) = [\log(R^\top S)]^\vee \in \R^3$, where $\log$ is the matrix logarithm and $[\cdot]^\vee$ the inverse hat map. $\Exp{R}(v) = R\,\exp([v]^\wedge)$.
Parallel transport (Levi-Civita connection of the bi-invariant metric, in body-frame coordinates): $\PT{R}{S}(v) = \exp([-\xi/2]^\wedge)\,v$ with $\xi=\Log{R}(S)$.
Both the left-Cartan and bi-invariant Levi-Civita transports are linear isometries that preserve tangent norms, so the geometric guarantees of Theorem~\ref{thm:cutlocus} and the Rauch-comparison consistency bound (Theorem~\ref{thm:consistency}) apply under either convention; numerical coefficient values depend on the chosen convention.
Bi-invariant metric with $\kappa = 1/4$ and $\inj(\mathrm{SO}(3)) = \pi$.

\textbf{Flat torus $\mathbb T^d = \R^d/\Lambda$ (lattice $\Lambda$).} $\dM(p,q) = \min_{n \in \Lambda} \|p - q + n\|$; $\Log{p}(q) = q - p + n^*$ with $n^* = \argmin_{n \in \Lambda}\|q - p + n\|$; $\Exp{p}(v) = (p + v) \bmod \Lambda$.
The connection is flat, so parallel transport is the identity in coordinates, $\PT{p}{q}(v) = v$, and the holonomy of every loop is trivial.
Sectional curvature $\kappa = 0$ and $\inj(\mathbb T^d) = \tfrac12\,\mathrm{sys}(\mathbb T^d)$.
Trivial holonomy means parallel transport on $\mathbb T^d$ does not depend on the path.

\textbf{$\mathrm{Sym}^+(k)$ (affine-invariant metric).} $\dM(P,Q) = \|\log(P^{-1/2}QP^{-1/2})\|_F$. $\Log{P}(Q) = P^{1/2}\log(P^{-1/2}QP^{-1/2})P^{1/2}$; $\Exp{P}(V) = P^{1/2}\exp(P^{-1/2}VP^{-1/2})P^{1/2}$.
Parallel transport: $\PT{P}{Q}(V) = E_0\,V\,E_0^\top$, where $E_0 = (QP^{-1})^{1/2}$~\citep{Pennec2006}.
Non-positive curvature ($\kappa_+ = 0$), $\inj(\mathrm{Sym}^+(k)) = \infty$ (Hadamard manifold).

\textbf{Poincar\'e ball $\mathbb{H}^d$ (curvature $\kappa = -1$).} $\dM(x,y) = \mathrm{arcosh}\!\bigl(1 + \frac{2\|x-y\|^2}{(1-\|x\|^2)(1-\|y\|^2)}\bigr)$.
The conformal factor $\lambda_x = 2/(1-\|x\|^2)$ rescales the Euclidean inner product: $g_x(u,v) = \lambda_x^2 \langle u,v\rangle$.
Exponential and logarithm maps follow the M\"obius gyrovector formalism~\citep{Ganea2018}.
Parallel transport uses the gyroisometry: $\PT{x}{y}(v) = \frac{\lambda_x}{\lambda_y}\,\mathrm{gyr}[y,-x]\,v$.
Sectional curvature $\kappa = -1$ and $\inj(\mathbb{H}^d)=\infty$ (Hadamard manifold).
The experiments generate data in the ball and fit in the isometric Lorentz model (Appendix~\ref{sec:S7-dgp}).

\textbf{Wasserstein space $\mathcal P_2(\R)$.}
Probability measures on $\R$ with finite second moment under the $W_2$ metric.
Via the quantile-function map $\mu \mapsto F_\mu^{-1}: [0,1] \to \R$, $\mathcal P_2(\R)$ is isometric to the convex cone of non-decreasing functions in $L^2([0,1])$ and inherits its flat $L^2$ geometry~\citep{PetersenMuller2019wasserstein}. $\dM(\mu,\nu) = \|F_\mu^{-1} - F_\nu^{-1}\|_{L^2([0,1])}$; $\Log{\mu}(\nu) = F_\nu^{-1} - F_\mu^{-1}$ (in the quantile representation); $\Exp{\mu}(\xi) = (F_\mu^{-1} + \xi)_{\#}\mathrm{Unif}[0,1]$ when $F_\mu^{-1} + \xi$ is non-decreasing.
Parallel transport is the identity in the quantile representation: $\PT{\mu}{\nu}(\xi) = \xi$, since the embedded cone is flat in $L^2([0,1])$.
Sectional curvature $\kappa = 0$ and $\inj(\mathcal P_2(\R)) = \infty$.

\section{Fixed-Base Residual Bootstrap (Oracle-Variance Diagnostic)}
\label{sec:S6}

The nonparametric pairs bootstrap after Theorem~\ref{thm:inference} estimates the full covariance $\Sigma_{\bb\beta}$. Each replicate resamples whole observations and recomputes the Fr\'echet means, transports, and ROLS coefficients, so the covariance includes the cross term between the tangent-space score and the Fr\'echet-mean influence functions.
This appendix records a simpler pivotal bootstrap that conditions on the empirical Fr\'echet mean $\hat{\FM{Y}}_n$ and the transported tangent design, resampling only the tangent-space residuals.
By construction this fixed-base scheme targets the oracle tangent-space component $\Sigma_{\mathrm{tan}}$ of the asymptotic variance.

\begin{algorithm}[]
  \caption{Pivotal Residual Bootstrap for ROLS}
  \label{alg:bootstrap}
  \begin{algorithmic}[1]
  \REQUIRE Original data $\{(Y_i, X_{i1},\ldots,X_{ip}, z_{i1},\ldots,z_{iq})\}_{i=1}^n$,
           ROLS fit $(\hat{\bb\beta}, \hat w_i, \hat\varepsilon_i)$ at the
           empirical Fr\'echet mean $\hat{\FM{Y}}_n$, tangent design
           $(u_{ij}, \tilde z_{ik})$, number of bootstrap replicates $B$
  \FOR{$b = 1, \ldots, B$}
    \STATE Draw $\tilde\varepsilon_i^{*(b)} \stackrel{\mathrm{iid}}{\sim}
           \mathrm{Uniform}\{\hat\varepsilon_1,\ldots,\hat\varepsilon_n\}$
           and explicitly center: $\varepsilon_i^{*(b)} = \tilde\varepsilon_i^{*(b)} - n^{-1}\sum_{i'=1}^n \tilde\varepsilon_{i'}^{*(b)}$
    \STATE Construct bootstrap responses in $\Tan{\hat{\FM{Y}}_n}$:
           $w_i^{*(b)} = \hat w_i + \varepsilon_i^{*(b)}$
    \STATE Refit the tangent-space normal equations with
           $w_i^{*(b)}$ as responses and the
           same transported predictors $u_{ij}$ and centered
           covariates $\tilde z_{ik}$ to obtain
           $\hat{\bb\beta}^{*(b)} = (\hat\alpha_j^{*(b)}, \hat v_k^{*(b)})$
           and residual scale $\hat\sigma^{*(b)}$
    \STATE Compute the bootstrap pivotal statistic:
           \[
             T_j^{*(b)} = \frac{\hat\alpha_j^{*(b)} - \hat\alpha_j}
                                {\hat{\mathrm{se}}(\hat\alpha_j^{*(b)})}
             \quad\text{or}\quad
             W_k^{*(b)} = n\,
             (\hat v_k^{*(b)} - \hat v_k)^\top
             (\hat\Sigma_{\bb\beta}^{kk,*(b)})^{-1}
             (\hat v_k^{*(b)} - \hat v_k)
           \]
  \ENDFOR
  \STATE Compute the bootstrap $p$-value: for the $t$-test,
         $p_j^{\mathrm{boot}} = \bigl(1+\sum_{b=1}^B
         \mathbf{1}\bigl\{|T_j^{*(b)}| \geq |T_j|\bigr\}\bigr)/(B+1)$,
         where $T_j = \hat\alpha_j / \hat{\mathrm{se}}(\hat\alpha_j)$
         is the original test statistic.
         Analogously for the Wald test using $W_k^{*(b)} \geq W_k$.
         Both $T_j$ and $T_j^{*(b)}$ (and $W_k$, $W_k^{*(b)}$) use the
         model-based covariance of their own fit.
  \STATE \textit{Pairs variant.}  Replace lines~2--4 by resampling whole
         observations with replacement and refitting the Fr\'echet means,
         transports, and coefficients; parallel-transport
         $\hat v_k^{*(b)}$ from $\Tan{\hat{\FM{Y}}{}^{*(b)}}$ to
         $\Tan{\hat{\FM{Y}}_n}$ before forming $\hat v_k^{*(b)}-\hat v_k$.
  \end{algorithmic}
  \end{algorithm}
  
Under (D2) of Appendix~\ref{sec:S4}, the oracle tangent score is uncorrelated with the Fr\'echet-mean influence functions, and the variance splits as $\Sigma_{\bb\beta}=\Sigma_{\mathrm{tan}}+\Sigma_{\mathrm{curv}}$.
Under (D1) the tangent noise is isotropic, and $\Sigma_{\mathrm{tan}}$ is proportional to $\Sigma_{\bb Z\bb Z}^{-1}$.
Under (D3) the Fr\'echet-mean covariance is block diagonal, and $\Sigma_{\mathrm{curv}}$ is a sum over the $p+1$ means.
On a flat manifold $\Sigma_{\mathrm{curv}}=0$. Otherwise $\Sigma_{\mathrm{curv}}$ is controlled by the curvature and the Afsari radius.

Suppose we wish to test $H_0\colon \alpha_j = 0$ (or $H_0\colon v_k = 0$).
We have the ROLS fit on the original data: estimated parameters $\hat{\bb\beta} = (\hat\alpha_1,\ldots,\hat\alpha_p, \hat v_1,\ldots,\hat v_q)$, fitted values $\hat w_i = \sum_j \hat\alpha_j u_{ij} + \sum_k \tilde z_{ik} \hat v_k$, and residuals $\hat\varepsilon_i = w_i - \hat w_i \in \Tan{\FM{Y}}$.

The pivotal construction recomputes both the numerator and denominator of $T_j^{*(b)}$ on the bootstrap sample, so the statistic is pivotal and its distribution does not depend on $\sigma^2$.

Conditional on $\hat{\FM{Y}}_n$ and the transported tangent design, and under isotropic homoskedastic noise, the conditional law of $\sqrt n(\hat{\bb\beta}^{*(b)} - \hat{\bb\beta})$ converges to $\mathcal{N}(0, \Sigma_{\mathrm{tan}})$, and the conditional distribution of $T_j^{*(b)}$ converges in probability to $N(0,1)$.
The argument keeps the Fr\'echet-mean terms $J_m H_m^{-1}\psi_m$ in~\eqref{eq:boot-if} fixed.
Proposition~\ref{prop:pairs-boot} treats the pairs bootstrap, which recomputes those terms.

Condition throughout on the empirical Fr\'echet mean $\hat{\FM{Y}}_n$, which is held fixed across all bootstrap replicates in Algorithm~\ref{alg:bootstrap}.
Given $\hat{\FM{Y}}_n$, the tangent responses $w_i \in \Tan{\hat{\FM{Y}}_n}$, transported predictors $u_{ij}$, and centered covariates $\tilde z_{ik}$ all live in the fixed Euclidean space $\Tan{\hat{\FM{Y}}_n} \cong \R^d$, and the ROLS normal equations reduce to an ordinary least squares system for $\hat{\bb\beta} = (\hat\alpha_1,\ldots,\hat\alpha_p,\hat v_1^\top,\ldots,\hat v_q^\top)^\top$ with design matrix $\mathbf{X} \in \R^{nd \times (p+qd)}$ whose rows encode $(u_{ij}, \tilde z_{ik} e_\ell)$.
The resampled responses $w_i^{*(b)} = \hat w_i + \varepsilon_i^{*(b)}$ are generated by Efron's residual bootstrap applied to this Euclidean system with fixed design $\mathbf{X}$.
Algorithm~\ref{alg:bootstrap} recenters the resampled residuals. The coefficient $\hat{\bb\beta}^{*(b)}$ is unchanged, because $\sum_i\bb Z_i=0$ at the empirical Fr\'echet means with centered covariates.
\citet[Theorem 2.2, p.~1223]{Freedman1981} then applies directly: under the regression-model assumptions (1.1)--(1.4) of that paper (fixed design, finite error variance, and $n^{-1}\mathbf{X}^\top\mathbf{X} \to V$ positive definite, verified here by the full-rank and moment conditions underlying Theorem~\ref{thm:inference}), along almost all sample sequences,
\begin{enumerate}
\item[\textup{(a)}] the conditional law of $\sqrt{n}(\hat{\bb\beta}^{*(b)} - \hat{\bb\beta})$ given $\{(Y_i,X_{i1},\ldots,z_{iq})\}_{i=1}^n$ converges weakly to $N(0, \sigma^2 V^{-1})$;
\item[\textup{(b)}] the bootstrap variance estimator satisfies $\hat\sigma^{*(b)} \to \sigma$ in conditional probability; and
\item[\textup{(c)}] consequently, the conditional law of the pivotal quantity $(\mathbf{X}^\top\mathbf{X})^{1/2}(\hat{\bb\beta}^{*(b)} - \hat{\bb\beta}) /\hat\sigma^{*(b)}$ converges to the standard normal distribution on $\R^{p+qd}$.
\end{enumerate}
Applying (c) to the $j$-th coordinate yields $\sup_t |P^*(T_j^{*(b)} \leq t \mid \mathrm{data}) - \Phi(t)| \xrightarrow{P} 0$, where $\Phi$ is the standard normal distribution function.
The same continuous-mapping argument gives the Wald pivot $W_k^{*(b)}$.
\citet{MammenBoot1993} treats the corresponding Euclidean bootstrap when the dimension grows with $n$.

The Fr\'echet mean is held fixed, so this conditional variance is $\Sigma_{\mathrm{tan}}$.
The term $\Sigma_{\mathrm{curv}} = J_\theta\Sigma_\theta J_\theta^\top$ comes from the $O_P(n^{-1/2})$ fluctuation of $\hat{\FM{Y}}_n$ \citep{BhattacharyaLin2017} and appears when the mean is recomputed, as in the pairs bootstrap.
On a flat manifold, Appendix~\ref{sec:S4} gives $\Sigma_{\mathrm{curv}}=0$, and the residual bootstrap is then consistent for $\Sigma_{\bb\beta}$.

Each bootstrap replicate requires one ROLS fit, which is $O(n(p+qd)^2)$ (dominated by assembling the Gram matrix).
The Riemannian operations (log, exp, parallel transport) are shared with the original fit.
The total cost is therefore $B$ times that of one refit.

\section{Additional Simulation Results}
\label{sec:S7}

This appendix provides the full numerical simulation results and details for the theorem-verification experiments summarized in Section~\ref{sec:sims}.

\newcommand{\V}[2]{\hbox{$#1{\scriptstyle_{\pm#2}}$}}
\newcommand{\B}[2]{\hbox{$\mathbf{#1}{\scriptstyle_{\pm#2}}$}}

\subsection{Data-Generating Process and Baseline Implementations}
\label{sec:S7-dgp}

All synthetic experiments use Model~\eqref{eq:model} with manifold predictors $X_{i1},\ldots,X_{ip}$ sampled near per-manifold centers $\FM X_j$, scalar covariates $z_{ik}\stackrel{\mathrm{iid}}{\sim}\mathcal N(0,1)$, response base point $\FM Y$, and additive tangent noise of scale $\tau$.

\paragraph{Coefficients.}
The manifold-predictor coefficients are taken from the fixed pool $(0.5,-0.3,0.2,-0.15,0.1,-0.08,0.06,-0.04)$, retaining the first $p$ entries.
For each scalar predictor $k$, the tangent-vector coefficient $v_k\in\Tan{\FM Y}$ is drawn as an isotropic Gaussian direction projected to $\Tan{\FM Y}$ and rescaled to $\|v_k\|=0.2$; on $\mathrm{Sym}^+(3)$, the random vector is first symmetrized and then rescaled in the Frobenius norm.

\paragraph{Per-manifold base points and predictor spreads.}
Table~\ref{tab:S7-dgp-fixtures} lists $\FM Y$, the predictor centers $\FM X_1,\ldots,\FM X_4$ used cyclically when $p>4$, and the geodesic spread $r_X$ that controls how far the $X_{ij}$ deviate from $\FM X_j$.
Sampling at $\FM X_j$ uses a centered, scaled tangent draw followed by $\Exp{\FM X_j}$, with a $0.95$-norm projection on $\mathbb H^3$ to keep samples in the open ball.
In every experiment, the $\mathbb H^3$ samples are mapped isometrically to the Lorentz model and fitted there with closed-form logarithm, exponential, and parallel-transport maps.
The Fr\'echet mean is computed by the damped fixed-point iteration $\mu\leftarrow\Exp{\mu}\bigl(n^{-1}\sum_i\Log{\mu}(Y_i)\bigr)$ until the gradient reaches machine precision, and the error of $\hat v$ is measured in orthonormal frames after parallel transport.

\begin{table}[h]
\centering
\caption{Synthetic-experiment fixtures.  Predictor centers
$\FM X_1,\ldots,\FM X_4$ are reused cyclically when $p>4$.
$r_X$ is the predictor sampling spread (in radians for $\mathbb S^2$,
$\mathrm{SO}(3)$; in coordinate units for $\mathbb T^2$, $\mathbb H^3$;
and in the $\mathrm{Sym}^+(3)$ tangent norm at $\FM X_j$).}
\label{tab:S7-dgp-fixtures}
\small
\setlength{\tabcolsep}{4pt}
\renewcommand{\arraystretch}{1.2}
\begin{tabular}{l c c p{0.58\textwidth}}
\toprule
Manifold & $\FM Y$ & $r_X$ & Predictor centers $\FM X_1,\ldots,\FM X_4$ \\
\midrule
$\mathbb S^2$ & $(0,0,1)$ & $0.6$ &
$e_1,\ e_2,\ \tfrac{1}{\sqrt2}(e_1{+}e_2),\ \tfrac{1}{\sqrt2}(-e_1{+}e_2)$ \\
$\mathbb T^2$ & $(5.0,1.2)$ & $0.45$ &
$(4.8,1.0),\ (5.4,1.6),\ (4.4,1.8),\ (5.8,0.8)$ \\
$\mathrm{SO}(3)$ & $I_3$ & $0.5$ &
$\exp(\hat\omega_j)$ for $\omega_j \in \{(0.5,0.3,0.2),$ $(-0.4,0.6,-0.1),$ $(0.2,-0.5,0.4),$ $(-0.3,-0.2,0.7)\}$ \\
$\mathrm{Sym}^+(3)$ & $1.5\,I_3$ & $0.3$ &
$\Exp{c_j I_3}(\tfrac12(A_j{+}A_j^\top))$ for $c_j \in \{1.2, 0.8, 1.0, 1.4\}$, $A_j \sim 0.1\,\mathcal N(0,I_{3\times 3})$ \\
$\mathbb H^3$ & $(0.2,0,0)$ & $0.15$ &
$0.3\,d_j$ for $d_j\in$ $\{(0.3,0.1,0),$ $(-0.1,0.3,0),$ $(0,-0.1,0.3),$ $(0.2,0.2,-0.1)\}$ \\
\bottomrule
\end{tabular}
\end{table}

\paragraph{Tangent noise.}
The additive tangent noise $\varepsilon_i$ is built from independent $\mathcal N(0,\tau^2)$ entries.
On $\mathbb S^2$ and $\mathbb T^2$ it has intrinsic dimension $2$ and is expressed in an orthonormal frame at $\FM Y$.
On $\mathrm{SO}(3)$ and $\mathbb H^3$ it has intrinsic dimension $3$ and is expressed in the standard frame, with $\mathrm{SO}(3)$ noise realized as a skew matrix and $\mathbb H^3$ noise applied in ambient Poincar\'e coordinates so that the Riemannian magnitude scales with the conformal factor $\lambda_{\FM Y}=2/(1-\|\FM Y\|^2)\approx 2.083$.
On $\mathrm{Sym}^+(3)$ a $3\times 3$ Gaussian matrix $A$ with iid $\mathcal N(0,\tau^2)$ entries is symmetrized as $\tfrac12(A+A^\top)$, giving a $6$-dimensional symmetric tangent draw.
In every case the linear predictor $m_i \coloneqq m_{\bb\beta}(X_i,z_i)$ is formed in $\Tan{\FM Y}$ and the response is $Y_i=\Exp{\FM Y}(m_i+\varepsilon_i)$.

\paragraph{Configurations and replications.}
The default configuration is $(n,p,q,\tau)=(200,2,3,0.15)$; the prediction-accuracy table sweeps $\tau\in\{0.05,0.15,0.30\}$, $q\in\{0,3,6\}$, $n\in\{100,200,500\}$, and $p\in\{1,2,4\}$, giving $9$ cells per manifold ($45$ in total) with $100$ Monte Carlo replications per cell and $n=100$ held-out test observations.

\paragraph{Baseline implementations.}
All baselines are evaluated on the same train/test splits as ROLS.
The six baselines used in Fig.~\ref{fig:headline}(d) and Table~\ref{tab:comp-sigma-q} are as follows.
Fr\'echet mean predicts the empirical Karcher mean of $\{Y_i\}$.
Geodesic regression fits a multi-scalar tangent-linear model in $\Tan{\FM Y}$, the $p{=}0$, $q{\geq}1$ special case of ROLS (the original single-scalar geodesic regression of \citet{Fletcher2013} extends to multiple scalars by stacking tangent-vector coefficients $v_k$).
Fr\'echet regression~\citep{PetersenMuller2019frechet} predicts the weighted Fr\'echet mean $\arg\min_y\sum_i w_i(z)\,d^2(Y_i,y)$ with the global linear weights $w_i(z)=n^{-1}\{1+(z_i-\bar z)^\top\hat S_z^{-1}(z-\bar z)\}$, where $\hat S_z$ is the sample covariance of the scalar covariates and the weights may be negative.
Single-index Fr\'echet (SI-Fr.)~\citep{BhattacharjeeMueller2023} searches $500$ random unit directions $\theta$, selects a bandwidth for each by $5$-fold cross-validation, and keeps the direction with the smallest in-sample loss.
Its prediction is the local linear weighted Fr\'echet mean on the index $z^\top\theta$ with a Gaussian kernel.
We compute the loss on the whole sample, which \citet{BhattacharjeeMueller2023} allow in place of their binned loss, and we evaluate the loss and the cross-validation error with tangent residuals at $\FM Y$ to keep the search feasible.
NL-Fr\'echet (NL-Fr.)~\citep{BhattacharjeeLiXue2025} predicts the weighted Fr\'echet mean with weights $w(x)=n^{-1}\mathbf 1+G(G+\epsilon I)^{-1}Q(K+\epsilon I)^{-1}Q\,k(x)$, where $K$ is the kernel matrix, $k(x)$ is the vector of kernel values at $x$, $Q=I-n^{-1}\mathbf 1\mathbf 1^\top$, and $G=QKQ$.
The kernel is the Gaussian radial basis function $\exp(-\gamma\|f-f'\|^2)$ on the features $f$ (manifold-predictor logarithms at each $\hat{\bar X}_j$, stacked with $z$), with $\gamma=(2\sigma^2)^{-1}$ and $\sigma^2$ the mean squared pairwise distance.
We use these features because \citet{BhattacharjeeLiXue2025} require a positive definite kernel, and the Gaussian kernel of the geodesic distance has negative eigenvalues on $\mathbb S^2$ and $\mathrm{SO}(3)$ in our simulations.
The constant $\epsilon$ minimizes their generalized cross-validation criterion over $\{10^{-6},\dots,10^{2}\}$, which extends their grid above $10^{-1}$ because the criterion often has its minimum there.
All three weighted Fr\'echet means are computed by gradient descent, started from the best, by objective value, of the weighted tangent average and $16$ sampled responses.
ROLS\,(m) is the manifold-only ablation that drops the scalar block.
Separately, an ROLS ablation called tangent linear ridge replaces each scalar coefficient $\alpha_j$ with a full $d\times d$ matrix $A_j$ in $\Tan{\FM Y}$ and adds a ridge penalty of strength $10^{-3}$, sharing ROLS's tangent design otherwise; its RMSE is reported inline in Section~\ref{sec:sims} rather than included in the figures or table.

\subsection{\texorpdfstring{Prediction Accuracy Across Manifolds}{Prediction Accuracy Across Manifolds}}

\begin{figure}[!htbp]
\centering
\includegraphics[width=\textwidth]{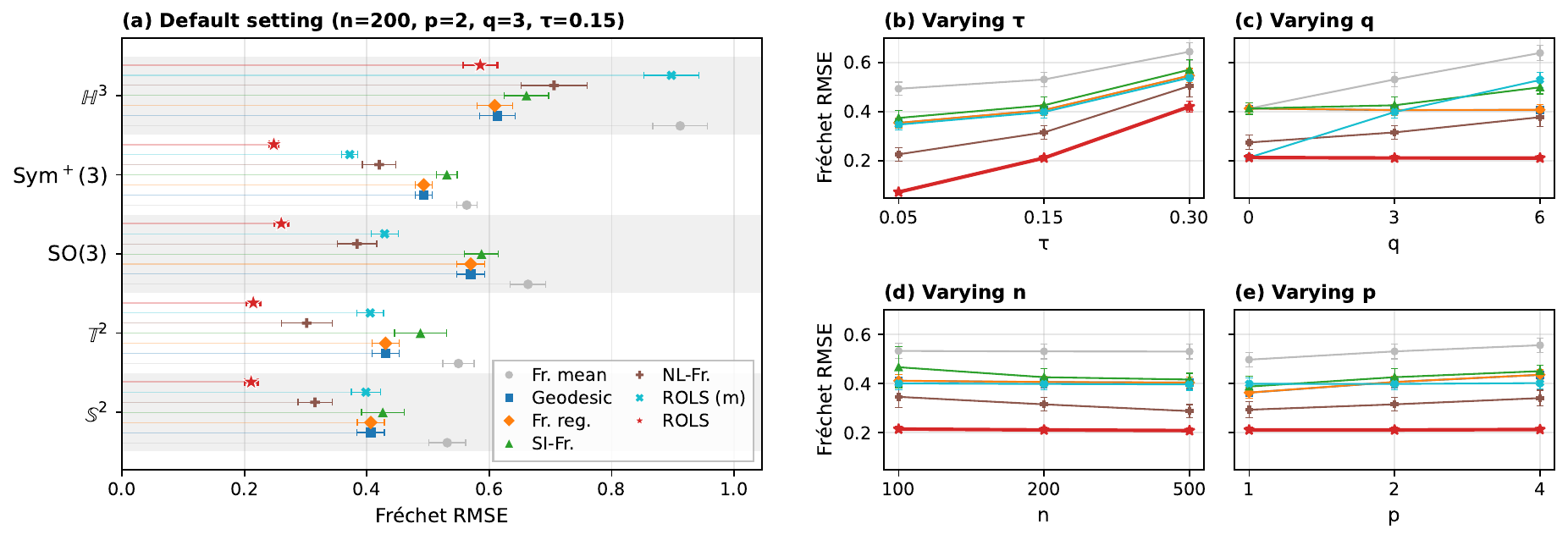}
\caption{Predictive accuracy across configurations. (a)~Default-cell Fr\'echet RMSE across the five response manifolds (same data as main-text Fig.~\ref{fig:headline}(d), shown here for reference). (b)--(e)~Sensitivity on $\mathbb S^2$ to noise level $\tau$, scalar count $q$, sample size $n$, and manifold-predictor count $p$, sharing a unified $y$-axis.
Full numerical entries are in Table~\ref{tab:comp-sigma-q}.}
\label{fig:sim-summary}
\end{figure}

Table~\ref{tab:comp-sigma-q} reports Fr\'echet RMSE on $\mathbb S^2$, $\mathrm{SO}(3)$, $\mathrm{Sym}^+(3)$, $\mathbb H^3$, and the flat torus~$\mathbb T^2$ under varying noise level~$\tau$, scalar-covariate count~$q$, sample size~$n$, and manifold-predictor count~$p$, with $100$ Monte Carlo replications per cell.
The default configuration is $n{=}200$, $p{=}2$, $q{=}3$, $\tau{=}0.15$.
ROLS is compared against the same six baselines as in Fig.~\ref{fig:headline}(d) (Fr\'echet mean, geodesic regression, Fr\'echet regression, single-index Fr\'echet (SI-Fr.), nonlinear Fr\'echet (NL-Fr.), and the manifold-only ablation ROLS\,(m)).
Figure~\ref{fig:sim-summary} shows the same grid, with panel~(a) the default cell from Fig.~\ref{fig:headline}(d) and panels~(b)--(e) the sweeps of $\tau$, $q$, $n$, and $p$ on $\mathbb S^2$.

\begin{landscape}
\begin{table}[p]
\centering
\caption{Full simulation grid: Fr\'echet RMSE\, (mean$_{\pm\text{std}}$,
100 trials) under varying $\tau$, $q$, $n$, and $p$ on
$\mathbb S^2$, $\mathrm{SO}(3)$, $\mathrm{Sym}^+(3)$, $\mathbb H^3$, and the
flat torus~$\mathbb T^2$ (rightmost block).
Default: $n{=}200$, $p{=}2$, $q{=}3$, $\tau{=}0.15$.
SI-Fr.\ $=$ single-index Fr\'echet regression;
NL-Fr.\ $=$ nonlinear Fr\'echet regression.
\textbf{Bold}: best per column within each three-column group.}
\label{tab:comp-sigma-q}
\scriptsize
\setlength{\tabcolsep}{0.85pt}
\renewcommand{\arraystretch}{1.0}
\begin{tabular}{@{}l *{15}{c}@{}}
\toprule
& \multicolumn{3}{c}{$\mathbb{S}^2$} & \multicolumn{3}{c}{$\mathrm{SO}(3)$} & \multicolumn{3}{c}{$\mathrm{Sym}^+(3)$} & \multicolumn{3}{c}{$\mathbb{H}^3$} & \multicolumn{3}{c}{$\mathbb{T}^2$} \\
\cmidrule(lr){2-4}\cmidrule(lr){5-7}\cmidrule(lr){8-10}\cmidrule(lr){11-13}\cmidrule(lr){14-16}
$\tau$ & .05 & .15 & .30 & .05 & .15 & .30 & .05 & .15 & .30 & .05 & .15 & .30 & .05 & .15 & .30 \\
\midrule
Fr.\ mean  & \V{.49}{.03} & \V{.53}{.03} & \V{.64}{.04} & \V{.62}{.03} & \V{.66}{.03} & \V{.80}{.03} & \V{.51}{.02} & \V{.56}{.02} & \V{.71}{.02} & \V{.76}{.04} & \V{.91}{.04} & \V{1.31}{.06} & \V{.51}{.03} & \V{.55}{.03} & \V{.66}{.03} \\
Geodesic    & \V{.35}{.02} & \V{.41}{.02} & \V{.55}{.03} & \V{.51}{.02} & \V{.57}{.02} & \V{.73}{.03} & \V{.43}{.01} & \V{.49}{.01} & \V{.65}{.02} & \V{.27}{.01} & \V{.61}{.03} & \V{1.17}{.06} & \V{.38}{.02} & \V{.43}{.02} & \V{.57}{.03} \\
Fr.\ reg    & \V{.35}{.02} & \V{.41}{.02} & \V{.55}{.03} & \V{.51}{.02} & \V{.57}{.02} & \V{.73}{.03} & \V{.43}{.01} & \V{.49}{.01} & \V{.65}{.02} & \V{.27}{.01} & \V{.61}{.03} & \B{1.15}{.05} & \V{.38}{.02} & \V{.43}{.02} & \V{.57}{.03} \\
SI-Fr.      & \V{.37}{.03} & \V{.43}{.03} & \V{.57}{.04} & \V{.53}{.03} & \V{.59}{.03} & \V{.75}{.05} & \V{.48}{.01} & \V{.53}{.02} & \V{.69}{.04} & \V{.38}{.02} & \V{.66}{.04} & \V{1.18}{.06} & \V{.44}{.04} & \V{.49}{.04} & \V{.63}{.06} \\
NL-Fr.      & \V{.23}{.03} & \V{.32}{.03} & \V{.50}{.04} & \V{.26}{.05} & \V{.38}{.03} & \V{.62}{.04} & \V{.29}{.07} & \V{.42}{.03} & \V{.63}{.03} & \V{.38}{.06} & \V{.71}{.05} & \V{1.21}{.05} & \V{.19}{.05} & \V{.30}{.04} & \V{.51}{.05} \\
ROLS\,(m)   & \V{.35}{.02} & \V{.40}{.02} & \V{.54}{.03} & \V{.35}{.02} & \V{.43}{.02} & \V{.62}{.03} & \V{.29}{.01} & \V{.37}{.01} & \V{.57}{.02} & \V{.74}{.04} & \V{.90}{.05} & \V{1.30}{.06} & \V{.35}{.02} & \V{.41}{.02} & \V{.55}{.03} \\
ROLS        & \B{.07}{.00} & \B{.21}{.01} & \B{.42}{.02} & \B{.09}{.00} & \B{.26}{.01} & \B{.52}{.02} & \B{.08}{.00} & \B{.25}{.01} & \B{.50}{.01} & \B{.20}{.01} & \B{.59}{.03} & \V{1.16}{.05} & \B{.07}{.00} & \B{.21}{.01} & \B{.43}{.02} \\
\midrule
$q$ & 0 & 3 & 6 & 0 & 3 & 6 & 0 & 3 & 6 & 0 & 3 & 6 & 0 & 3 & 6 \\
\midrule
Fr.\ mean  & \V{.41}{.02} & \V{.53}{.03} & \V{.64}{.03} & \V{.57}{.03} & \V{.66}{.03} & \V{.75}{.03} & \V{.49}{.01} & \V{.56}{.02} & \V{.63}{.02} & \V{.57}{.02} & \V{.91}{.04} & \V{1.17}{.05} & \V{.43}{.02} & \V{.55}{.03} & \V{.65}{.04} \\
Geodesic    & \V{.41}{.02} & \V{.41}{.02} & \V{.41}{.02} & \V{.57}{.03} & \V{.57}{.02} & \V{.57}{.03} & \V{.49}{.01} & \V{.49}{.01} & \V{.50}{.01} & \V{.57}{.02} & \V{.61}{.03} & \V{.66}{.03} & \V{.43}{.02} & \V{.43}{.02} & \V{.43}{.02} \\
Fr.\ reg    & \V{.41}{.02} & \V{.41}{.02} & \V{.41}{.02} & \V{.57}{.03} & \V{.57}{.02} & \V{.57}{.03} & \V{.49}{.01} & \V{.49}{.01} & \V{.50}{.01} & \V{.57}{.02} & \V{.61}{.03} & \V{.66}{.03} & \V{.43}{.02} & \V{.43}{.02} & \V{.43}{.02} \\
SI-Fr.      & \V{.41}{.02} & \V{.43}{.03} & \V{.50}{.03} & \V{.57}{.03} & \V{.59}{.03} & \V{.65}{.03} & \V{.49}{.01} & \V{.53}{.02} & \V{.58}{.02} & \V{.57}{.02} & \V{.66}{.04} & \V{.91}{.05} & \V{.43}{.02} & \V{.49}{.04} & \V{.53}{.03} \\
NL-Fr.      & \V{.27}{.03} & \V{.32}{.03} & \V{.38}{.04} & \V{.33}{.03} & \V{.38}{.03} & \V{.45}{.05} & \V{.32}{.02} & \V{.42}{.03} & \V{.47}{.03} & \V{.57}{.02} & \V{.71}{.05} & \V{.84}{.11} & \V{.26}{.03} & \V{.30}{.04} & \V{.36}{.05} \\
ROLS\,(m)   & \B{.21}{.01} & \V{.40}{.02} & \V{.53}{.03} & \B{.26}{.01} & \V{.43}{.02} & \V{.55}{.03} & \B{.25}{.01} & \V{.37}{.01} & \V{.47}{.02} & \B{.55}{.02} & \V{.90}{.05} & \V{1.16}{.05} & \B{.21}{.01} & \V{.41}{.02} & \V{.53}{.03} \\
ROLS        & \V{.21}{.01} & \B{.21}{.01} & \B{.21}{.01} & \V{.26}{.01} & \B{.26}{.01} & \B{.26}{.01} & \V{.25}{.01} & \B{.25}{.01} & \B{.25}{.01} & \V{.55}{.02} & \B{.59}{.03} & \B{.63}{.03} & \V{.21}{.01} & \B{.21}{.01} & \B{.22}{.01} \\
\midrule
$n$ & 100 & 200 & 500 & 100 & 200 & 500 & 100 & 200 & 500 & 100 & 200 & 500 & 100 & 200 & 500 \\
\midrule
Fr.\ mean  & \V{.53}{.03} & \V{.53}{.03} & \V{.53}{.03} & \V{.67}{.03} & \V{.66}{.03} & \V{.66}{.03} & \V{.56}{.02} & \V{.56}{.02} & \V{.56}{.02} & \V{.91}{.04} & \V{.91}{.04} & \V{.91}{.04} & \V{.55}{.03} & \V{.55}{.03} & \V{.55}{.03} \\
Geodesic    & \V{.41}{.02} & \V{.41}{.02} & \V{.40}{.02} & \V{.58}{.03} & \V{.57}{.02} & \V{.57}{.02} & \V{.50}{.01} & \V{.49}{.01} & \V{.49}{.01} & \V{.62}{.03} & \V{.61}{.03} & \V{.61}{.03} & \V{.44}{.02} & \V{.43}{.02} & \V{.43}{.02} \\
Fr.\ reg    & \V{.41}{.02} & \V{.41}{.02} & \V{.40}{.02} & \V{.58}{.03} & \V{.57}{.02} & \V{.57}{.02} & \V{.50}{.01} & \V{.49}{.01} & \V{.49}{.01} & \V{.62}{.03} & \V{.61}{.03} & \V{.61}{.03} & \V{.44}{.02} & \V{.43}{.02} & \V{.43}{.02} \\
SI-Fr.      & \V{.47}{.08} & \V{.43}{.03} & \V{.42}{.03} & \V{.61}{.05} & \V{.59}{.03} & \V{.58}{.02} & \V{.54}{.02} & \V{.53}{.02} & \V{.53}{.01} & \V{.68}{.06} & \V{.66}{.04} & \V{.66}{.03} & \V{.54}{.09} & \V{.49}{.04} & \V{.47}{.02} \\
NL-Fr.      & \V{.35}{.04} & \V{.32}{.03} & \V{.29}{.03} & \V{.44}{.05} & \V{.38}{.03} & \V{.34}{.03} & \V{.45}{.04} & \V{.42}{.03} & \V{.35}{.02} & \V{.69}{.05} & \V{.71}{.05} & \V{.66}{.03} & \V{.34}{.05} & \V{.30}{.04} & \V{.27}{.03} \\
ROLS\,(m)   & \V{.40}{.02} & \V{.40}{.02} & \V{.40}{.02} & \V{.43}{.02} & \V{.43}{.02} & \V{.43}{.02} & \V{.37}{.01} & \V{.37}{.01} & \V{.37}{.01} & \V{.90}{.04} & \V{.90}{.05} & \V{.90}{.04} & \V{.41}{.02} & \V{.41}{.02} & \V{.40}{.02} \\
ROLS        & \B{.21}{.01} & \B{.21}{.01} & \B{.21}{.01} & \B{.26}{.01} & \B{.26}{.01} & \B{.26}{.01} & \B{.25}{.01} & \B{.25}{.01} & \B{.25}{.01} & \B{.59}{.03} & \B{.59}{.03} & \B{.58}{.03} & \B{.22}{.01} & \B{.21}{.01} & \B{.21}{.01} \\
\midrule
$p$ & 1 & 2 & 4 & 1 & 2 & 4 & 1 & 2 & 4 & 1 & 2 & 4 & 1 & 2 & 4 \\
\midrule
Fr.\ mean  & \V{.50}{.03} & \V{.53}{.03} & \V{.56}{.03} & \V{.61}{.03} & \V{.66}{.03} & \V{.70}{.03} & \V{.48}{.02} & \V{.56}{.02} & \V{.59}{.02} & \V{.91}{.04} & \V{.91}{.04} & \V{.92}{.05} & \V{.51}{.02} & \V{.55}{.03} & \V{.57}{.03} \\
Geodesic    & \V{.36}{.02} & \V{.41}{.02} & \V{.44}{.02} & \V{.51}{.02} & \V{.57}{.02} & \V{.61}{.02} & \V{.39}{.01} & \V{.49}{.01} & \V{.52}{.02} & \V{.60}{.03} & \V{.61}{.03} & \V{.62}{.03} & \V{.38}{.02} & \V{.43}{.02} & \V{.46}{.02} \\
Fr.\ reg    & \V{.36}{.02} & \V{.41}{.02} & \V{.44}{.02} & \V{.51}{.02} & \V{.57}{.02} & \V{.61}{.02} & \V{.39}{.01} & \V{.49}{.01} & \V{.52}{.02} & \V{.60}{.03} & \V{.61}{.03} & \V{.61}{.03} & \V{.38}{.02} & \V{.43}{.02} & \V{.46}{.02} \\
SI-Fr.      & \V{.39}{.04} & \V{.43}{.03} & \V{.45}{.02} & \V{.53}{.02} & \V{.59}{.03} & \V{.63}{.03} & \V{.44}{.02} & \V{.53}{.02} & \V{.56}{.02} & \V{.65}{.03} & \V{.66}{.04} & \V{.67}{.04} & \V{.44}{.04} & \V{.49}{.04} & \V{.52}{.04} \\
NL-Fr.      & \V{.29}{.03} & \V{.32}{.03} & \V{.34}{.03} & \V{.35}{.03} & \V{.38}{.03} & \V{.43}{.03} & \V{.36}{.03} & \V{.42}{.03} & \V{.45}{.03} & \V{.69}{.05} & \V{.71}{.05} & \V{.70}{.05} & \V{.28}{.03} & \V{.30}{.04} & \V{.33}{.04} \\
ROLS\,(m)   & \V{.40}{.02} & \V{.40}{.02} & \V{.40}{.02} & \V{.43}{.02} & \V{.43}{.02} & \V{.43}{.02} & \V{.37}{.01} & \V{.37}{.01} & \V{.37}{.01} & \V{.90}{.04} & \V{.90}{.05} & \V{.90}{.05} & \V{.40}{.02} & \V{.41}{.02} & \V{.41}{.02} \\
ROLS        & \B{.21}{.01} & \B{.21}{.01} & \B{.21}{.01} & \B{.26}{.01} & \B{.26}{.01} & \B{.26}{.01} & \B{.25}{.01} & \B{.25}{.01} & \B{.25}{.01} & \B{.58}{.03} & \B{.59}{.03} & \B{.59}{.03} & \B{.21}{.01} & \B{.21}{.01} & \B{.21}{.01} \\
\bottomrule
\end{tabular}

\end{table}
\end{landscape}

\subsection{Cut-Locus Sweep and Convergence Rate}

\begin{figure}[h]
\centering
\includegraphics[width=\textwidth]{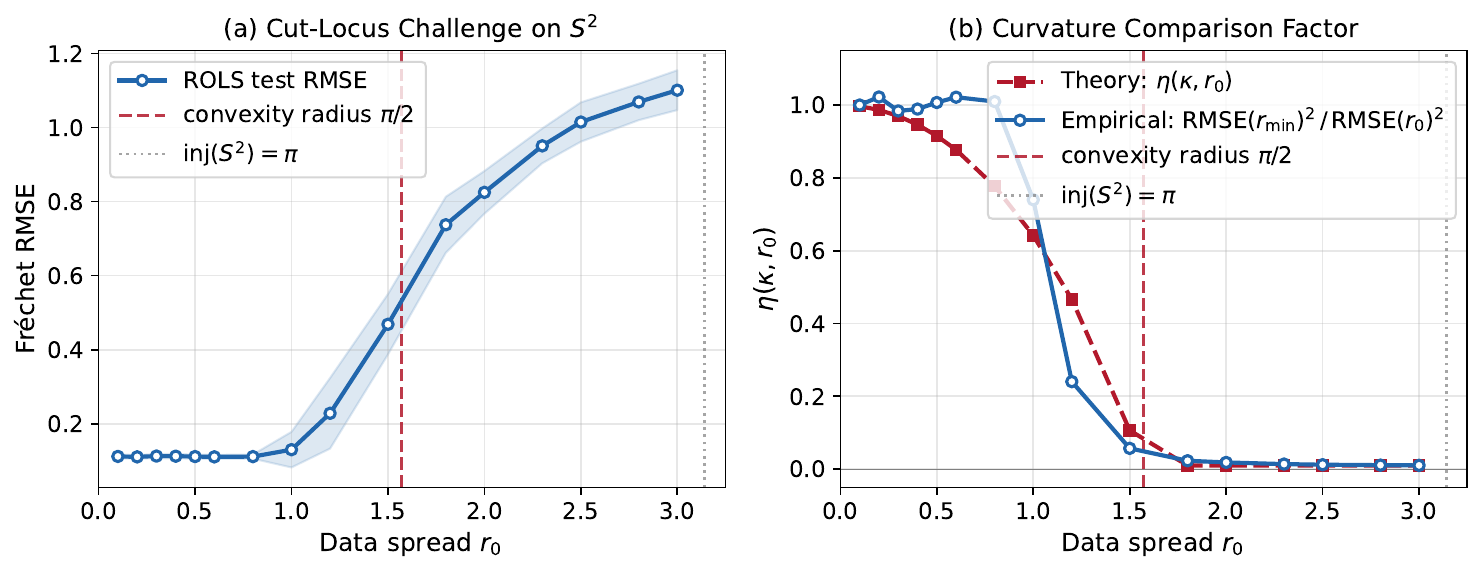}
\caption{Cut-locus challenge on $S^2$ (Theorems~\ref{thm:cutlocus} \& \ref{thm:consistency}). (a) ROLS test RMSE versus data spread $r_0$; the red dashed line marks the Karcher convexity radius $\pi/2$ and the gray dotted line marks $\mathrm{inj}(S^2) = \pi$. (b) Hessian comparison factor $\eta(1, r_0)$ from~\eqref{eq:eta} (red) and empirical ratio $\mathrm{RMSE}(r_{\min})^2/\mathrm{RMSE}(r_0)^2$ (blue), with the same two reference lines, illustrating the degradation of the tangent-space approximation as data spread beyond the origin.}
\label{fig:cutlocus}
\end{figure}

On $S^2$ ($\kappa=1$, $\inj(S^2)=\pi$), we probe the injectivity condition~\eqref{eq:inj-cond} by varying the data spread $r_0 \in [0.1, 3.0]$.
Figure~\ref{fig:cutlocus}(a) shows that RMSE is stable at approximately $0.11$ for $r_0 \le 0.8$ and rises as $r_0$ approaches the Karcher convexity radius $\pi/2$ (where Afsari support fails and the squared distance ceases to be convex), well before the cut locus at $\pi$.
This behavior illustrates Theorem~\ref{thm:cutlocus}.

Panel (b) plots the Hessian comparison factor $\eta(\kappa_+, r_0)$ from~\eqref{eq:eta} alongside the empirical ratio $\mathrm{RMSE}(r_{\min})^2/\mathrm{RMSE}(r_0)^2$.
On $S^2$, $\kappa_-=\kappa_+=1$, so the factor $\tilde\eta$ in Theorem~\ref{thm:consistency} equals one and does not grow with $r_0$.
The rise in RMSE tracks the misspecification bias $b^2$ as the tangent-linear approximation picks up higher-order curvature terms.
The empirical ratio stays near one for $r_0\le0.8$ while $\eta$ declines, falls below $\eta$ by $r_0=1.2$, and both approach zero near $\pi/2$.
The lower bound is $\dM^2 \geq \eta\norm{w-\hat{w}}^2$.
This bound controls how faithfully the tangent-space residual $\norm{w-\hat{w}}$ represents the geodesic residual $\dM(\hat Y, Y)$, and this relationship deteriorates as data spread further from the Fr\'echet mean.

\begin{figure}[h]
\centering
\includegraphics[width=\textwidth]{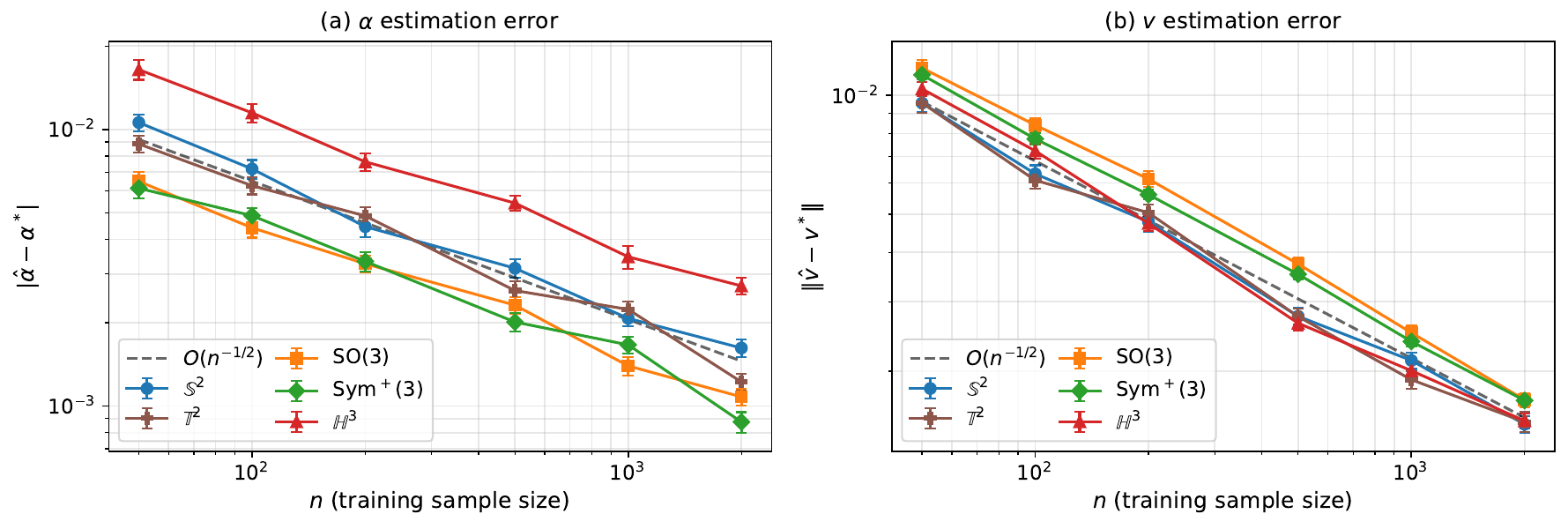}
\caption{Convergence of ROLS parameter estimates as $n$ grows (up to $2{,}000$): (a) error $|\hat\alpha - \alpha^*|$; (b) error $\|\hat v - v^*\|$.
Curves: $\mathbb{S}^2$, $\mathbb{T}^2$, $\mathrm{SO}(3)$, $\mathrm{Sym}^+(3)$, $\mathbb{H}^3$ (fitted in the exact Lorentz model, Appendix~\ref{sec:S7-dgp}).
Dashed line: reference $O(n^{-1/2})$ slope in log--log space.}
\label{fig:convergence}
\end{figure}

Figure~\ref{fig:convergence} plots parameter estimation errors versus $n$ on a log--log scale for all five manifolds used elsewhere in the simulations, with training sample size $n$ up to $2{,}000$ (100 replications at each~$n$).
Table~\ref{tab:conv-slopes} reports per-manifold log--log slopes and goodness-of-fit~($R^2$).

\begin{table}[h]
\centering
\footnotesize
\caption{Log--log regression slopes for parameter estimation errors vs.\ $n$
  on $\mathbb{S}^2$, the flat torus~$\mathbb{T}^2$, $\mathrm{SO}(3)$,
  $\mathrm{Sym}^+(3)$, and $\mathbb{H}^3$
  (100 trials at each~$n$; $n \in \{50,100,200,500,1000,2000\}$; $\mathbb{H}^3$ is fitted in the exact Lorentz model with a converged Fr\'echet mean, Appendix~\ref{sec:S7-dgp}).
  Theory predicts slope $-0.50$ for~$\hat\alpha$ and~$\hat v$ at this rate.}
\label{tab:conv-slopes}
\begin{tabular}{lccccc}
\toprule
& $\mathbb{S}^2$ & $\mathbb{T}^2$ & $\mathrm{SO}(3)$ & $\mathrm{Sym}^+(3)$ & $\mathbb{H}^3$ \\
\midrule
$\hat\alpha$ slope & $-0.51$ & $-0.52$ & $-0.49$ & $-0.51$ & $-0.49$ \\
$\hat\alpha$ $R^2$  & $0.99$ & $0.98$ & $0.99$ & $0.98$ & $1.00$ \\
$\hat v$ slope      & $-0.50$ & $-0.51$ & $-0.53$ & $-0.51$ & $-0.54$ \\
$\hat v$ $R^2$      & $1.00$ & $0.99$ & $1.00$ & $1.00$ & $0.99$ \\
\bottomrule
\end{tabular}
\end{table}

On all five manifolds, both $\hat\alpha$ and $\hat v$ log--log slopes are close to $-0.50$ with $R^2 \geq 0.98$, in line with the $O(n^{-1/2})$ rate in Corollary~\ref{cor:beta-rate}.

\subsection{Type~I Error Calibration and Power}

We evaluate the inference procedure of Theorem~\ref{thm:inference} via Monte Carlo on five manifolds ($S^2$, $\mathbb{T}^2$, $\mathrm{SO}(3)$, $\mathrm{Sym}^+(3)$, $\mathbb{H}^3$) with $n=200$, $\tau=0.15$, and $500$ replications.
Effect size is varied from $0$ (null) to $0.5$.

\begin{figure}[h]
\centering
\includegraphics[width=\textwidth]{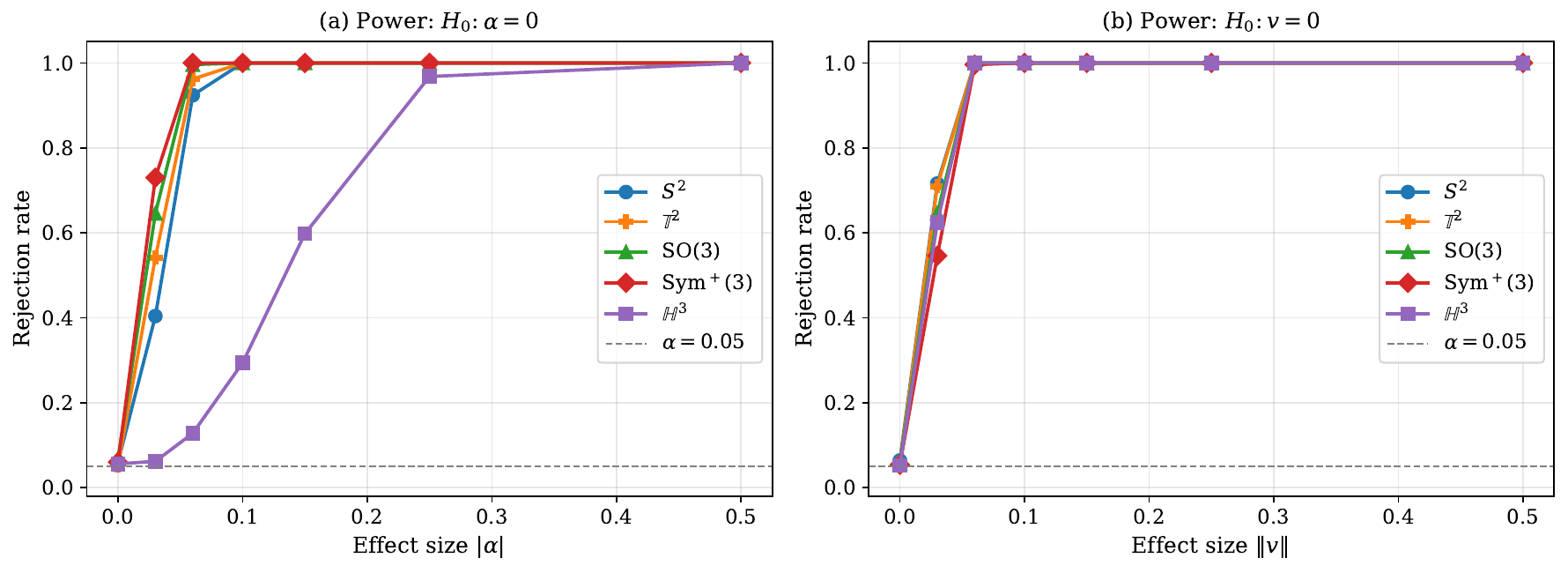}
\caption{Power curves of ROLS hypothesis tests ($n=200$, $\tau=0.15$, $500$ replications) on five manifolds. (a) $t$-test for $H_0\colon \alpha=0$. $\mathbb{H}^3$ shows reduced power because the Poincar\'e ball conformal factor inflates the effective Riemannian noise relative to the predictor variation. (b) Wald test for $H_0\colon v=0$, which converges quickly across all manifolds since the model is linear in $v$ in the tangent space.
Dashed line at nominal $\alpha=0.05$.}
\label{fig:power}
\end{figure}

Figure~\ref{fig:power} shows the power curves.
Table~\ref{tab:bootstrap} reports Type~I error calibration.

\begin{table}[h]
\centering
\caption{Type~I error of the $v$-Wald test at nominal level $0.05$ ($300$
trials, $B{=}199$ bootstrap replicates, $\tau=0.5$).
Asym.: asymptotic Wald test (Theorem~\ref{thm:inference}) with the model-based covariance.
Pairs-$t$: nonparametric pairs bootstrap with full Fr\'echet-mean refit on
each replicate, with the observed and replicate statistics both studentized
by the model-based covariance and aligned by parallel transport.
Pairs-emp: the same bootstrap with the observed statistic studentized by
the empirical bootstrap covariance.
Resid.: fixed-base residual bootstrap of Appendix~\ref{sec:S6},
conditioning on the empirical Fr\'echet mean and resampling tangent-space
residuals, targeting the oracle $\Sigma_{\mathrm{tan}}$ component.}
\label{tab:bootstrap}
\footnotesize
\setlength{\tabcolsep}{3pt}
\resizebox{\textwidth}{!}{%
\begin{tabular}{lcccccccccccccccc}
\toprule
 & \multicolumn{4}{c}{$n=50$} & \multicolumn{4}{c}{$n=100$} & \multicolumn{4}{c}{$n=200$} & \multicolumn{4}{c}{$n=500$} \\
\cmidrule(lr){2-5} \cmidrule(lr){6-9} \cmidrule(lr){10-13} \cmidrule(lr){14-17}
Manifold & Asym. & Pairs-$t$ & Pairs-emp & Resid. & Asym. & Pairs-$t$ & Pairs-emp & Resid. & Asym. & Pairs-$t$ & Pairs-emp & Resid. & Asym. & Pairs-$t$ & Pairs-emp & Resid. \\
\midrule
$S^2$ & $.063$ & $.063$ & $.097$ & $.057$ & $.047$ & $.043$ & $.057$ & $.043$ & $.053$ & $.043$ & $.063$ & $.040$ & $.060$ & $.053$ & $.063$ & $.053$ \\
$\mathbb{T}^2$ & $.037$ & $.027$ & $.063$ & $.023$ & $.053$ & $.053$ & $.073$ & $.043$ & $.047$ & $.057$ & $.070$ & $.050$ & $.057$ & $.043$ & $.043$ & $.047$ \\
$\mathbb{H}^3$ & $.053$ & $.053$ & $.097$ & $.040$ & $.057$ & $.033$ & $.060$ & $.033$ & $.027$ & $.033$ & $.053$ & $.030$ & $.060$ & $.050$ & $.067$ & $.043$ \\
$\mathrm{SO}(3)$ & $.063$ & $.060$ & $.123$ & $.037$ & $.070$ & $.060$ & $.087$ & $.060$ & $.063$ & $.053$ & $.067$ & $.053$ & $.057$ & $.040$ & $.053$ & $.037$ \\
$\mathrm{Sym}^+(3)$ & $.053$ & $.040$ & $.163$ & $.033$ & $.053$ & $.043$ & $.113$ & $.040$ & $.047$ & $.037$ & $.083$ & $.040$ & $.037$ & $.033$ & $.060$ & $.033$ \\
\bottomrule
\end{tabular}%
}
\end{table}

\begin{figure}[h]
\centering
\includegraphics[width=\textwidth]{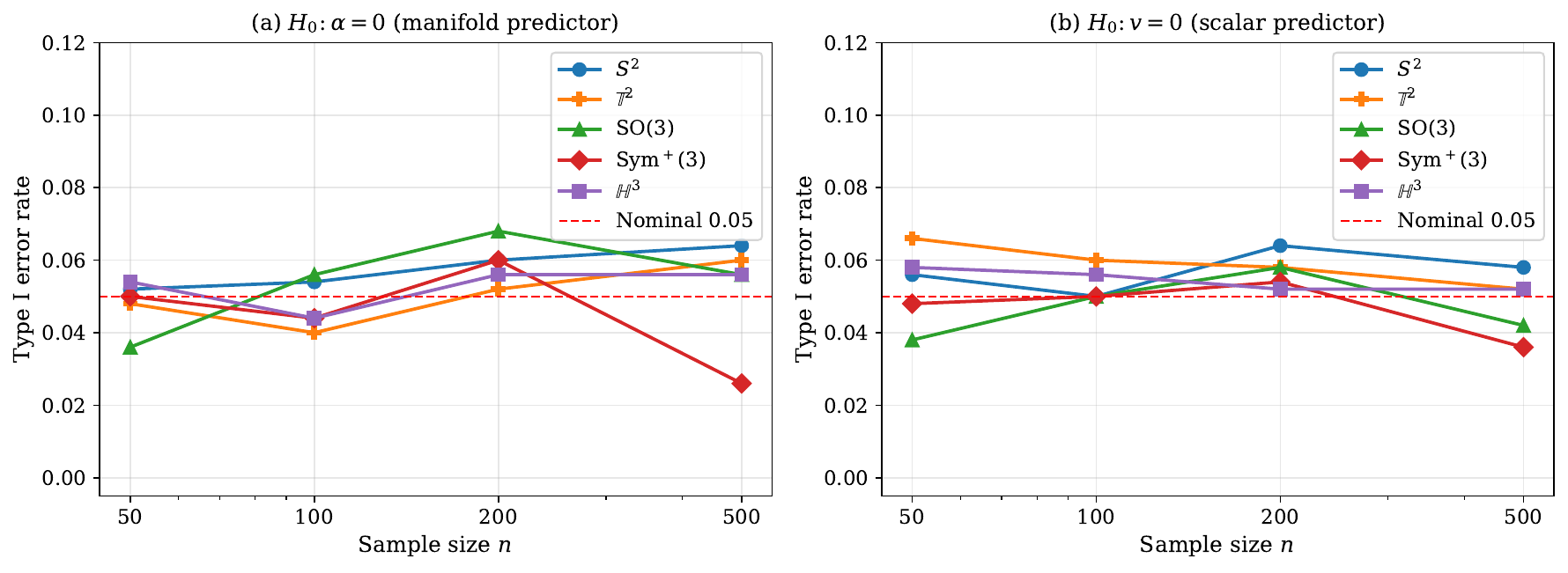}
\caption{Type~I error of the asymptotic ROLS tests at $\tau=0.15$ across $500$ trials per cell on five manifolds. (a) $H_0\colon \alpha=0$ (manifold predictor, $t$-test). (b) $H_0\colon v=0$ (scalar predictor, Wald test).
Of the $40$ entries, $39$ lie within the binomial $95\%$ envelope of nominal $0.05$ (red dashed).
The corresponding bootstrap-versus-asymptotic comparison at $\tau=0.5$ is in Table~\ref{tab:bootstrap}.}
\label{fig:typeI-tau015}
\end{figure}

In panel~(a) of Figure~\ref{fig:power}, the $t$-test for $H_0\colon \alpha=0$ reaches near-saturation by $|\alpha|=0.06$ on $\mathrm{SO}(3)$, $\mathrm{Sym}^+(3)$, and $\mathbb{T}^2$ ($\geq 96\%$ rejection), with $S^2$ at $92\%$ and $\mathbb{H}^3$ trailing at $13\%$. $\mathbb{H}^3$ requires $|\alpha|\!\approx\!0.25$ to attain $95\%$ rejection, because the Poincar\'e conformal factor at the response base point inflates the effective Riemannian noise by roughly $2\times$ while simultaneously reducing the Riemannian variation of the predictor relative to its ambient spread; the resulting effective signal-to-noise ratio is about an order of magnitude smaller than on $S^2$.
In panel~(b), the Wald test for $H_0\colon v=0$ saturates by $\|v\|=0.06$ on all five manifolds, since the model is linear in $v$ in tangent coordinates and this coefficient is unaffected by the curvature-induced distortions of the manifold predictor.
In Table~\ref{tab:bootstrap}, Pairs-$t$ stays at or below $0.063$, against the nominal $0.05$, in all $20$ settings.
Pairs-emp studentizes the observed statistic with the empirical bootstrap covariance and each replicate with the model-based covariance.
The resulting errors are largest on $\mathrm{Sym}^+(3)$, where $d=6$, for $n\le200$.
On $S^2$, $\mathbb T^2$, and $\mathrm{Sym}^+(3)$ the pairs-bootstrap standard error is $1.02$, $1.00$, and $0.97$ times the fixed-base standard error ($n=200$, $B=200$, $60$ trials).
A companion check at $\tau=0.15$, the noise level of Figure~\ref{fig:power}, uses $500$ trials per cell.
All $40$ rejection rates of the asymptotic tests lie in $[0.026, 0.068]$, on all five manifolds, including $\mathbb{T}^2$.
The residual bootstrap of Appendix~\ref{sec:S6} estimates $\Sigma_{\mathrm{tan}}$.
On the flat torus $\Sigma_{\mathrm{curv}}=0$, so that estimate is the full variance, in agreement with the $\mathbb{T}^2$ row of Table~\ref{tab:bootstrap}.
Figure~\ref{fig:typeI-tau015} summarizes the $\tau=0.15$ companion check, plotting Type~I rejection rate against $n$ for each manifold and each null hypothesis.

\subsection{Robustness Under Model Misspecification}
\label{sec:S7-misspec}

\begin{figure}[h]
\centering
\includegraphics[width=0.55\textwidth]{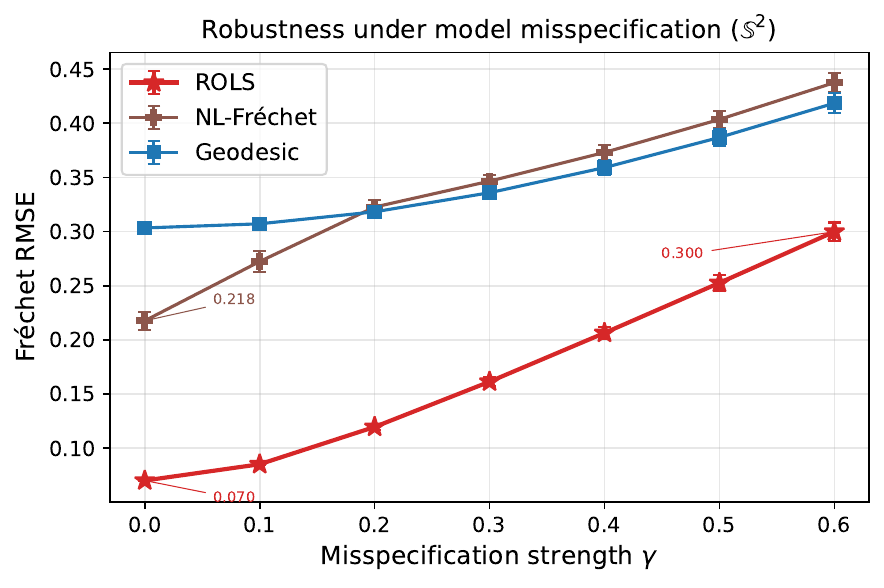}
\caption{ROLS RMSE under increasing model misspecification ($\gamma$) on $S^2$.
At $\gamma = 0$, ROLS has RMSE $0.070$ and NL-Fr\'echet has RMSE $0.218$.
At $\gamma = 0.6$ the two RMSEs are $0.300$ and $0.438$.}
\label{fig:misspec}
\end{figure}

To probe ROLS's sensitivity to the tangent-space linearity assumption, we generate data on $S^2$ via the nonlinear tangent-space model
\[
  \Log{\bar Y}(Y_i) = \alpha\,u_i + z_i\,v + \gamma\,\|u_i\|^2\,\hat n
  + \varepsilon_i,
\]
where $\hat n$ is a fixed tangent direction and $\gamma \in \{0, 0.1, 0.2, 0.3, 0.4, 0.5, 0.6\}$ controls the degree of model misspecification ($\gamma = 0$ is correctly specified).
We compare ROLS, NL-Fr\'echet (kernel regression), and geodesic regression over 30 trials with $n = 300$.
Figure~\ref{fig:misspec} shows that ROLS has lower RMSE than both baselines at every value of $\gamma$, and its margin over the better baseline narrows as $\gamma$ grows beyond $0.2$.

In this design the omitted term contributes a squared bias of order $\gamma^2$, the misspecification term $b^2$ in Theorem~\ref{thm:consistency}.

\paragraph{Inference under misspecification.}
We rerun the design of Figure~\ref{fig:misspec} with $\alpha=0$, a second scalar covariate, $n=300$, and $500$ trials.
Under $b>0$, $\hat\alpha$ stays centered at its projection value $0$, but its sampling standard deviation is about twice the model standard error, because the omitted quadratic term adds design-correlated variance.
Table~\ref{tab:misspec-calib} reports the Type~I error of the model-based Wald test and of the same test with a sandwich standard error, together with the rejection rate of the targeted lack-of-fit diagnostic.
The diagnostic appends $\|u\|^2$, $z_1^2$, and $z_2^2$ to the tangent regression as additional scalar covariates and tests their coefficients jointly.

\begin{table}[h]
\centering
\caption{Rejection rates at nominal level $0.05$ under the misspecification
of Figure~\ref{fig:misspec} ($H_0\colon\alpha=0$, $n=300$, $500$ trials).}
\label{tab:misspec-calib}
\footnotesize
\begin{tabular}{lccccccc}
\toprule
$\gamma$ & $0$ & $0.1$ & $0.2$ & $0.3$ & $0.4$ & $0.5$ & $0.6$ \\
\midrule
Model Wald & $.046$ & $.092$ & $.208$ & $.266$ & $.290$ & $.330$ & $.360$ \\
Robust $\alpha$ & $.050$ & $.038$ & $.074$ & $.076$ & $.054$ & $.062$ & $.084$ \\
Lack-of-fit & $.050$ & $1.0$ & $1.0$ & $1.0$ & $1.0$ & $1.0$ & $1.0$ \\
\bottomrule
\end{tabular}
\end{table}

The leverage-corrected heteroskedasticity-consistent (HC3) version~\citep{MacKinnonWhite1985} of the sandwich covariance of Section~\ref{sec:stats-inference} replaces $r_i$ in $\hat\Omega$ by $(I_d-H_{ii})^{-1}r_i$ with $H_{ii}=\bb Z_i^\top\hat A^{-1}\bb Z_i$.
Table~\ref{tab:hc3-calib} applies this estimator to all blocks.
The omitted term is centered, standardized, and placed either in $u$ or in $z_2$ ($n=200$, $1500$ trials).
The block targeted by the omitted term shows the dominant distortion, reaching $0.32$ for $\alpha$ and $0.37$ for $v_2$, while the untargeted block stays much closer to nominal.
HC3 brings both targeted blocks back to $0.05$--$0.07$.

\begin{table}[h]
\centering
\caption{Type~I error at nominal level $0.05$ of the model-based and HC3
Wald tests when the omitted term involves $u$ or $z_2$ ($n=200$, $1500$
trials).}
\label{tab:hc3-calib}
\footnotesize
\begin{tabular}{lllcccc}
\toprule
Omitted term & Block & Test & $\gamma=0$ & $0.1$ & $0.2$ & $0.3$ \\
\midrule
in $u$ & $\alpha$ & model & $.043$ & $.131$ & $.252$ & $.324$ \\
in $u$ & $\alpha$ & HC3 & $.043$ & $.063$ & $.055$ & $.069$ \\
in $u$ & $v_2$ & model & $.052$ & $.055$ & $.065$ & $.077$ \\
\addlinespace
in $z_2$ & $v_2$ & model & $.052$ & $.168$ & $.319$ & $.369$ \\
in $z_2$ & $v_2$ & HC3 & $.051$ & $.059$ & $.066$ & $.063$ \\
in $z_2$ & $\alpha$ & model & $.043$ & $.043$ & $.050$ & $.053$ \\
\bottomrule
\end{tabular}
\end{table}

The lack-of-fit diagnostic is targeted.
Its basis matches the alternative of Figure~\ref{fig:misspec}, and its power against alternatives outside the basis is lower at the same standardized strength (Table~\ref{tab:lof-power}).

\begin{table}[h]
\centering
\caption{Rejection rate of the targeted lack-of-fit diagnostic at nominal
level $0.05$.}
\label{tab:lof-power}
\footnotesize
\begin{tabular}{lccccc}
\toprule
Alternative & In basis & $\gamma=0$ & $0.1$ & $0.2$ & $0.3$ \\
\midrule
Quadratic in $u$ (Figure~\ref{fig:misspec}) & yes & $.045$ & $1.000$ & $.999$ & $.997$ \\
Cubic in $z_1$ & no & $.045$ & $.228$ & $.501$ & $.675$ \\
$u\times z_1$ interaction & no & $.045$ & $.189$ & $.353$ & $.475$ \\
\bottomrule
\end{tabular}
\end{table}

\subsection{Sensitivity to the Base Point}
\label{sec:S7-base}

On $\mathrm{Sym}^+(3)$ with the affine-invariant metric ($200$ trials), we refit the response and the predictor at four kinds of base point, namely the Fr\'echet mean, the Fr\'echet mean displaced by geodesic distance $\delta\cdot(\text{spread})$ in a random direction, the projected ambient mean, and a random observation.
Table~\ref{tab:base-point} reports the coefficient error $|\hat\alpha-\alpha^*|$ and the held-out Fr\'echet RMSE.
The coefficient error stays near the Fr\'echet-mean value at $\delta=0.1$ and rises once the base point moves by a quarter of the spread.
In the expansion of Appendix~\ref{sec:S4}, the population coefficient $\bb\beta^*(\theta)$ depends on the base point through $J_\theta$, so a displacement of that size changes the target of the estimator.

\begin{table}[h]
\centering
\caption{Coefficient error and held-out Fr\'echet RMSE on
$\mathrm{Sym}^+(3)$ under different base points ($200$ trials).}
\label{tab:base-point}
\footnotesize
\begin{tabular}{lccccccc}
\toprule
Base & Fr\'echet & $\delta=0.1$ & $\delta=0.25$ & $\delta=0.5$ & $\delta=1.0$ & Ambient & Random \\
\midrule
Coeff.\ error & $0.0153$ & $0.0158$ & $0.0373$ & $0.1116$ & $0.285$ & $0.0217$ & $0.256$ \\
RMSE & $0.495$ & $0.501$ & $0.533$ & $0.624$ & $0.865$ & $0.500$ & $0.858$ \\
\bottomrule
\end{tabular}
\end{table}

\subsection{Higher Tangent Dimension}
\label{sec:S7-highdim}

We grow the tangent dimension on $\mathrm{Sym}^+(k)$ up to $d=55$ with one manifold predictor and one scalar covariate.
At $d=21$ the log--log slope of the parameter error is $-0.495$.
The leading variance term of Theorem~\ref{thm:consistency} is $C\sigma^2(p+qd)/(n\lambda_*)$, so the rate depends on $n/(p+qd)$. The geometric factor is $\tilde\eta^2$, and the rate is the same.
Table~\ref{tab:highdim} reports the Type~I error of the $v$-Wald test.
Calibration follows $n/d$, because the $(p+qd)$-dimensional Gram matrix is poorly estimated when $n/d$ is small, as the $\chi^2$ entry of the $d=6$ row shows at the smallest sample size in that row, $n=11$.
The $F$ reference brings that entry to $.062$ and changes the others by at most $.006$.

\begin{table}[h]
\centering
\caption{Type~I error of the $v$-Wald test on $\mathrm{Sym}^+(k)$ at
nominal level $0.05$ ($500$ trials, $n=\max\{\mathrm{round}(md),\,d+5\}$ for $m\in\{1.5,3,6,12\}$), with the $\chi^2_d$ and $F$ references.}
\label{tab:highdim}
\footnotesize
\begin{tabular}{lcccc}
\toprule
$d$ ($\chi^2$ / $F$) & $n\approx1.5d$ & $n\approx3d$ & $n\approx6d$ & $n\approx12d$ \\
\midrule
$6$ & $.086$ / $.062$ & $.062$ / $.056$ & $.064$ / $.062$ & $.062$ / $.060$ \\
$55$ & $.060$ / $.058$ & $.044$ / $.044$ & $.040$ / $.040$ & $.040$ / $.040$ \\
\bottomrule
\end{tabular}
\end{table}

\subsection{Computational Cost}
\label{sec:S7-cost}

ROLS evaluates the logarithm and transport maps only at the $p+1$ estimated means, which gives $O(np)$ map evaluations, whereas kernel and Fr\'echet regression need $O(n^2)$ pairwise evaluations.
A full refit on $\mathrm{Sym}^+(3)$ with $p=q=1$ takes $0.002$\,s at $n=100$ and $0.14$\,s at $n=10^4$ (median over repeated single-threaded runs on one machine; timings depend on the implementation).
A batch update requires a full refit, and there is no exact $O(1)$ streaming update, because a change in any of the $p+1$ means changes the logarithm maps and transports of all earlier observations.
The Fr\'echet means themselves can be updated recursively with a Karcher estimator~\citep{Ho2013,Cheng2016}, but an approximate online ROLS would need a separate analysis.

\section{Real-Data Case Study Details}
\label{sec:S8}

This appendix provides complete descriptions of all three real-data case studies in Section~\ref{sec:casestudies}.

For every dataset the ROLS model specializes the canonical form
\begin{equation}\label{eq:rols-canonical-app}
  \Log{\FM{Y}}(Y_i)
  = \sum_{j=1}^p \alpha_j \, \PT{\FM{X}_j}{\FM{Y}}\!\bigl(\Log{\FM{X}_j}(X_{ij})\bigr)
  + V (z_i - \bar z) + \varepsilon_i,
\end{equation}
where $\FM{Y}, \FM{X}_j$ are empirical Fr\'echet means, $\bar z=\E z$, and $V \in \R^{d \times q}$ collects the scalar-covariate tangent vectors $v_k\in\Tan{\FM Y}$ as columns.
All reported RMSEs use the Fr\'echet distance on the respective manifold as the evaluation metric.
Nominal categorical covariates are one-hot encoded with the first level dropped as the reference category, and periodic covariates are $(\sin, \cos)$-encoded.

Table~\ref{tab:hc3-casestudy} reports HC3 Wald $p$-values at the full-data fit.
These tests treat observations as independent.
For DTI-CC, the $p$-values that respect within-subject clustering are the subject-blocked permutation $p$-values in Table~\ref{tab:fulldata_inference}.

\begin{table}[h]
  \centering
  \caption{HC3 Wald $p$-values at the full-data fit for the three case studies, treating observations as independent. Predictors are as in Table~\ref{tab:fulldata_inference}.}
  \label{tab:hc3-casestudy}
  \footnotesize
  \begin{tabular}{@{}lc@{\hskip 16pt}lc@{\hskip 16pt}lc@{}}
  \toprule
  \multicolumn{2}{c}{\textbf{DTI-CC}}
   & \multicolumn{2}{c}{\textbf{Protein backbone}}
   & \multicolumn{2}{c}{\textbf{Wind}} \\
  \cmidrule(r){1-2} \cmidrule(lr){3-4} \cmidrule(l){5-6}
  Predictor & HC3 & Predictor & HC3 & Predictor & HC3 \\
  \midrule
  Tensor neigh.\ 1 & $<\!10^{-5}$ & Offset $-2$ & $<\!10^{-5}$ & Grid pt.\ 1 & $<\!10^{-5}$ \\
  Tensor neigh.\ 2 & $<\!10^{-5}$ & Offset $-1$ & $<\!10^{-5}$ & Grid pt.\ 2 & $<\!10^{-5}$ \\
  CC subregion & $<\!10^{-5}$ & Offset $+1$ & $<\!10^{-5}$ & Grid pt.\ 3 & $<\!10^{-5}$ \\
  Age & $<\!10^{-5}$ & Kidera PC1 & $<\!10^{-5}$ & Pressure & $0.700$ \\
  Sex & $3.1{\times}10^{-5}$ & Kidera PC2 & $<\!10^{-5}$ & Temperature & $<\!10^{-5}$ \\
  ASD diagnosis & $1.3{\times}10^{-3}$ & Kidera PC3 & $0.118$ & Season & $0.007$ \\
  FIQ & $4.6{\times}10^{-2}$ & $\alpha$-helix & $<\!10^{-5}$ & & \\
   & & Solvent-accessibility & $1.8{\times}10^{-3}$ & & \\
  \bottomrule
  \end{tabular}
\end{table}

\subsection{Diffusion Tensor Field on the Corpus Callosum in
  $\mathrm{Sym}^+(3)$}
\label{sec:S8-dti}

\paragraph{Scientific question.}
Diffusion tensors summarize local water-diffusion anisotropy in white matter and are naturally represented as $3\times3$ symmetric positive-definite (SPD) matrices under the affine-invariant metric~\citep{Pennec2006}.
The corpus callosum is the largest commissural tract, and its tensor field varies systematically along the genu--body axis.
We ask whether a local tensor can be predicted from neighboring along-tract tensors after adjusting for subject demographics and anatomical subregion.

\paragraph{Data and preprocessing.}
We use ABIDE-II diffusion magnetic resonance imaging (MRI) data~\citep{DiMartino2014} from three sites with compatible single-shell acquisitions.
Preprocessing comprises motion and eddy-current correction, tensor fitting, atlas-based corpus-callosum extraction, site harmonization in the matrix-log embedding, and removal of subjects with excessive head motion.
The final cohort contains $141$ subjects and $n=112{,}694$ corpus-callosum voxels, with approximately $800$ voxels sampled per subject.

\paragraph{Variables and model.}
The response $Y_i\in\mathrm{Sym}^+(3)$ is the diffusion tensor at a corpus-callosum voxel.
The two manifold predictors are along-tract neighbor tensors $X_{i1}$ and $X_{i2}$ sampled at $\pm2$\,mm by deterministic tractography.
Scalar covariates $z_i$ comprise age, sex, ASD diagnosis, FIQ, and a within-subject indicator for genu versus body.
The fitted ROLS model is the canonical specification \eqref{eq:rols-canonical-app} with $p=2$ manifold predictors and $q=5$ scalar covariates.

\paragraph{Riemannian structure and CV design.}
On $\mathrm{Sym}^+(3)$ with the affine-invariant metric, Log, Exp, and parallel transport admit closed forms via the matrix exponential.
Empirical Fr\'echet means lie in the interior of the cone, so injectivity at $\FM{Y},\FM{X}_1,\FM{X}_2$ holds without restriction and the hypotheses of Theorem~\ref{thm:cutlocus} are satisfied.
Cross-validation is subject-blocked so that voxels from the same subject do not co-occur in training and test folds, which prevents within-subject leakage.
Because voxel observations are clustered within subjects, we report both asymptotic Wald tests and block-restricted permutation $p$-values with $B=500$ and subject-level exchangeability blocks; the scheme shuffles between-subject covariates at the subject level and within-subject covariates within subject blocks.

\paragraph{Findings.}
ROLS reduces Fr\'echet RMSE from $1.128$ for the strongest scalar-only baseline to $.400$ when along-tract tensor predictors are included (Table~\ref{tab:casestudy}).
Both manifold predictors attain the permutation floor $p=0.002$, and the genu-vs-body indicator is also significant at $p=0.002$ (Table~\ref{tab:fulldata_inference}).
The analytic Wald test treats the $112{,}694$ voxels as independent and gives $p$-values $<10^{-5}$, $1.9\times10^{-5}$, $5.1\times10^{-3}$, and $1.6\times10^{-3}$ for age, sex, ASD diagnosis, and FIQ.
HC3 uses the same independence assumption and gives $<10^{-5}$, $3.1\times10^{-5}$, $1.3\times10^{-3}$, and $4.6\times10^{-2}$.
The subject-blocked permutation $p$-value for FIQ is $0.447$.
Benjamini--Hochberg adjustment of the permutation $p$-values is $0.005$ for the two tensor neighbors and the genu-vs-body indicator, and at least $0.42$ for the four demographic covariates.
The lack-of-fit diagnostic of Appendix~\ref{sec:S7-misspec} attributes about $10\%$ of the residual variation to omitted quadratic terms.

\subsection{Protein Backbone Dihedral Angles on $\mathbb T^2$}
\label{sec:S8-protein}

\paragraph{Scientific question.}
Protein backbone conformation depends on the residue-level dihedral angles $(\varphi,\psi)$ on the flat torus $\mathbb T^2=S^1\times S^1$.
The torus is flat, so the tangent-space fit has no curvature contribution, and the remaining geometric feature is the periodicity.
We ask whether local backbone geometry is driven primarily by neighboring dihedral angles, residue chemistry, or basin-level secondary-structure indicators.

\paragraph{Data and preprocessing.}
We extract $n=6{,}141$ residue windows from 30 high-resolution chains in the Protein Data Bank~\citep{Berman2000}.
For each central residue with a complete local window we record its dihedral pair, neighboring pairs at offsets $-2,-1,+1$, the Kidera physicochemical descriptors of the central residue, an $\alpha$-helix-basin indicator derived from secondary-structure assignment, and a solvent-accessibility proxy based on neighboring $\mathrm C_\alpha$ distances.

\paragraph{Variables and model.}
The response $Y_i=(\varphi_i,\psi_i)\in\mathbb T^2$ is the central-residue dihedral pair.
The three manifold predictors are neighboring dihedral pairs at offsets $-2,-1,+1$; the five scalar predictors are Kidera PC1--PC3, the $\alpha$-helix indicator, and the solvent-accessibility proxy.
The fitted ROLS model is the canonical specification \eqref{eq:rols-canonical-app} with $p=3$ manifold predictors and $q=5$ scalar covariates.

\paragraph{Riemannian structure and CV design.}
On $\mathbb T^2$ the logarithm and exponential maps are component-wise shortest-arc operations on $S^1$ and parallel transport is the identity, so \eqref{eq:rols-canonical-app} reduces to vector addition in $\R^2$-tangent representation.
The injectivity radius is $\pi$ in each $S^1$ component; the empirical Fr\'echet means are well-defined, and the shortest-arc logarithm is single-valued off the cut locus.
Prediction is evaluated by three-fold cross-validation repeated three times with Fr\'echet RMSE in radians, and the fitted model is compared with the Fr\'echet mean, geodesic regression, scalar Fr\'echet regression, single-index Fr\'echet regression, nonlinear Fr\'echet regression, and a manifold-only ROLS ablation.

\paragraph{Findings.}
The Ramachandran conditional distribution has distinct $\alpha$-helix and $\beta$-sheet basins.
Full ROLS attains Fr\'echet RMSE $1.177$, geodesic regression on the scalar predictors attains $1.215$, and manifold-only ROLS attains $1.395$ (Table~\ref{tab:casestudy}).
The neighboring angles and the residue-level descriptors each reduce the error relative to the fit that uses the other set alone.
All three neighboring dihedral predictors, Kidera PC1--PC2, the $\alpha$-helix indicator, and the solvent-accessibility proxy are significant in Table~\ref{tab:fulldata_inference}.
Kidera PC3, kept as a null control, has Wald $p=0.345$ and HC3 $p=0.118$ at the full-data fit.
Benjamini--Hochberg adjustment leaves Kidera PC3 at $0.345$ and the solvent-accessibility proxy at $3.2\times10^{-4}$.
The lack-of-fit diagnostic attributes about $1.6\%$ of the residual variation to omitted quadratic terms.
On the Ramachandran plot (Figure~\ref{fig:protein_ramachandran_main}) the full ROLS estimator places predictions in both the $\alpha$-helix and $\beta$-sheet basins. Scalar-only methods concentrate in a point or narrow clusters, nonlinear Fr\'echet smooths across the two dominant basins, and manifold-only ROLS spreads predictions over the $\varphi<0$ half-plane without separating the two basins.

\subsection{\texorpdfstring{Atmospheric Winds on $S^2$}{Atmospheric Winds on S2}}
\label{sec:S8-wind}

\paragraph{Scientific question.}
Mid-latitude weather over central Europe is organized by synoptic systems that translate eastward across the continent, so winds at nearby grid points should share a common spatial signal once seasonal and thermodynamic covariates are removed.
We ask whether the wind at a central European grid point is predictable from the winds at three neighboring grid points after adjusting for sea-level pressure, temperature, and day-of-year.

\paragraph{Data and preprocessing.}
We use $850$\,hPa winds and surface meteorological covariates from the NCEP/NCAR Reanalysis~1~\citep{Kalnay1996} for the year 2023, sampled at four grid points ($50^\circ$N, $10^\circ$E for the response; $52.5^\circ$N, $5^\circ$E; $47.5^\circ$N, $5^\circ$E; and $50^\circ$N, $15^\circ$E for the predictors) at daily resolution.
Each daily horizontal wind $(u,v)$ is mapped to the point on $S^2$ with azimuth $\mathrm{atan2}(v,u)$ and polar angle $(\pi/2)\,s/s_{\max}$, where $s$ is the wind speed and $s_{\max}$ its annual maximum at that grid point.
Sea-level pressure and near-surface temperature are extracted from the same reanalysis product at the response grid point.

\paragraph{Variables and model.}
The response $Y_i\in S^2$ is the daily wind at the response grid point.
The three manifold predictors are the winds at the three neighboring grid points, each in $S^2$.
Day-of-year is encoded by two orthogonal features $(\sin 2\pi\,\text{doy}/365,\cos 2\pi\,\text{doy}/365)$, yielding four scalar predictors (pressure, temperature, and the two seasonal features).
A combined ``season'' $p$-value pools the two seasonal-feature coefficient tests via the joint Wald test on these two coefficients.
The fitted ROLS model is the canonical specification \eqref{eq:rols-canonical-app} with $p=3$ and $q=4$.

\paragraph{Riemannian structure and CV design.}
$S^2$ is equipped with the round metric inherited from $\R^3$.
The exponential map uses the Rodrigues formula, the logarithm map is its inverse, and parallel transport is rotation about the normal to the geodesic plane.
The injectivity radius is $\pi$ and all daily wind points lie in the same hemisphere, so the injectivity condition~\eqref{eq:inj-cond} of Theorem~\ref{thm:cutlocus} is satisfied at every empirical Fr\'echet mean.
The 365 daily observations are strongly autocorrelated because days within the same week share the same synoptic weather system, so random day-level folds would leak synoptic information across train and test.
We therefore use blocked-temporal CV in the sense of \citet{BergmeirBenitez2012}, with five contiguous day-blocks, no random shuffling, and a single pass.

\paragraph{Findings.}
The sample has $n=365$ days and manifold dimension $d=2$.
The four grid points share synoptic systems, and including the neighboring winds produces the RMSE reduction in Table~\ref{tab:casestudy}.
All three predictor winds are significant in Table~\ref{tab:fulldata_inference}.
Temperature and the combined seasonal factor remain significant after Benjamini--Hochberg adjustment and under HC3.
Sea-level pressure does not.
The directional signal comes from the neighboring winds, and the seasonal and thermodynamic covariates account for smaller adjustments.

\newpage
\section*{NeurIPS Paper Checklist}

\begin{enumerate}

\item {\bf Claims}
    \item[] Question: Do the main claims made in the abstract and introduction accurately reflect the paper's contributions and scope?
    \item[] Answer: \answerYes{}
    \item[] Justification: The abstract and introduction state the cut-locus well-definedness condition (Theorem~\ref{thm:cutlocus}), a closed-form estimator (Theorem~\ref{thm:normal}), convergence rates (Theorem~\ref{thm:consistency}), and asymptotic inference (Theorem~\ref{thm:inference}).
    All are proved in the appendix and validated empirically in Sections~\ref{sec:sims}--\ref{sec:casestudies}.

\item {\bf Limitations}
    \item[] Question: Does the paper discuss the limitations of the work performed by the authors?
    \item[] Answer: \answerYes{}
    \item[] Justification: Sections~\ref{sec:model}--\ref{sec:conclusion} discuss limitations including the scalar coefficient restriction ($\alpha_j \in \mathbb{R}$ rather than a full linear map), the injectivity and tangent-linearity assumptions, the need for a fixed transport convention, and the requirement for closed-form geometric primitives, together with the same-manifold restriction, retraction-based approximations, multimodal responses, fixed dimension, and batch updates.

\item {\bf Theory assumptions and proofs}
    \item[] Question: For each theoretical result, does the paper provide the full set of assumptions and a complete (and correct) proof?
    \item[] Answer: \answerYes{}
    \item[] Justification: All theorems are stated with explicit assumptions (Assumption~\ref{ass:main} and conditions in each theorem statement).
    Complete proofs are given in the appendix.

    \item {\bf Experimental result reproducibility}
    \item[] Question: Does the paper fully disclose all the information needed to reproduce the main experimental results of the paper to the extent that it affects the main claims and/or conclusions of the paper (regardless of whether the code and data are provided or not)?
    \item[] Answer: \answerYes{}
    \item[] Justification: The model and data-generating process are specified in Eq.~\eqref{eq:model} (Model) and described in detail in Appendix~\ref{sec:S7-dgp} with all parameter values.
    Section~\ref{sec:model-estimation} gives the complete estimation procedure.
    Appendix~\ref{sec:S7} provides full experimental details for all simulations and Appendix~\ref{sec:S8} for all case studies, including data sources, preprocessing, and hyperparameters.

\item {\bf Open access to data and code}
    \item[] Question: Does the paper provide open access to the data and code, with sufficient instructions to faithfully reproduce the main experimental results, as described in supplemental material?
    \item[] Answer: \answerYes{}
    \item[] Justification: The core computer programs implementing ROLS are provided at \url{https://github.com/XiaoyuChenPub/ROLS}.
    The paper provides algorithmic details and experimental settings in the main text and appendix, and all datasets used are publicly available and cited.

\item {\bf Experimental setting/details}
    \item[] Question: Does the paper specify all the training and test details (e.g., data splits, hyperparameters, how they were chosen, type of optimizer) necessary to understand the results?
    \item[] Answer: \answerYes{}
    \item[] Justification: Section~\ref{sec:sims} specifies default parameters ($n$, $p$, $q$, $\tau$, true coefficients).
    Appendix~\ref{sec:S8} details data splits (subject-blocked, repeated, and blocked-temporal CV as appropriate) and all preprocessing steps.

\item {\bf Experiment statistical significance}
    \item[] Question: Does the paper report error bars suitably and correctly defined or other appropriate information about the statistical significance of the experiments?
    \item[] Answer: \answerYes{}
    \item[] Justification: Simulation results report mean $\pm$ standard deviation across Monte Carlo trials, with 100 trials for the prediction-accuracy and convergence-rate experiments (Tables~\ref{tab:comp-sigma-q},~\ref{tab:conv-slopes}), 500 trials for Type~I error and power, 300 trials for the bootstrap calibration (Table~\ref{tab:bootstrap}), and 30 trials for the misspecification probe.
    Case study RMSEs in Table~\ref{tab:casestudy} report mean $\pm$ standard deviation across cross-validation folds.
    Error bars in Figure~\ref{fig:headline} are $\pm 1$ Monte Carlo standard deviation across trials.

\item {\bf Experiments compute resources}
    \item[] Question: For each experiment, does the paper provide sufficient information on the computer resources (type of compute workers, memory, time of execution) needed to reproduce the experiments?
    \item[] Answer: \answerNo{}
    \item[] Justification: The paper does not report specific compute resources.
    All experiments are lightweight (closed-form linear algebra on small-to-moderate datasets) and run on a standard laptop CPU in minutes.
    No GPU is required.
    Single-threaded timings of one ROLS fit are reported in Appendix~\ref{sec:S7-cost}.

\item {\bf Code of ethics}
    \item[] Question: Does the research conducted in the paper conform, in every respect, with the NeurIPS Code of Ethics \url{https://neurips.cc/public/EthicsGuidelines}?
    \item[] Answer: \answerYes{}
    \item[] Justification: The research is theoretical and methodological with no human subjects, no personally identifiable data, and no foreseeable negative societal impact.

\item {\bf Broader impacts}
    \item[] Question: Does the paper discuss both potential positive societal impacts and negative societal impacts of the work performed?
    \item[] Answer: \answerNA{}
    \item[] Justification: This is foundational statistical methodology for regression on Riemannian manifolds.
    It is not tied to a specific application that poses foreseeable societal risks.
    The case studies use publicly available scientific datasets.

\item {\bf Safeguards}
    \item[] Question: Does the paper describe safeguards that have been put in place for responsible release of data or models that have a high risk for misuse (e.g., pre-trained language models, image generators, or scraped datasets)?
    \item[] Answer: \answerNA{}
    \item[] Justification: The paper does not release pre-trained models, generative models, or scraped datasets.
    The method is a statistical estimator with no dual-use risk.

\item {\bf Licenses for existing assets}
    \item[] Question: Are the creators or original owners of assets (e.g., code, data, models), used in the paper, properly credited and are the license and terms of use explicitly mentioned and properly respected?
    \item[] Answer: \answerYes{}
    \item[] Justification: All datasets are cited with original references or public repositories (ABIDE-II diffusion MRI, Protein Data Bank, NCEP/NCAR Reanalysis~1).
    Data sources and preprocessing are provided in Appendix~\ref{sec:S8}.

\item {\bf New assets}
    \item[] Question: Are new assets introduced in the paper well documented and is the documentation provided alongside the assets?
    \item[] Answer: \answerYes{}
    \item[] Justification: The ROLS code is released at \url{https://github.com/XiaoyuChenPub/ROLS} under the CC BY-NC 4.0 license, with installation instructions and a usage example.

\item {\bf Crowdsourcing and research with human subjects}
    \item[] Question: For crowdsourcing experiments and research with human subjects, does the paper include the full text of instructions given to participants and screenshots, if applicable, as well as details about compensation (if any)?
    \item[] Answer: \answerNA{}
    \item[] Justification: The paper does not involve crowdsourcing or research with human subjects.

\item {\bf Institutional review board (IRB) approvals or equivalent for research with human subjects}
    \item[] Question: Does the paper describe potential risks incurred by study participants, whether such risks were disclosed to the subjects, and whether Institutional Review Board (IRB) approvals (or an equivalent approval/review based on the requirements of your country or institution) were obtained?
    \item[] Answer: \answerNA{}
    \item[] Justification: The paper does not involve human subjects research.

\item {\bf Declaration of LLM usage}
    \item[] Question: Does the paper describe the usage of LLMs if it is an important, original, or non-standard component of the core methods in this research?
    Note that if the LLM is used only for writing, editing, or formatting purposes and does \emph{not} impact the core methodology, scientific rigor, or originality of the research, declaration is not required.
    \item[] Answer: \answerNA{}
    \item[] Justification: LLMs are not used as a component of the core methodology.

\end{enumerate}

\end{document}